\documentclass[
    aps,
    prd,
    reprint,
    superscriptaddress,
    nofootinbib,
    longbibliography,
    floatfix
]{revtex4-2}

\usepackage{amsmath,amssymb, amsthm}

\usepackage{placeins}
\usepackage{tcolorbox}
\usepackage{scalerel}
\usepackage{dblfloatfix}
\usepackage[normalem]{ulem}
\usepackage[english]{babel}
\usepackage{graphicx}
\usepackage{dcolumn}
\usepackage{bm}
\usepackage{blindtext}
\usepackage{verbatim}
\usepackage{relsize}
\usepackage{mathrsfs}
\usepackage{musicography}
\usepackage{amsmath}
\usepackage{blindtext}
\usepackage{cancel}
\usepackage{physics}
\usepackage{epstopdf}
\usepackage{mathtools}
\usepackage{blindtext}
\usepackage{tensor}
\usepackage{color}
\usepackage{epsfig}
\usepackage{hyperref}
\usepackage{verbatim}
\usepackage{slashed}
\usepackage{dsfont}
\usepackage[english]{babel}
\usepackage{leftindex,enumerate}
\usepackage{enumitem, kantlipsum}

\newcommand{\proj}[2]{\left| {#1} \right\rangle\!\left\langle {#2} \right|}
\newcommand{\mf}{\mathsf}

\usepackage[normalem]{ulem}
\definecolor{eduviolet}{RGB}{105,45,160}

\newcommand{\eduadd}[1]{#1} 

\newcommand{\ii}{\mathrm{i}}

\newcommand{\tc}[1]{\textsc{#1}}

\definecolor{goldenrod}{rgb}{0.85, 0.65, 0.13}

\allowdisplaybreaks[1]

\begin{document}

\title{Ultraviolet structure of entanglement harvesting from energy density and other quadratic couplings}

\author{Matheus H. Zambianco}
\email{mhzambia@uwaterloo.ca}

\affiliation{Department of Applied Mathematics, University of Waterloo, Waterloo, Ontario, N2L 3G1, Canada}
\affiliation{Institute for Quantum Computing, University of Waterloo, Waterloo, Ontario, N2L 3G1, Canada}
\affiliation{Perimeter Institute for Theoretical Physics, Waterloo, Ontario, N2L 2Y5, Canada}

\author{Boris Ragula}
\email{bragula@uwaterloo.ca}

\affiliation{Department of Applied Mathematics, University of Waterloo, Waterloo, Ontario, N2L 3G1, Canada}
\affiliation{Institute for Quantum Computing, University of Waterloo, Waterloo, Ontario, N2L 3G1, Canada}
\affiliation{Perimeter Institute for Theoretical Physics, Waterloo, Ontario, N2L 2Y5, Canada}

\author{Eduardo Mart\'{i}n-Mart\'{i}nez}
\email{emartinmartinez@uwaterloo.ca}

\affiliation{Department of Applied Mathematics, University of Waterloo, Waterloo, Ontario, N2L 3G1, Canada}
\affiliation{Institute for Quantum Computing, University of Waterloo, Waterloo, Ontario, N2L 3G1, Canada}
\affiliation{Perimeter Institute for Theoretical Physics, Waterloo, Ontario, N2L 2Y5, Canada}

\begin{abstract}

We study entanglement harvesting with particle detectors coupled to the renormalized energy density of a massless scalar field. Our analysis identifies the distributional mechanism underlying the persistent ultraviolet divergences previously observed for quadratic detector couplings and \eduadd{shows} that the energy density coupling exhibits the same underlying structure. We show that these divergences are  controlled by the switching correlation at coincident interaction times \eduadd{and, for smeared detectors, by local derivatives of the spatial overlap}. Consequently, pointlike detectors are UV finite whenever the switchings do not overlap. \eduadd{For smooth, rapidly decreasing switchings, they are also UV finite at any fixed nonzero separation.} For Gaussian-smeared detectors, the harvesting correlations are automatically finite in 1+1 and 2+1 dimensions, while in higher dimensions the remaining divergences are removed by non-overlapping switchings.  Finally, we derive general expressions for arbitrary zero-mean Gaussian \eduadd{Hadamard} states and illustrate the formalism with a thermal field state.

\end{abstract}

\maketitle

\section{Introduction}

Entanglement harvesting refers to a family of relativistic quantum information protocols in
which localized probes extract entanglement from the correlations of a quantum
field. It is most often studied using particle-detector models, in particular
Unruh--DeWitt detectors~\cite{Unruh1976,DeWitt,Unruh-Wald,Schlicht,Jorma}.
The idea goes back to the work of Valentini and Reznik
\cite{Valentini1991,Reznik2003}, and has since become a useful operational
probe of the correlation structure of quantum fields, both in flat spacetime \cite{Valentini1991, Reznik2003, Pozas-Kerstjens:2015, Pozas2016, morotebalboa2026} 
and in curved backgrounds \cite{Nick,HarvestingBHLaura, KeithRobEdu2018, ericksonBH, ampEntBH2020, hectorMass, Membrere_2023, boris, MatheusAdamEdu2026, wurtz2026}
Although entanglement harvesting has not yet been experimentally observed, several proposals have explored how the protocol could be implemented, especially in superconducting circuits and Bose-Einstein condensates
\cite{Sabin2010Dynamics,Sabin2011Fermi,Sabin2012PastFuture,HarvestingMultiqubitCQE,Adametalexperimental,Gooding_2024,TalesFrancescoMarkus}.

Most of the entanglement harvesting literature focuses on setups where the detectors couple linearly with the field amplitude \cite{Reznik2003, Reznik2005BellInequalities, Salton:2014jaa, Pozas-Kerstjens:2015, Pozas2016, Petar}, or with the field's momentum \cite{AdamEduderivativeharvesting, ericksonBH}. Beyond those standard setups, interaction models with additional structure,
such as detectors coupled to fermionic fields \cite{Takagi, carol, FermionicHarvestingSCZ}
and to the gravitational field \cite{boris, HarvestingGravitationalWaves}, have also been considered.  In particular, Ref.~\cite{SachsMannEdu} (see also Ref.~\cite{sachs2018entanglement}) showed that, when the detectors couple to the squared field amplitude with a Gaussian switching, persistent ultraviolet (UV) divergences arise in the harvesting protocol and cannot be removed by spatially smearing the detectors. This raises the question of whether such divergences are a more general feature of entanglement harvesting protocols involving quadratic field operators whose definition requires a renormalization prescription.

On the other hand, since particle-detector models provide operational tools for probing the local structure of quantum fields \cite{Unruh1976, Jorma, fewster1}, it is natural to ask what entanglement harvesting protocols can reveal about field observables themselves. Among these
observables, the energy density is of particular interest. The expectation value of the renormalized stress-energy tensor is the source term in the semiclassical Einstein equations and therefore plays a central role at the interface between quantum field theory and gravity. Moreover, quantum fields are known to admit states that violate the weak energy condition, giving rise to average negative energy densities \cite{EGlaserJaffe1965,FordRoman1997, FORD_2010}. Such violations are closely connected with the possibility of exotic gravitational phenomena, including traversable wormholes \cite{MorrisThorneWormholes} and warp-drives \cite{WarpDriveAlcubierre}. 

Motivated by both the physical relevance of the energy density and the UV issues that can arise in entanglement harvesting with detectors coupled non-linearly to the field, in this work, we consider detector models in which the detectors couple to the renormalized energy density of the field. This interaction model was considered for the first time in Ref.~\cite{MatheusEduFirewallArxiv}, where the authors consider a single particle detector across the Rindler firewall.


When the field is prepared in the vacuum state, the energy-density coupling can exhibit persistent UV divergences, but these depend on the temporal and spatial sampling. For pointlike detectors at fixed nonzero separation, the correlation term is UV finite for smooth, rapidly decreasing switchings. At spatial coincidence, however, persistent divergences can arise for overlapping switchings. By carrying out the calculation in position space using distributional methods, we show that Gaussian spatial smearing makes the non-local correlation term UV finite in $1+1$ and $2+1$ dimensions. In higher dimensions, this term is generically divergent for overlapping switchings. We identify two exceptional detector separations for which the leading singularity cancels, making the correlation term finite in $3+1$ and $4+1$ dimensions. Compactly supported switchings with non-overlapping supports remove the remaining divergences in all dimensions considered, including for spatially coincident pointlike detectors.


This paper is organized as follows. In section Sec.~\ref{sec:setup}, we present the setup for entanglement harvesting with energy density coupling. In Sec.~\ref{sec:vazio_pointlike}, we evaluate the entanglement acquired between pointlike detectors in $n + 1$ dimensions when the field is prepared in the vacuum state. The study with smeared detectors appears in Sec.~\ref{sec:vaccum_smeared}. In Sec.~\ref{sec:general_hadamard}, we present general expressions for evaluating the negativity in scenarios where the field is prepared in any zero-mean Gaussian state. Our concluding remarks appear in Sec.~\ref{sec:conclusions}.

\section{Setup}
\label{sec:setup}

Consider two Unruh--DeWitt detectors~\cite{Unruh1976,DeWitt,Unruh-Wald,Jorma,Schlicht}, labeled A and B. Assuming Fermi-Walker rigid detectors (see, for instance, \cite{EduTalesBruno2021, TalesBrunoEdu2020}), we can unambiguously define the proper times $\tau_{\tc{a}}$ and $\tau_{\tc{b}}$ associated with their respective centre-of-mass timelike trajectories. The detectors are taken as two-level quantum systems with energy gap $\Omega_{j}$, the time evolution of their internal degrees of freedom is dictated by the Hamiltonian (for $j = \text{A}, \text{B}$)
\begin{equation}
    \hat{H}_{\tc{d}, j} = \Omega_{j} \hat{\sigma}_j^{+} \hat{\sigma}_j^{-}, 
    \label{eq:H_detectors}
\end{equation}
where $\hat{\sigma}_j^{\pm}$ are the usual ladder operators. We denote  the ground and excited states of the $j-th$ detector as $\ket{g_j}$ and $\ket{e_j}$ respectively, those can be written as
$\hat{\sigma}_j^{+} = | e_{j} \rangle \langle g_{j}|$, and $\hat{\sigma}_j^{-} = | g_{j} \rangle \langle e_{j}|$.
We consider a massless, scalar quantum field, whose dynamics is given by the Klein-Gordon equation,
\begin{equation}
    \partial^{\mu}\partial_{\mu}\hat{\phi}(\mf x) = 0.
    \label{eq:KG_eq}
\end{equation}
Let us consider that the two detectors are comoving with four-velocity $\mathsf{u} = u^{\mu}\partial_{\mu}$ in a Minkowski background. The coupling between the quantum field and the detectors is described by the following interaction Hamiltonian density
\begin{equation}
    \hat{h}_{I}(\mf x) = \lambda \big[\Lambda_{\tc{a}}(\mf x) \hat{\mu}_{\tc{a}}(\tau_\tc{a}) + \Lambda_{\tc{b}}(\mf x) \hat{\mu}_{\tc{b}}(\tau_\tc{b})\big]u^{\mu}u^{\nu}\mathopen{:}\hat{T}_{\mu \nu}(\mf x)\mathopen{:},
    \label{eq:H_I_harvesting}
\end{equation}
where the detectors' monopoles read
\begin{equation}
    \hat{\mu}_{j}(\tau_j) = e^{\ii \Omega_{j} \tau_{j}} \hat{\sigma}^{+}_{j} + e^{-\ii \Omega_{j} \tau_j} \hat{\sigma}^{-}_{j}.\label{eq:defMonopole}
\end{equation}
and the field energy-momentum tensor is
\begin{equation}
    \hat{T}_{\mu \nu} = \partial_{\mu}\hat{\phi}\partial_{\nu}\hat{\phi} - \frac{1}{2}\eta_{\mu \nu}\partial^{\beta}\hat{\phi}\partial_{\beta}\hat{\phi},
\end{equation}
subject to the renormalization prescription 
\begin{equation}
      \mathopen{:} \hat{T}_{\mu\nu}(\mf x) \mathopen{:}  \,\equiv\, \hat{T}_{\mu\nu}(\mf x)- \bra{0} \hat{T}_{\mu\nu}(\mf x)  \ket{0}\openone.
    \label{eq:mean_value_zero}
\end{equation}
In the interaction model of Eq.~\eqref{eq:H_I_harvesting}, the detectors couple to the renormalized energy density of the field.

From now on, let us assume the detectors are inertial and choose coordinates $\mf x = (t,\boldsymbol{x}) = (x^{\mu})$, so that $\tau_{\tc{a}} = \tau_{\tc{b}} = t$ and $\mathsf{u}=\partial_t$. Then, the interaction Hamiltonian
density of Eq.~\eqref{eq:H_I_harvesting} becomes
\begin{equation}
    \hat{h}_{\tc{i}}(\mf x)
    =
    \lambda \hat{M}_{\tc{ab}}(\mf x)\mathopen{:}\hat{T}_{00}(\mf x)\mathopen{:},
\end{equation}
where, for convenience, we introduced
\begin{equation}
    \hat{M}_{\tc{ab}}(\mf x)
    =
    \Lambda_{\tc{a}}(\mf x)\hat{\mu}_{\tc{a}}(t)
    +
    \Lambda_{\tc{b}}(\mf x)\hat{\mu}_{\tc{b}}(t),
\end{equation}
and the energy density operator reads
\begin{equation}
    \hat{T}_{00}(\mf x)
    =
   \frac{1}{2}\sum_{\mu=0}^{n}
\left(\partial_{\mu}\hat{\phi}(\mf x)\right)^2.
\end{equation}
As usual in entanglement harvesting setups, we assume an initially uncorrelated state of detectors and field,
\begin{equation}
    \hat{\rho}_{0}
    =
    \hat{\rho}_{0,\tc{ab}}\otimes \hat{\rho}_{\phi}.
\end{equation}
The evolution of the state $\hat{\rho}_{0}$ is given by the operator
\begin{equation}
    \hat{U}_{\tc{i}}
    =
    \mathcal{T}
    \exp\left(
        -\ii \int \dd V\, \hat{h}_{\tc{i}}(\mf x)
    \right),
    \label{eq:U_I_section}
\end{equation}
where $\dd V=\dd t\,\dd^n\boldsymbol{x}$, and $\mathcal{T}$exp denotes the time ordered exponential. The time-ordering is to be understood with respect to the coordinate $t$  (see \cite{EduTalesBruno2021} for the subtleties of time-ordering with smeared detectors). Expanding perturbatively in the coupling constant $\lambda$, the reduced state of the detectors after the interaction can be written as
\begin{align}
    \hat{\rho}_{\tc{ab}}
    &=
    \Tr_{\phi}\!\left[
        \hat{U}_{\tc{i}}\hat{\rho}_{0}\hat{U}_{\tc{i}}^{\dagger}
    \right]
    \nonumber\\
    &=
    \hat{\rho}_{0,\tc{ab}}
    +
    \hat{\rho}^{(1)}_{\tc{ab}}
    +
    \hat{\rho}^{(2)}_{\tc{ab}}
    +
    \mathcal{O}(\lambda^3).
\end{align}
where the superindex $(i)$ notates the perturbative order of the correction. Let us now assume the concrete case in which both detectors are initially in their ground
state,
\begin{equation}
    \hat{\rho}_{0,\tc{ab}}
    =
    \proj{g_{\tc{a}}}{g_{\tc{a}}}
    \otimes
    \proj{g_{\tc{b}}}{g_{\tc{b}}}.
\end{equation}
The final state of the detectors depends on the renormalized energy density of the field state,
\begin{equation}
\mathcal{E}_{\phi}(\mf x) \equiv \text{Tr}[\hat{\rho}_{\phi}\mathopen{:} \hat{T}_{00}(\mf x) \mathopen{:}]
\end{equation}
and on the energy density two-point function
\begin{equation}
    \Xi_{\phi}(\mf x, \mf x') \equiv \text{Tr}[\hat{\rho}_{\phi}\mathopen{:} \hat{T}_{00}(\mf x) \mathopen{:} \mathopen{:} \hat{T}_{00}(\mf x') \mathopen{:}].
    \label{eq:energy_density_two_point_func}
\end{equation}
We will particularize our setup to the case where the detectors have identical proper energy gaps $\Omega_\tc{a} = \Omega_\tc{b} = \Omega$. Then, the explicit evaluation of the reduced state of the detectors $\hat{\rho}_{\tc{ab}}$ up to order $\mathcal{O}(\lambda^2)$ is given in Appendix~\ref{sec:derivation_matrix}. In the basis
$\{
\ket{g_{\tc{a}}g_{\tc{b}}},
\ket{g_{\tc{a}}e_{\tc{b}}},
\ket{e_{\tc{a}}g_{\tc{b}}},
\ket{e_{\tc{a}}e_{\tc{b}}}
\}$,
this matrix takes the form (see \cite{Petar_Edu_coherent,Petar,carol} for a similar derivation)
\begin{align}
    \hat{\rho}_{\tc{ab}}
    &=
    \begin{bmatrix}
        1-\lambda^2(\mathcal{L}_{\tc{aa}}+\mathcal{L}_{\tc{bb}})
        & \ii\lambda \mathcal{X}_{\tc{b}}^{*}
        & \ii\lambda \mathcal{X}_{\tc{a}}^{*}
        & -\lambda^2 \mathcal{M}^{*}
        \\
        -\ii\lambda \mathcal{X}_{\tc{b}}
        & \lambda^2 \mathcal{L}_{\tc{bb}}
        & \lambda^2 \mathcal{L}_{\tc{ab}}
        & 0
        \\
        -\ii\lambda \mathcal{X}_{\tc{a}}
        & \lambda^2 \mathcal{L}_{\tc{ba}}
        & \lambda^2 \mathcal{L}_{\tc{aa}}
        & 0
        \\
        -\lambda^2 \mathcal{M}
        & 0
        & 0
        & 0
    \end{bmatrix}
    \nonumber\\
    &\quad
    +
    \mathcal{O}(\lambda^3),
    \label{eq:rho_AB_harvesting_section}
\end{align}
where
\begin{equation}
    \mathcal{L}_{ij}
    =
    \int \dd V \dd V'\,
    e^{-\ii \Omega (t-t')}
    \Lambda_{i}(\mf x)\Lambda_{j}(\mf x')
    \Xi_{\phi}(\mf x,\mf x'),
\end{equation}
\begin{equation}
    \mathcal{X}_{j}
    =
    \int \dd V\,
    e^{\ii \Omega t}
    \Lambda_{j}(\mf x)\mathcal{E}_{\phi}(\mf x),
\end{equation}
\begin{equation}
    \mathcal{M}
    =
    \int \dd V \dd V'\,
    e^{\ii \Omega (t+t')}
    \Lambda_{\tc{a}}(\mf x)\Lambda_{\tc{b}}(\mf x')
    \mathcal{G}_{\phi}(\mf x, \mf x'),
\end{equation}
with
\begin{equation}
    \mathcal{G}_{\phi}(\mf x,\mf x')
    \equiv
    \Theta(t-t')\Xi_{\phi}(\mf x,
    \mf x')
    +
    \Theta(t'-t)\Xi_{\phi}(
    \mf x',\mf x).
\end{equation}
To quantify the entanglement acquired between the two detectors, one can use the {\it negativity}~\cite{VidalNegativity,Plenio_2005}, which is a faithful entanglement monotone for a system of two qubits that is of widespread use in the literature of entanglement harvesting. One reason for its widespread use is that it remains a valid entanglement measure even when the bipartite state is mixed. For a two-qubit system, the negativity is a faithful measure of entanglement. It is defined as the negative sum of the negative eigenvalues of the partially transposed density matrix. In particular, for the state $\hat{\rho}_{\tc{ab}}$ given in Eq.~\eqref{eq:rho_AB_harvesting_section}, one finds (see \cite{Petar_Edu_coherent}, also reproduced in Appendix \ref{sec:derivation_matrix}) 
\begin{equation}
    \mathcal{N}(\hat{\rho}_{\tc{ab}}) = \lambda^2 \text{max}\{0, \mathcal{V}\} + \mathcal{O}(\lambda^3),
    \label{eq:negativity_again_again_and_again}
\end{equation}
with
\begin{equation}
    \mathcal{V}
    =
    \frac{
        \sqrt{
            (\widetilde{\mathcal L}_{\tc{aa}}-\widetilde{\mathcal L}_{\tc{bb}})^2
            +
            4|\widetilde{\mathcal M}|^2
        }
        -
        \widetilde{\mathcal L}_{\tc{aa}}
        -
        \widetilde{\mathcal L}_{\tc{bb}}
    }{2}
    \label{eq:ItisHarvestingTime}
\end{equation}
and
\begin{equation}
   \widetilde{\mathcal L}_{jj} = \mathcal{L}_{jj} - |\mathcal{X}_{j}|^2,
\end{equation}
\begin{equation}
    \widetilde{\mathcal{M}} = \mathcal{M} - \mathcal{X}_{\tc{a}}\mathcal{X}_{\tc{b}}.
\end{equation}
By absorbing the smeared energy-density contributions $\mathcal{X}_{j}$ into redefined local noise terms $\mathcal{L}_{jj}$ and correlation term $\mathcal{M}$, Eq.~\eqref{eq:ItisHarvestingTime} can be cast in the same form that appears in standard entanglement-harvesting setups (see, for example, \cite{Pozas-Kerstjens:2015}). 


To finish the presentation of the entanglement harvesting setup considered here, we required the initial field state, $\hat{\rho}_{\phi}$, to be a Hadamard state \cite{fullingHadamard,fewsterNecessityHadamard,HadamardBible}, so that the energy-momentum tensor can be renormalized, yielding a finite one-point function $\mathcal{E}_{\phi}(\mf x)$. In this case,
the Wightman of the field over the state $\hat{\rho}_{\phi}$ can be decomposed as
\begin{equation}
    W(\mf x,\mf x')
    \equiv
    \Tr\!\left[
        \hat{\rho}_{\phi}\hat{\phi}(\mf x)\hat{\phi}(\mf x')
    \right]
    =
    W_{0}(\mf x,\mf x')
    +
    w(\mf x,\mf x'),
\end{equation}
where $W_{0}(\mf x,\mf x')$ is the (massless) Minkowski vacuum Wightman function and $w(\mf x,\mf x')$ is
regular in the coincidence limit. For $n>1$, (see for instance \cite{ericksonNew})
\begin{equation}
    W_{0}(\mf x,\mf x')
    =
    \frac{(-\ii)^{n-1}\Gamma\!\left(\frac{n-1}{2}\right)}
    {4\pi^{\frac{n+1}{2}}
    \left[(\Delta t-\ii\epsilon)^2-|\Delta\boldsymbol{x}|^2\right]^{\frac{n-1}{2}}},
    \label{eq:Wightmanvacuumgeneral}
\end{equation}
where $\Delta t=t-t'$, $\Delta\boldsymbol{x}=\boldsymbol{x}-\boldsymbol{x}'$, and the limit $\epsilon \to 0^{+}$ should be evaluated distributionally. For $n=1$, a small infrared (IR) regulator $\Lambda$ (akin to a small mass) is required, and the vacuum Wightman function becomes (see for instance~\cite{Louko2014})
\begin{equation}
    W_{0}(\mf x,\mf x')
    =
    -\frac{1}{4\pi}
    \log\!\left(
        -\Lambda^2\left[(\Delta t-\ii\epsilon)^2-\Delta x^2\right]
    \right).
    \label{eq:Wightman_IR_section}
\end{equation}
    Further assuming that $\hat{\rho}_{\phi}$ is also a zero-mean Gaussian state~\cite{Hollands_2002}, Wick's theorem can be applied to write the two-point function of Eq.~\eqref{eq:energy_density_two_point_func} in terms of the Wightman function over the state $\hat{\rho}_{\phi}$. The details are
presented in Appendix~\ref{sec:two_point_appendix}, and the result is
\begin{equation}
    \Xi_{\phi}(\mf x,\mf x')
    =
    \frac{1}{2}\sum_{\mu,\nu=0}^{n}
    \bigl(
        \partial_{\mu}\partial_{\nu'}W(\mf x,\mf x')
    \bigr)^2
    +
    \mathcal{E}_{\phi}(\mf x)\mathcal{E}_{\phi}(\mf x').
    \label{eq:two_point_energy_density_concrete}
\end{equation}
Similarly, the renormalized mean energy density is determined entirely by the
regular part $w(\mf x,\mf x')$:
\begin{equation}
    \mathcal{E}_{\phi}(\mf x)
    =
    \frac{1}{2}
    \lim_{\mf x'\to \mf x}
    \sum_{\mu=0}^{n}
    \partial_{\mu}\partial_{\mu'}w(\mf x,\mf x').
    \label{eq:energy_density_w}
\end{equation}
Therefore, once the state-dependent regular contribution $w(\mf x, \mf x')$ is known,
both the one-point and two-point functions of the energy density necessary to evaluate the negativity, Eq.~\eqref{eq:negativity_again_again_and_again}, are fixed.

\section{Field prepared in the vacuum state, pointlike detectors}
\label{sec:vazio_pointlike}

In Ref.~\cite{SachsMannEdu}, entanglement harvesting with quadratically coupled detectors was shown to exhibit persistent UV divergences. It is therefore natural to ask whether the same phenomenon occurs in the present model. To address this question, we first consider pointlike detectors coupled to the renormalized energy density of a scalar field prepared in the vacuum state. Using a distributional approach, we will show that the local noise terms are well defined, while possible persistent divergences arise from the correlation terms. This allows us to identify the precise switching-dependent quantities that control the UV behaviour of the protocol.

Concretely, consider the entanglement harvesting protocol presented in Sec.~\ref{sec:setup} for the case where the field is prepared in the vacuum:
\begin{equation}
    \hat{\rho}_{\phi} = \proj{0}{0}.
\end{equation}
In this case, the renormalized energy density \mbox{$\mathcal{E}_{\phi}(\mf x)\equiv \mathcal{E}_{0}(\mf x)$} vanishes, so that
\begin{equation}
    \mathcal{X}_{\tc{a}} = \mathcal{X}_{\tc{b}} = 0,
\end{equation}
and, consequently,
\begin{equation}
    \widetilde{\mathcal{L}}_{jj} = \mathcal{L}_{jj},
\qquad     \widetilde{\mathcal{M}} = \mathcal{M}.
\end{equation}
We consider an $(n+1)$-dimensional spacetime with coordinates $\mf x = (t, \bm x)$. If we consider two pointlike detectors following inertial trajectories $\mf z_{j}(t) = (t, \boldsymbol{x}_{j})$, the spacetime smearing becomes
\begin{equation}
    \Lambda_{j}(t, x) = \chi_{j}(t)\delta(x - \boldsymbol{x}_{j}),
\end{equation}
where $\chi_{j}(t)$ are switching functions that control the time
dependence of the interaction between each detector and the renormalized
energy density of the field. In this case, the local noise terms can be written as
\begin{equation}
    \mathcal{L}_{jj} = \int_{-\infty}^{\infty}\dd t \int_{-\infty}^{\infty}\dd t' e^{-\ii \Omega \Delta t}\chi_{j}(t)\chi_{j}(t')\Xi_{0}(t, t'),
    \label{eq:localnoise_jj_vacuum_pointlike}
\end{equation}
where $\Xi_{0}(t, t')$ is the pullback of the two-point function $\Xi_{0}(\mf x, \mf x')$ to the trajectory $\mf z_{j}(t)$. Using Eq.~\eqref{eq:Wightmanvacuumgeneral} in the definition of the renormalized energy density two-point function, Eq.~\eqref{eq:two_point_energy_density_concrete}, and setting $\boldsymbol{x} = \boldsymbol{x'} = \boldsymbol{x}_{j}$, one obtains
\begin{equation}
    \Xi_{0}(t, t')
    = \frac{C_{n}}{(\Delta t - \ii \epsilon)^{2(n + 1)}},
    \label{eq:WightmanDecay}
\end{equation}
where
\begin{equation}
    C_{n}
    =
    \begin{cases}
        \dfrac{1}{4\pi^{2}}, & n = 1, \\[0.8em]
        \dfrac{n(n+1)(n-1)^{2}(-\ii)^{2n-2}}{32 \pi^{n + 1}}
        \Gamma\left(\dfrac{n-1}{2}\right)^{2}, & n > 1.
    \end{cases}
\end{equation}
To evaluate $\mathcal{L}_{jj}$, we perform the change of variables  $(t, t') \to (u, s)$, with $u = \Delta t$, $s = t$. Then, we can write
\begin{equation}
    \mathcal{L}_{jj} = C_{n}\lim_{\epsilon \to 0^{+}}\int_{-\infty}^{\infty}\dd u \frac{\Phi_{j}(u)}{(u - \ii \epsilon)^{2(n + 1)}},
\end{equation}
where we introduced notation for the switching auto-correlation function
\begin{equation}\label{eq:Phij}
    \Phi_{j}(u) \equiv e^{-\ii \Omega u}\int_{-\infty}^{\infty}\dd s \, \chi_{j}(s)\chi_{j}(s - u).
\end{equation}
In Eq.~\eqref{eq:localnoise_jj_vacuum_pointlike}, the limit $\epsilon \to 0^{+}$ needs to be evaluated distributionally. To that end, we recall that for any $n \in \mathbb{N}$, the following identity holds \cite{GelfandShilov2016}
\begin{equation}
    \lim_{\epsilon \to 0}\frac{1}{(x \mp \ii \epsilon)^n} = \operatorname{FP}\left(\frac{1}{x^n} \right) \pm \frac{(-1)^{n - 1}}{(n - 1)!} \ii \pi \delta^{(n - 1)}(x),
    \label{eq:generalizedSP_main_text}
\end{equation}
where $\operatorname{FP}$ denotes the Hadamard finite part distribution and $\delta^{(n - 1)}$ is the distributional derivative of the Dirac delta. Hadamard's finite part bears the following relation with the principal value distribution ($\operatorname{PV}$) when acting on a test function $f$:
\begin{equation}
    \left \langle \operatorname{FP}\left(\frac{1}{x^{n}}\right), f  \right \rangle = \frac{1}{(n - 1)!}\left \langle \operatorname{PV}\left(\frac{1}{x}\right), f^{(n-1)}  \right \rangle. 
\label{eq:FP_and_PV_main_text}
\end{equation}
Also, recall that the derivative of the Dirac delta as a distribution is defined by its action on a test function $f$ as
\begin{equation}
    \int_{-\infty}^{\infty}\dd x f(x)\delta^{(n)}(x) = (-1)^{n}f^{(n)}(0).
\end{equation}
Therefore, for any $n \in \mathbb{N}$ and any function $f$ of class $\mathcal{C}^{n - 1}(\mathbb{R})$, we have
\begin{align}
    \lim_{\epsilon \to 0^{+}}\int_{-\infty}^{\infty}\dd x \, \frac{f(x)}{(x \mp\ii \epsilon)^n} & = \pm\frac{\ii \pi f^{(n - 1)}(0)}{(n - 1)!}  \nonumber \\ & + \operatorname{FP}\int_{-\infty}^{\infty}\dd x \, \frac{f(x)}{x^n}.
    \label{eq:ladyD}
\end{align}
This identity will be crucial for obtaining the main results of this paper. Applying this to our case, the local noise of each detector for any switching function of class at least $\mathcal{C}^{(2n + 1)}(\mathbb{R})$ can be evaluated as follows:
\begin{align}
    \mathcal{L}_{jj} & = \frac{C_{n}}{(2n + 1)!}\left[\ii \pi \Phi_{j}^{(2n + 1)}(0) + \operatorname{PV} \int_{-\infty}^{\infty} \dd u \frac{\Phi_{j}^{(2n + 1)}(u)}{u} \right] \nonumber \\ & = \frac{C_{n}}{(2n + 1)!}\Biggl[\ii \pi \Phi_{j}^{(2n + 1)}(0) \nonumber \\ & \hspace{1.0cm}+ \int_{0}^{\infty} \dd u \frac{\Phi_{j}^{(2n + 1)}(u) - \Phi_{j}^{(2n + 1)}(-u)}{u} \Biggr].
    \label{eq:Fjj_explicit_pointlike}
\end{align}
Next, we evaluate the correlation terms.

The pullback of the vacuum correlator to the two detector trajectories depends on the time difference $u=t-t'$ and their spatial separation $d=|\boldsymbol{x}_{\tc a}-\boldsymbol{x}_{\tc b}|$. Substituting the vacuum Wightman function into Eq.~\eqref{eq:two_point_energy_density_concrete}, we obtain
\begin{align}
 \Xi_0(u,d)
 &=\frac{A_n}{[d^2-u_\epsilon^2]^{n+3}}
 \Bigl[n(u_\epsilon^4+d^4)
 \nonumber\\
 &\hspace{2.1cm}+2(n+2)u_\epsilon^2d^2\Bigr],
 \label{eq:edu_pointlike_separated_kernel}
\end{align}
where $u_\epsilon=u-\ii\epsilon$ and $A_n=(-1)^{n-1}C_n/n$. This expression holds for $n\geq1$, including $A_1=1/(4\pi^2)$, and reduces to Eq.~\eqref{eq:WightmanDecay} when $d=0$. Introduce the temporal correlation

\begin{equation}
    \Phi_{\tc{ab}}(u) \equiv e^{-\ii \Omega u}\int_{-\infty}^{\infty}\dd s \, e^{2\ii \Omega s}\chi_{\tc{a}}(s)\chi_{\tc{b}}(s - u),
    \label{eq:Phi_ab_def}
\end{equation}

and let $f(u)=\Phi_{\tc{ab}}(u)+\Phi_{\tc{ab}}(-u)$. The time-ordered correlation term becomes
\begin{align}
 \mathcal M(d)
 &=\lim_{\epsilon\to0^+}\int_{-\infty}^{\infty}\dd u\,
 \Phi_{\tc{ab}}(u)
 \nonumber\\
 &\quad\times\bigl[\Theta(u)\Xi_0(u,d)
 +\Theta(-u)\Xi_0(-u,d)\bigr]
 \nonumber\\
 &=\lim_{\epsilon\to0^+}\int_0^\infty\dd u\,f(u)\Xi_0(u,d).
 \label{eq:edu_pointlike_separated_M}
\end{align}
The step functions distinguish the two time orderings, $t>t'$ ($u>0$) and $t<t'$ ($u<0$). The last equality follows by changing $u\mapsto-u$ in the negative half-line contribution. Both contributions are then integrated over $u>0$, with equal interaction times, $u=0$, as the lower endpoint. We call this endpoint the time-ordering boundary. To evaluate the regulator limit, we assume that the switchings are smooth and compactly supported, or belong to the Schwartz class. Then $f$ is also smooth and rapidly decreasing.

At fixed $d>0$, the real singularities occur at the null separations $u=\pm d$. In particular, the kernel is smooth near the time-ordering boundary, where $\lim_{\epsilon\to0^+}\Xi_0(0,d)=nA_n/d^{2n+2}$. On the positive half-line, only the pole at $u=d$ needs a distributional prescription. Factoring the denominator gives, in a neighbourhood of that point,
\begin{equation}
 \Xi_0(u,d)=\frac{H_\epsilon(u,d)}{(u-d-\ii\epsilon)^{n+3}},
 \label{eq:edu_pointlike_interior_pole}
\end{equation}
where $H_\epsilon(u,d)=(u-d-\ii\epsilon)^{n+3}\Xi_0(u,d)$ is smooth there and converges with all its derivatives as $\epsilon\to0^+$. The other denominator factor, $u+d-\ii\epsilon$, does not vanish in this neighbourhood.

Choose a smooth cutoff with compact support inside $(0,\infty)$ and equal to one near $u=d$. Multiplying the integrand by this cutoff isolates the pole, and the resulting integral can be extended to the whole line because its support stays away from the endpoint. Equation~\eqref{eq:ladyD}, with $x=u-d$, then gives a finite-part integral and a finite term involving the $(n+2)$th derivative of the cutoff-weighted product $fH_0$, evaluated at $u=d$, where $H_0=\lim_{\epsilon\to0^+}H_\epsilon$. The remaining integral has no singularity on its domain and converges ordinarily. This proves that $\mathcal M(d)$ has a finite regulator limit for every fixed $d>0$, even when the switchings overlap in time. Finiteness alone does not establish a positive negativity or exclude communication.

For example, in $1+1$ dimensions the separated kernel takes the particularly simple form
\begin{equation}
 \Xi_0(u,d)=\frac{1}{8\pi^2}
 \left[\frac{1}{(u-d-\ii\epsilon)^4}
 +\frac{1}{(u+d-\ii\epsilon)^4}\right].
 \label{eq:edu_pointlike_separated_1d}
\end{equation}
Its two terms merge into Eq.~\eqref{eq:WightmanDecay} at $d=0$. For fixed nonzero separation, they are treated by the same distributional identities used for the local noise.

The case of spatially coincident detectors, $\boldsymbol{x}_{\tc a}=\boldsymbol{x}_{\tc b}\equiv\boldsymbol{x}_0$, requires separate treatment. At $d=0$, the two null singularities meet at the time-ordering boundary $u=0$, so a cutoff around the pole can no longer avoid the endpoint. The singularity is then sampled from only one side of the half-line integral, and the preceding argument for finiteness does not apply. To determine whether $\mathcal M$ is well defined in this case, and which conditions on the switchings ensure finiteness, we evaluate the regulated integral before taking $\epsilon\to0^+$. Writing $\mathcal M=\mathcal M(0)$ and leaving the two switching functions independent, we have

\begin{align}
    \mathcal{M} & = C_{n}\int_{-\infty}^{\infty}\dd t \int_{-\infty}^{\infty}\dd t' e^{\ii \Omega(t + t')}\chi_{\tc{a}}(t)\chi_{\tc{b}}(t') \times \nonumber \\ & \hspace{2.0cm}\Biggl[\frac{\Theta(t - t')}{(\Delta t - \ii \epsilon)^{2(n + 1)}} + \frac{\Theta(t' - t)}{(\Delta t + \ii \epsilon)^{2(n + 1)}} \Biggr],
\end{align}
and applying the same change of variables as before, we obtain
\begin{align}
    \mathcal{M} & = C_{n}\int_{-\infty}^{\infty}\dd u \, \Phi_{\tc{ab}}(u) \Biggl[\frac{\Theta(u)}{(u - \ii \epsilon)^{2(n + 1)}} + \frac{\Theta(-u)}{(u + \ii \epsilon)^{2(n + 1)}}\Biggr] \nonumber \\ & = C_{n}\int_{0}^{\infty}\dd u \, \frac{\Phi_{\tc{ab}}(u) + \Phi_{\tc{ab}}(-u)}{(u - \ii \epsilon)^{2(n + 1)}},
    \label{eq:term_M_half_line_pointlike}
\end{align}
\eduadd{where $\Phi_{\tc{ab}}$ is defined in Eq.~\eqref{eq:Phi_ab_def}.}
Because the integrand of Eq.~\eqref{eq:term_M_half_line_pointlike} is only defined on the half-line, the identity in Eq.~\eqref{eq:ladyD} cannot be applied. Of course, one could extend the integrand to the whole line by introducing $\Theta(u)$, but the resulting function would no longer be a valid test function, as it is not differentiable at the origin. Instead, the correct way to proceed is to derive the corresponding distributional identity for the half-line. That is, we need to consider integrals of the form
\begin{equation}
    \mathcal{I}^{m}_{\epsilon}(f) = \int_{0}^{\infty}\dd x \, \frac{f(x)}{(x \pm \ii \epsilon)^m}.
    \label{eq:defintegral_n_main}
\end{equation}
In Appendix \ref{sec:half_line_appendix}, we show that the distributional limit $\epsilon \to 0^{+}$ of $\mathcal{I}^{m}_{\epsilon}(f)$ contains persistent divergences. Those are completely characterized in Eq.~\eqref{eq:brazilianBBQ}. Adapting this result for the integral in Eq.~\eqref{eq:term_M_half_line_pointlike} with $f(x) = \Phi_{\tc{ab}}(x) + \Phi_{\tc{ab}}(-x)$ and $m = 2n + 2$, we can write
\begin{equation}
    \lim_{\epsilon \to 0^{+}}(\mathcal{M} - \mathcal{D}_{\epsilon})  =  \mathcal{M}_{\tc{f}}
\end{equation}
where the finite part is
\begin{equation}
    \mathcal{M}_{\tc{f}}
    =
    \frac{C_{n}}{(2n + 1)!}
    \operatorname{FP}\int_{0}^{\infty}\dd u \,
    \frac{\Phi^{(2n + 1)}_{\tc{ab}}(u) - \Phi^{(2n + 1)}_{\tc{ab}}(-u)}{u}.
    \label{eq:it_is_finite}
\end{equation}
and the divergences are of the form
\begin{equation}
    \mathcal{D}_{\epsilon} =    2C_n \ii
    \sum_{q=0}^{n}
    (-1)^{n-q}
    \frac{(2n-2q)!}{(2n+1)!}
    \frac{
        \Phi_{\textsc{ab}}^{(2q)}(0)
    }{
        \epsilon^{2n+1-2q}
    } .
    \label{eq:UV_divergences_general_pointlike}
\end{equation}
In deriving
Eq.~\eqref{eq:UV_divergences_general_pointlike}, we used
$f(u)=\Phi_{\tc{ab}}(u)+\Phi_{\tc{ab}}(-u)$, which implies
$f^{(2q)}(0)=2\Phi_{\tc{ab}}^{(2q)}(0)$ and
$f^{(2q+1)}(0)=0$. In particular, since $m=2n+2$, the
logarithmic divergence in Eq.~\eqref{eq:brazilianBBQ}
is proportional to $f^{(2n+1)}(0)$ and therefore vanishes.

 \eduadd{For the coincident configuration, the divergent coefficients are controlled by $\Phi_{\tc{ab}}^{(2q)}(0)$, for $q=0,\ldots,n$. At fixed nonzero separation, the light-cone singularities instead have the finite distributional action established in Eq.~\eqref{eq:edu_pointlike_separated_M}.} In particular, when both switchings are compactly supported and have non-overlapping
supports, all of the terms $\Phi_{\tc{ab}}^{(2q)}(0)$ vanish, yielding a finite, divergence-free correlations term:
\begin{equation}
    \mathcal{M} = \mathcal{M}_{\tc{f}}.
\end{equation}
In such cases, the entanglement acquired by the detectors is computable from Eq.~\eqref{eq:negativity_again_again_and_again}, with
\begin{equation}
    \mathcal{V} =
    \frac{
        \sqrt{
            (\mathcal{L}_{\tc{aa}}-\mathcal{L}_{\tc{bb}})^2
            +
            4|\mathcal {M}_{\tc{f}}|^2
        }
        -
        \mathcal{L}_{\tc{aa}}
        -
        \mathcal{L}_{\tc{bb}}
    }{2},
    \label{eq:negativity_dimitrescu}
\end{equation}
where $\mathcal{L}_{jj}$, with $j\in\{\text{A},\text{B}\}$, is given by
Eq.~\eqref{eq:Fjj_explicit_pointlike}, and $\mathcal{M}_{\tc f}$ by
Eq.~\eqref{eq:it_is_finite}. This formula holds for compactly supported
switching functions $\chi_j(t)$ with non-overlapping supports, provided
$\chi_j\in\mathcal{C}^{(2n + 1)}(\mathbb{R})$.

The negativity in Eq.~\eqref{eq:negativity_again_again_and_again} can be further simplified by using properties of the convolution 
\begin{equation}
    R_j(u) \equiv \int_{-\infty}^{\infty}\dd s\, \chi_j(s)\chi_j(s-u).
\end{equation}
Specifically, if $\chi_\textsc{b}(t) = \chi_\textsc{a}(t+t_{c})$ where $t_{c}$ is some fixed constant, then $R_\textsc{a}(u) = R_\textsc{b}(u)$. This can be seen from the definition above, where we have 
\begin{align}\label{eq:autocorrelationprop}
    R_\textsc{b}(u) =& \int_{-\infty}^{\infty}\dd s\, \chi_\textsc{b}(s)\chi_\textsc{b}(s-u)\nonumber\\
    =&\int_{-\infty}^{\infty}\dd s\, \chi_\textsc{a}(s+t_{c})\chi_\textsc{a}(s+t_{c}-u)\nonumber\\
    =&\int_{-\infty}^{\infty}\dd r\,\chi_\textsc{a}(r)\chi_\textsc{a}(r-u) = R_\textsc{a}(u).
\end{align}
This means that $\mathcal{L}_{\tc{aa}} = \mathcal{L}_{\tc{bb}} \equiv \mathcal{L}$. 
In this case, the negativity reduces to 
\begin{equation}
   \mathcal{N}(\hat{\rho}_{\tc{ab}})= \lambda^2 \text{max}\{0,  |\mathcal{M}_\textsc{f}| - \mathcal{L} \},
   \label{eq:negativity_vanilla}
\end{equation}
which has the same form as the negativity obtained in standard entanglement harvesting protocols with identical detectors in backgrounds with translational symmetries (see, e.g., \cite{Pozas-Kerstjens:2015}).

Before illustrating the protocol with an explicit example in $1+1$ dimensions, we discuss why non-overlapping switchings remove the divergences for spatially coincident pointlike detectors. The correlation term is obtained by pulling back the time-ordered energy-density two-point function to the detector trajectories and integrating against the switchings. For $d=0$, its singularity lies at the boundary $u=0$ of the half-line integral. Compactly supported switchings with disjoint supports make $\Phi_{\tc{ab}}$ vanish in a neighbourhood of that boundary, so the singularity is not sampled. For $d>0$, the light-cone poles lie away from the boundary and have a finite distributional action under the smoothness assumptions stated above, even when the switchings overlap.


\eduadd{At spatial coincidence, the same half-line analysis applies to other quadratic couplings whenever their pulled-back correlator has the corresponding power-law form.} Indeed, for a quadratic coupling to the amplitude, the renormalized vacuum two-point function is
\begin{equation}
    \langle 0 | \mathopen{:}\hat{\phi}(\mf x)^2 \mathopen{:}\mathopen{:}\hat{\phi}(\mf x')^2 \mathopen{:} | 0 \rangle = 2 W_{0}(\mf x, \mf x')^2,
\end{equation}
so that, for \eduadd{$n>1$}, Eq.~\eqref{eq:WightmanDecay} should be replaced by
\begin{equation}
    \Xi^{(\phi^2)}_{0}(t, t') =
    \frac{
        C^{(\phi^2)}_n
    }{
        (\Delta t-\ii\epsilon)^{2(n-1)}
    },
    \label{eq:WightmanDecay_phi2}
\end{equation}
where
\begin{equation}
    C^{(\phi^2)}_n
    =
    \frac{
        (-1)^{n-1}
        \Gamma\!\left(\frac{n-1}{2}\right)^2
    }{
        8\pi^{n+1}
    } .
\end{equation}

For spatially coincident pointlike detectors with this amplitude coupling, the half-line integral has a different pole order but the same endpoint singularity. Generic overlapping Gaussian switchings do not make all the required temporal derivatives vanish, and persistent divergences can remain. At nonzero pointlike separation, the light-cone poles lie away from the time-ordering boundary. For spatially smeared detectors, as in Refs.~\cite{SachsMannEdu,sachs2018entanglement}, the correlation term also involves a spatial average, so its singular behaviour depends on the profile overlap as well as the switching functions.


\subsection*{Example in \texorpdfstring{$1 + 1$}{1 + 1} dimensions}
\label{sec:example_vacuum_pointlike}

We now illustrate the previous result with a concrete example in $1 + 1$ dimensions using compactly supported $C^3$ switching functions. \eduadd{We take the detectors to be spatially coincident, $d=0$, to examine whether non-overlapping switchings can remove the UV divergence without preventing entanglement harvesting.} Namely, we take
\begin{equation}\label{eq:pointlikeswitch}
    \chi_j(t)
    =
    \begin{cases}
        \displaystyle
        \sin^4\!\left(\frac{\pi(t-t_j)}{T}\right),
        & t_j \leq t \leq t_j+T, \\[1ex]
        0,
        & \text{otherwise}.
    \end{cases}
\end{equation}
Without loss of generality, let us assume that detector B switches on the interaction after detector A. We notate the switching delay between the two detectors as \mbox{$\delta t \equiv t_{\tc{b}} - t_{\tc{a}} > 0$}. Then, the condition $\text{supp} \, \chi_{\tc{a}} \cap \text{supp} \,  \chi_{\tc{b}}  = \emptyset$ is equivalent to
\begin{equation}
    \delta t > T.
    \label{eq:ladyD_condition}
\end{equation}
Recall that the detectors are spatially coincident.
The condition $\delta t>T$ therefore makes their interaction regions
strictly timelike-separated. In general, entanglement acquired in
such configurations can include contributions from field-mediated
communication~\cite{ericksonNew}. In the present example, however,
this contribution vanishes because the vacuum expectation value of
the energy-density commutator at coincident spatial positions is
supported only at equal times. Indeed, in the distributional sense,
\begin{align}
    &\langle 0|
    [\,:\!\hat{T}_{00}(t,x_0)\!:,\,
       :\!\hat{T}_{00}(t',x_0)\!:\,]
    |0\rangle
    \nonumber\\
    &\quad=
    \frac{1}{4\pi^2}
    \lim_{\epsilon\to0^+}
    \left[
        \frac{1}{(\Delta t-\ii\epsilon)^4}
        -
        \frac{1}{(\Delta t+\ii\epsilon)^4}
    \right]
    \nonumber\\
    &\quad=
    -\frac{\ii}{12\pi}\delta^{(3)}(\Delta t).
\end{align}
Since $\delta t>T$, the switching product vanishes
in a neighbourhood of $\Delta t=0$, and hence the commutator
contribution to $\mathcal{M}$ vanishes. Thus, any nonzero negativity
obtained in this example represents genuine vacuum entanglement
harvesting at the perturbative order considered.

For the switching function in Eq.~\eqref{eq:pointlikeswitch}, the convolution
integral in Eq.~\eqref{eq:Phij} has support only for $|u|\leq T$.
\eduadd{We set $t_\textsc{a}=0$ and $t_\textsc{b}=\delta t$, with $\delta t>T$, and vary the delay when evaluating the negativity.}
Using Eq.~\eqref{eq:autocorrelationprop}, we find 
\begin{align}     
R(u)
&= \int_{-\infty}^{\infty}\dd s\,\chi_j(s)\chi_j(s-u)\nonumber\\
&=T\bigg[\frac{9}{64}\left(1-\frac{|u|}{T}\right)
+
\frac{1}{8}\left(1-\frac{|u|}{T}\right)
\cos\left(2\pi \frac{|u|}{T}\right)\nonumber\\
&+
\frac{1}{128}\left(1-\frac{|u|}{T}\right)
\cos\left(4\pi \frac{|u|}{T}\right)
+
\frac{5}{48\pi}
\sin\left(2\pi \frac{|u|}{T}\right)\nonumber\\
&+
\frac{25}{1536\pi}
\sin\left(4\pi \frac{|u|}{T}\right)\bigg],
\end{align}
where we now drop the subscript $j$ since the convolution integral is identical for both detectors. This leads to 
\begin{equation}\label{eq:pointlikePhi}
    \Phi(u) = e^{-\ii\Omega u}\,R(u),
\end{equation}
where again we drop the subscript to ease the notation. With this, the local noise can be written analytically as (with $\mathcal{L}_{\tc{aa}} = \mathcal{L}_{\tc{bb}} \equiv \mathcal{L}$)
\begin{equation}
\mathcal{L} =\frac{1}{24\pi^2 T^2}
\left[2J(\Omega T)-\pi\left(\frac{35}{128}\Omega^3 T^3 +\frac{15\pi^2}{8}\Omega T\right)\right],
\end{equation}
where
\begin{align}
J(x)
=
\frac{9}{64}A(x)
&+\frac{1}{16}
\left[A(x+2\pi)+A(x-2\pi)\right]\nonumber\\
&+\frac{1}{256}\left[A(x+4\pi)+A(x-4\pi)\right]\nonumber\\
&+\frac{5}{96\pi}\left[B(2\pi+x)+B(2\pi-x)\right]\nonumber\\
&+\frac{25}{3072\pi}\left[B(4\pi+x)+B(4\pi-x)\right],
\end{align}
\begin{align}
A(x)=&3x^2L(x)+x^3\operatorname{Si}(x)-x^2\left(1-\cos x\right),\\
B(x)=&-x^3L(x),\\
L(x)=&\operatorname{Ci}(|x|)-\gamma-\ln|x|.
\end{align}
Here, $\gamma$ is the Euler–Mascheroni constant, $\operatorname{Si}(x)$ is the Sine integral, and $\operatorname{Ci}(x)$ is the Cosine integral.

The correlation term $\mathcal{M}$ cannot be expressed in a comparably compact closed form. Instead, the term  $\Phi_{\tc{ab}}(u)$ in Eq.~\eqref{eq:Phi_ab_def} is evaluated numerically and then used to evaluate Eq.~\eqref{eq:it_is_finite} with $n = 1$.

\begin{figure}[h!]
    \centering
    \includegraphics[width=1\linewidth]{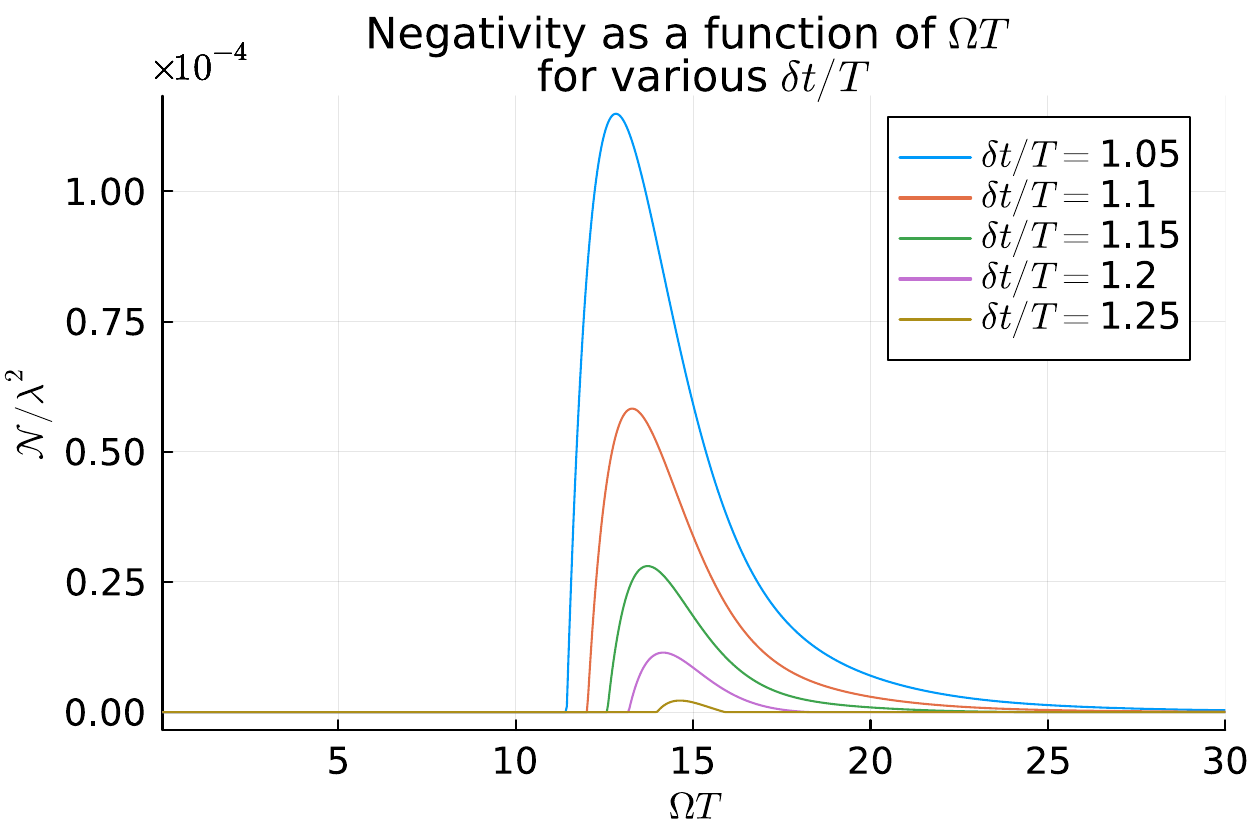}
    \caption{Negativity as a function of the detectors' gap $\Omega$ for multiple values of the switching function offset $\delta t$. \eduadd{The detectors are spatially coincident ($d=0$).} In each plot, we fix the scale by setting $T=1$. We find that for each choice of time delay, there is an optimal $\Omega$ commensurate with the results found in~\cite{hectorMass}.}
    \label{fig:PointlikeVacuumNegativity}
\end{figure}
In Fig.~\ref{fig:PointlikeVacuumNegativity}, we plot the negativity as a function of the detectors' energy gap for several values of the switching delay $\delta t$. For each delay, there exists a minimum value of $\Omega T$ above which the detectors become entangled. As the temporal separation between the switching functions increases, this threshold shifts to larger values of $\Omega T$. Moreover, for each value of $\delta t$ there is an optimal detector gap that maximizes the harvested entanglement, in agreement with the behaviour previously reported in Ref.~\cite{hectorMass}. As the switching delay increases, this optimal gap also shifts towards larger values of $\Omega T$, while the corresponding maximum negativity decreases. Numerically, we find that entanglement harvesting ceases for $\delta t/T \gtrsim 1.25$.
This behaviour can be understood from the decay of the vacuum energy density correlations. In $1+1$ dimensions, Eq.~\eqref{eq:WightmanDecay} shows that the pulled-back energy density two-point function decays as $|\Delta t|^{-4}$ for large temporal separations. Consequently, increasing the switching delay suppresses the correlation term responsible for harvesting, while the local noise remains essentially unchanged. Beyond a sufficiently large delay, the harvested correlations are no longer strong enough to overcome the local noise, and no entanglement is generated.


This coincident example shows that compactly supported non-overlapping switchings remove the endpoint divergences while still allowing vacuum entanglement harvesting for the delays and gaps shown in Fig.~\ref{fig:PointlikeVacuumNegativity}.

\section{Field prepared in the vacuum state, spatially smeared detectors}
\label{sec:vaccum_smeared}


Spatial smearing modifies the short-distance structure of the sampled energy-density correlator. For Gaussian profiles, it removes the persistent divergences in $1+1$ and $2+1$ dimensions. In higher dimensions, divergences generally remain for overlapping switchings, although specific detector separations can cancel the singular coefficients. We will identify these cancellations and explain how they depend on the spatial profiles.


Let us now consider smeared detectors with a separable spacetime profile of the form
\begin{equation}
    \Lambda_{j}(t, \boldsymbol{x}) = \chi_{j}(t)F_{j}(\boldsymbol{x}).
\end{equation}
The center of mass of each detector follows the inertial trajectory
$\mf z_{j}(t) = (t, \boldsymbol{x}_{j})$. For concreteness, we take the detectors'
spatial profiles to be Gaussian functions of width $\sigma$:
\begin{equation}
    F_j(\boldsymbol{x})
    =
    \frac{
        e^{
            -\frac{|\boldsymbol{x}-\boldsymbol{x}_j|^2}{2\sigma^2}}
    }{
        (2\pi\sigma^2)^{n/2}
    } .
    \label{eq:gaussian_and_friends}
\end{equation}
We use Gaussian spatial profiles as a tractable
example of finite-size detectors. This standard choice also
allows comparison with earlier studies of detectors coupled to composite field operators~\cite{SachsMannEdu,sachs2018entanglement}.

To write the local noise $\mathcal{L}_{jj}$, we introduce the change of
variables $(t,t')\to(s,u)$, with $u=\Delta t$ and $s=t$, and
$(\boldsymbol{x},\boldsymbol{x'})\to(\boldsymbol{z},\boldsymbol{y})$, with $\boldsymbol{y}=\Delta \boldsymbol{x} = \boldsymbol{x} - \boldsymbol{x'}$ and $\boldsymbol{z}=\boldsymbol{x}$. In these variables, the energy density two-point function reads
\begin{equation}
    \Xi_{0}(u, \boldsymbol{y}) = A_n
    \frac{
        n\left[(u-\ii\epsilon)^4+|\boldsymbol{y}|^4\right]
        +
        2(n+2)(u-\ii\epsilon)^2|\boldsymbol{y}|^2
    }{
        \left(            |\boldsymbol{y}|^2-(u-\ii\epsilon)^2
        \right)^{n+3}
    },
    \label{eq:energy_density_twopoint_general_vacuum}
\end{equation}
with
\begin{equation}
    A_n = \frac{
        (n+1)(n-1)^2
        \Gamma\!\left(\frac{n-1}{2}\right)^2
    }{
        32\pi^{n+1}
    } .
    \label{eq:def_An_main}
\end{equation}
\eduadd{For $n=1$, this expression is understood by continuity, giving $A_1=1/(4\pi^2)$ as in Sec.~\ref{sec:vazio_pointlike}.}
Now, by defining the auxiliary function
\begin{equation}
   \vartheta_{j}(\boldsymbol{y}) \equiv \int_{\mathbb{R}^{n}}\dd^{n} \boldsymbol{z} \, F_{j}(\boldsymbol{z})F_{j}(\boldsymbol{z} - \boldsymbol{y}),
\end{equation}
and the smeared distribution
\begin{equation}
    \mathfrak{W}_{j}(u) = \int_{\mathbb{R}^{n}}\dd^{n} \boldsymbol{y} \,\vartheta_{j}(\boldsymbol{y})\Xi_{0}(u, \boldsymbol{y}),
    \label{eq:mathfrack_and_friends}
\end{equation}
the local noise of each detector can be compactly written as
\begin{equation}
    \mathcal{L}_{jj} = \int_{-\infty}^{\infty}\dd u \,   \Phi_{j}(u) \mathfrak{W}_{j}(u).
    \label{eq:local_noise_smeared_vaccum_n_dim}
\end{equation}
By performing a similar change of variables, the non-local correlations term can be compactly written as
\begin{equation}
    \mathcal{M} = \int_{-\infty}^{\infty}\dd u \, \Phi_{\tc{ab}}(u) \mathfrak{W}_{\tc{ab}}(u),
    \label{eq:Mterm_smeared_vacuum_n_dim}
\end{equation}
with the definitions
\begin{align}
    \mathfrak{W}_{\tc{ab}}(u) & = \Theta(u)\int_{\mathbb{R}^{n}}\dd^{n} \boldsymbol{y} \,  \vartheta_{\tc{ab}}(\boldsymbol{y}) \Xi_{0}(u, \boldsymbol{y}) \nonumber \\ 
    & + \Theta(-u)\int_{\mathbb{R}^{n}}\dd^{n} \boldsymbol{y} \,  \vartheta_{\tc{ab}}(\boldsymbol{y}) \Xi_{0}(-u, -\boldsymbol{y}),
    \label{eq:smeared_distribution_ab_n_dim}
\end{align}
\begin{equation}
    \vartheta_{\tc{ab}}(\boldsymbol{y}) \equiv \int_{\mathbb{R}^{n}}\dd ^{n} \boldsymbol{z} \, F_{\tc{a}}(\boldsymbol{z})F_{\tc{b}}(\boldsymbol{z} - \boldsymbol{y}),
    \label{eq:edu_spatial_overlap_definition}
\end{equation}
where we recall that $\Phi_{\tc{ab}}(u)$ is given by Eq.~\eqref{eq:Phi_ab_def}.

For the Gaussian spatial profile of Eq.~\eqref{eq:gaussian_and_friends}, the convolution $\vartheta_{j}$ is the same for both detectors, namely
\begin{equation}
    \vartheta(\boldsymbol{y}) = \frac{
        e^{-\frac{|\boldsymbol{y}|^2}{4 \sigma^2}}
    }{
        (4\pi\sigma^2)^{n/2}
    } .
\end{equation}
Similarly,
\begin{equation}
    \vartheta_{\tc{ab}}(\boldsymbol{y})
    =
    \frac{e^{-\frac{|\boldsymbol{y} + \boldsymbol{d}|^2}{4\sigma^2}}}{(4\pi\sigma^2)^{n/2}}
    ,
\label{eq:vartheta_ab_gaussian_compact}
\end{equation}
with $\boldsymbol{d} \equiv \boldsymbol{x}_{\tc{b}} - \boldsymbol{x}_{\tc{a}}$.

Now, because the pointlike case is recovered in the limit $\sigma \to 0$, this suggests that the spatial profile should not introduce any additional obstruction in the evaluation of the local noise terms in Eq.~\eqref{eq:local_noise_smeared_vaccum_n_dim}. As in Eq.~\eqref{eq:Fjj_explicit_pointlike}, the quantities $\mathcal{L}_{jj}$ are expected to be free of UV divergences once the relevant distributions are treated carefully.

Indeed, let us consider the explicit form of the smeared distribution $\mathfrak{W}(u)$. Notice that the integrand in Eq.~\eqref{eq:mathfrack_and_friends} is purely radial. Thus, by letting $r = |\boldsymbol{y}|$, the integral can be cast into (see Appendix \ref{sec:appendix_smeared_distribution})
\begin{align}
    \mathfrak{W}(u)
    &=
    \frac{A_n 2 \pi^{n/2}}{(4\pi\sigma^2)^{n/2}\Gamma(n/2)}
    \int_{0}^{\infty}\dd r\,
    r^{n-1}
    e^{-r^2/(4\sigma^2)}
    \nonumber\\
    &\quad\times
    \frac{
        n(u_{\epsilon}^{4}+r^{4})
        +
        2(n+2)u_{\epsilon}^{2}r^{2}
    }{
        (r^{2}-u_{\epsilon}^{2})^{n+3}
    },
    \label{eq:smeared_distribution_single_detector}
\end{align}
with $u_{\epsilon} \equiv u - \ii \epsilon$, and we dropped the detector subscript $j$ as the smeared distribution is detector independent.  Evaluating this integral, we obtain (see Appendix \ref{sec:appendix_smeared_distribution})
\begin{align}
\mathfrak{W}(u)
&=
\frac{A_{n} n}{2^{n}\sigma^{n}}
\left(-u_{\epsilon}^{2}\right)^{-\frac{n}{2}-1}
\Biggl[
U\!\left(
\frac{n}{2},
-\frac{n}{2}-2,
-\frac{u_{\epsilon}^{2}}{4\sigma^{2}}
\right)
\nonumber\\
&\qquad
-(n+2)\,
U\!\left(
\frac{n}{2}+1,
-\frac{n}{2}-1,
-\frac{u_{\epsilon}^{2}}{4\sigma^{2}}
\right)
\nonumber\\
&\qquad
+\frac{n(n+2)}{4}\,
U\!\left(
\frac{n}{2}+2,
-\frac{n}{2},
-\frac{u_{\epsilon}^{2}}{4\sigma^{2}}
\right)
\Biggr],
\label{eq:smeared_tricomi}
\end{align}
with $U(a, b, c)$ the Tricomi confluent hypergeometric function~\cite{NIST:DLMF}. In general, after taking the limit $\epsilon\to 0$, the smeared
distribution $\mathfrak W(u)$ is not regular at $u=0$. The exceptional case is $n=1$, for which the spatial smearing renders the distribution regular at the origin. Indeed, the small $u$ behavior of Eq.~\eqref{eq:smeared_tricomi} for $n=1$ reads
\begin{equation}
    \mathfrak{W}(u) = \frac{1}{48 \pi^2\sigma^4} + \mathcal{O}(|u|).
    \label{eq:tricomi_stuff_n=1}
\end{equation}
 On the other hand, for $n = 2$ we have a logarithmic divergence,
\begin{equation}
    \mathfrak{W}(u) \sim -\frac{3}{2048 \pi^2 \sigma^6} \log\left(-\frac{u_{\epsilon}^2}{4\sigma^2} \right).
    \label{eq:tricomi_stuff_n=2}
\end{equation}
Finally, for $n \ge 3$, one can show that
\begin{equation}
    \mathfrak{W}(u) \sim \frac{B_{n}}{(u - \ii \epsilon)^{n -2}}
    \label{eq:another_day_another_distribution}
\end{equation}
in the distributional sense, where $B_{n}$ is a constant that depends on $A_{n}$ and $\sigma$.

Although the singular structure strongly depends on the dimension, all three cases define tempered distributions whose action on the smooth switching convolution is finite.
 For $n=1$, $\mathfrak{W}(u)$ is regular at $u=0$, so no distributional prescription is needed. For $n=2$, the singularity is only logarithmic and therefore locally integrable. For $n\geq 3$, the short distance behavior is given by the distribution in Eq.~\eqref{eq:another_day_another_distribution}, and, therefore, the integral yielding the local noise can be evaluated with the aid of identity \eqref{eq:ladyD}.

The resulting expressions are considerably more cumbersome than those obtained in Sec.~\ref{sec:vazio_pointlike} for the pointlike case. Rather than deriving explicit formulas for both the local noise $\mathcal{L}_{ii}$ and the correlation terms $\mathcal{M}$, let us focus on the UV behaviour of the entanglement harvesting setup with smeared detectors.


Since the local noise has a well-defined distributional limit, it is enough to analyse the correlation term $\mathcal M$. Spatial integration can sample arbitrarily small relative distances between the field operators even when the detector centres differ. The resulting singular contributions at the time-ordering boundary depend on the spatial overlap as well as the temporal sampling.

For Gaussian profiles, taking $\sigma\to0$ at fixed $\epsilon>0$ recovers the pointlike kernel at the chosen separation. This limit need not be uniform in $\epsilon$. At fixed nonzero width, overlapping profiles can retain a divergence as $\epsilon\to0^+$, whereas taking the pointlike limit first at fixed $d>0$ gives the finite result established in Sec.~\ref{sec:vazio_pointlike}. The UV behaviour at finite width must therefore be determined from the small-$u$ expansion of the smeared kernel.


To see this explicitly, let us evaluate the smeared distribution $\mathfrak{W}_{\tc{ab}}$. Because $\Xi_{0}(-u, -\boldsymbol{y}) = \Xi_{0}(u, \boldsymbol{y})^{*}$, we can rewrite Eq.~\eqref{eq:smeared_distribution_ab_n_dim} as
\begin{equation}
    \mathfrak{W}_{\tc{ab}}(u) = \Theta(u)\mathfrak{W}_{\boldsymbol{d}}(u) + \Theta(-u)\mathfrak{W}_{\boldsymbol{d}}(u)^{*},
    \label{eq:mathfrak_taxi_home}
\end{equation}
with
\begin{equation}
    \mathfrak{W}_{\boldsymbol{d}}(u) = \int_{\mathbb{R}^{n}} \dd^{n}\boldsymbol{y} \, \vartheta_{\tc{ab}}(\boldsymbol{y})\Xi_{0}(u, \boldsymbol{y}).
\end{equation}
In Appendix \ref{sec:appendix_smeared_distribution}, we show that
\begin{align}
    \mathfrak{W}_{\boldsymbol{d}}(u)
    &=
    \frac{
        A_{n}e^{-d^{2}/(4\sigma^{2})}
    }{
        2^{n-1}\sigma^{n}
    }
    \sum_{m=0}^{\infty}
    \frac{1}{m!}
    \left(
        \frac{d^{2}}{16\sigma^{4}}
    \right)^{m}
    (-u_{\epsilon}^{2})^{m-\frac{n}{2}-1}\nonumber
    \\
    &\quad \times
    \mathcal{U}_{m}\!\left(
        -\frac{u_{\epsilon}^{2}}{4\sigma^{2}}
    \right),
\end{align}
with $\alpha_{m} \equiv m + n/2$, and
\begin{align}
    \mathcal{U}_{m}(z)
    & =
    \frac{n}{2}
    U\!\left(
        \alpha_{m},
        m-\frac{n}{2}-2,
        z
    \right)
    \nonumber \\ & -
    (n+2)\alpha_{m}
    U\!\left(
        \alpha_{m}+1,
        m-\frac{n}{2}-1,
        z
    \right) \nonumber \\ & +
    \frac{n}{2}\alpha_{m}(\alpha_{m}+1)
    U\!\left(
        \alpha_{m}+2,
        m-\frac{n}{2},
        z
    \right).
    \label{eq:app_Um_def_main}
\end{align}
One can check that, in the limit $d \to 0$, 
$\mathfrak{W}_{\boldsymbol{d}}(u)$ reduces to
Eq.~\eqref{eq:smeared_tricomi}, since only the term $m=0$
contributes to the sum. For each fixed $d$ and $n$, the divergence structure of $ \mathfrak{W}_{\boldsymbol{d}}(u)$ and hence $\mathfrak{W}_{\tc{ab}}(u)$, can be determined by expanding in series around $u = 0$. The details of this calculation are presented in Appendix \ref{sec:new_appendix}. For $n=1$ the leading small $u$
behaviour of $\mathfrak{W}_{\tc{ab}}(u)$ is regular,
\begin{equation}
    \mathfrak{W}_{\tc{ab}}(u) = \frac{e^{-d^2/(4\sigma^2)}}{48\pi^2\sigma^4}
    {}_1F_1\!\left(-\frac32;\frac12;
        \frac{d^2}{4\sigma^2}\right) + \mathcal O(|u|).
    \label{eq:n=1}
\end{equation}
For $n=2$, one obtains 
\begin{align}
   \mathfrak W_{\tc{ab}}(u)
   & = -\frac{3e^{-d^2/(4\sigma^2)}}{1024\pi^2\sigma^6} \left( \frac{1}{2} -\frac{d^2}{4\sigma^2} + \frac{d^4}{64\sigma^4}\right) \nonumber \\ & 
    \times\Biggl[
    \Theta(u)\log\!\left(-\frac{(u-\ii\epsilon)^2}{4\sigma^2}\right)
    \nonumber \\ & +\Theta(-u)\log\!\left(-\frac{(u+\ii\epsilon)^2}{4\sigma^2}\right)
    \Biggr]+\mathcal O(1).
    \label{eq:n=2}
\end{align}
Finally, for any $n\geq 3$, 
\begin{align}
       \mathfrak W_{\tc{ab}}(u)
   & \sim e^{-d^2/(4\sigma^2)}
    \left(1-\frac{d^2}{n\sigma^2}+\frac{d^4}{4n(n+2)\sigma^4}\right) \times \nonumber \\ & \Biggl[
        \frac{B_n\Theta(u)}{(u-\ii\epsilon)^{n-2}}
        +\frac{B_n^*\Theta(-u)}{(u+\ii\epsilon)^{n-2}}
    \Biggr]
    \label{eq:n_3}
\end{align}
Therefore, for $n=1$ and $n=2$, the non-local correlation term $\mathcal{M}$ is free of UV divergences. The case $n=1$ is immediate, since $\mathfrak{W}_{\tc{ab}}(u)$ is regular at $u=0$; we will explicitly evaluate the negativity in this case in the next subsection. For $n=2$, the singularity is only logarithmic and hence locally integrable. Therefore, the limit $\epsilon \to 0^{+}$ can be taken before evaluating the integral in Eq.~\eqref{eq:Mterm_smeared_vacuum_n_dim}. Those cases are consistent with the
standard role of spatial profiles in UDW-like models, where the finite size of
the detector can act as a natural UV regulator
\cite{Schlicht, Jorma, Langlois_2006}.

For $n\geq3$, the coefficient of the leading
term in Eq.~\eqref{eq:n_3} vanishes at the two separations
\begin{equation}
    d_\pm=2\sigma\sqrt{\frac n2+1\pm\sqrt{\frac n2+1}}.
    \label{eq:exceptional_separations}
\end{equation}
At these separations, the small-$u$ behaviour
is determined by the subleading terms. For $n=3$, the remaining
term is bounded, while for $n=4$ it is at most logarithmically
singular. Thus, $\mathcal M$ is free of UV divergences in both
cases. For $n\geq5$, cancelling the displayed term alone is
generally insufficient, since the subleading terms can still
be nonintegrable.


At a generic separation, the leading term in Eq.~\eqref{eq:n_3} yields a half-line integral of the form in Eq.~\eqref{eq:defintegral_n_main}, with $m=n-2$ and
$f(u)=\Phi_{\tc{ab}}(u)+\Phi_{\tc{ab}}(-u)$. This follows from
$B_n^*=(-1)^{n-2}B_n$ and the change of variable $u\mapsto-u$ in the negative half-line contribution. The divergent coefficients are therefore determined by Eq.~\eqref{eq:brazilianBBQ}, or by Eq.~\eqref{eq:Harveydent} when $n=3$.

Although $f$ is even, the logarithmic coefficient now involves
$f^{(n-3)}(0)$. It vanishes for even $n$, but need not vanish for odd $n$. For example, at $n=3$,
\begin{equation}
 \int_0^\infty\dd u\,\frac{f(u)}{u-\ii\epsilon}
 =-f(0)\log\!\left(\frac{\epsilon}{\tau_0}\right)+\mathcal O(1),
 \label{eq:edu_smeared_log}
\end{equation}
where $\tau_0>0$ is a fixed reference time that only affects the finite part. Thus, the generic divergence of $\mathcal M$ in $3+1$ dimensions is logarithmic. In higher dimensions, the subleading singular terms must also be included when determining all divergent coefficients. Compactly supported switchings with non-overlapping supports make $\Phi_{\tc{ab}}$ vanish in a neighbourhood of $u=0$, eliminating every such contribution.


Spatial smearing averages the two-point function over the relative positions of the field operators, both in the present model and in the quadratic amplitude-coupling models of Refs.~\cite{SachsMannEdu,sachs2018entanglement}. The singularity of this spatial average determines whether the subsequent time-ordered integral is finite. For the energy-density coupling with Gaussian profiles, the remaining kernel is locally integrable in $1+1$ and $2+1$ dimensions. In higher dimensions, finiteness depends on the coefficients of the nonintegrable terms and on the switching functions.


The dependence on the spatial profile can be made explicit. For smooth, rapidly decreasing profiles, the coefficient of the leading singular term at $n\geq3$ is proportional to
$\Delta_{\boldsymbol y}^{\,2}\vartheta_{\tc{ab}}(\boldsymbol0)$, where $\Delta_{\boldsymbol y}$ denotes the spatial Laplacian and $\vartheta_{\tc{ab}}$ is the spatial overlap defined in Eq.~\eqref{eq:edu_spatial_overlap_definition}. Appendix~\ref{sec:edu_general_profiles} derives this statement and shows that, for the Gaussian profiles,
\begin{align}
 \Delta_{\boldsymbol y}^{\,2}\vartheta_{\tc{ab}}(\boldsymbol0)
 &=\frac{n(n+2)}{4\sigma^4}\vartheta_{\tc{ab}}(\boldsymbol0)
 \nonumber\\
 &\quad\times\left(1-\frac{d^2}{n\sigma^2}
 +\frac{d^4}{4n(n+2)\sigma^4}\right).
 \label{eq:edu_gaussian_bilaplacian}
\end{align}
The polynomial in Eq.~\eqref{eq:n_3} therefore expresses a cancellation of this spatial derivative. It can vanish even when the overlap $\vartheta_{\tc{ab}}(\boldsymbol0)$ is nonzero. For $n\geq5$, higher spatial derivatives can control further nonintegrable terms. Appendix~\ref{sec:edu_general_profiles} also derives sufficient conditions on these spatial derivatives for the correlation term to remain finite and shows that smooth compactly supported profiles separated by a positive distance satisfy them.

These observations give a common interpretation of the finite cases. A persistent divergence can arise if the spatially sampled kernel is not locally integrable near $u=0$, the lower endpoint of the half-line integral obtained by combining the two time orderings. In low dimensions, spatial smearing weakens that singularity sufficiently to make it integrable. A suitable choice of spatial profiles can also cancel its coefficients. If the profiles have compact supports separated by a positive distance, $\vartheta_{\tc{ab}}$ vanishes near $\boldsymbol y=\boldsymbol0$, and the sampled correlator is regular near $u=0$. Alternatively, non-overlapping compact switchings make the temporal sampling vanish near this endpoint. The relevant spatial criterion is therefore the local behaviour of the profile overlap, including its derivatives, rather than the presence of non-compact tails alone.

\subsection*{Example in \texorpdfstring{$1 + 1$}{1 + 1} dimensions}
\label{sec:vacuum_smeared_examples}

We now illustrate the previous general analysis with an explicit example in $1 + 1$ dimensions. In this case, the energy density two-point function is evaluated by taking the limit $n \to 1$ in Eq.~\eqref{eq:energy_density_twopoint_general_vacuum}. The resulting expression is
\begin{equation}
    \Xi_{0}(u, y) = \frac{1}{8\pi^2}\left[\frac{1}{(u - y - \ii \epsilon)^4} + \frac{1}{(u + y - \ii \epsilon)^4}\right].
    \label{eq:W0_uy}
\end{equation}
In what follows, we will need to evaluate integrals of the form (for any $a \in \mathbb{R}$)
\begin{align}
    \mathcal{J}_{\pm}(a) & = \lim_{\epsilon \to 0^{+}} \frac{1}{16 \pi^{5/2}\sigma} \int_{-\infty}^{\infty}\dd x  \, \frac{e^{-x^2/(4\sigma^2)}}{(x + a \pm \ii \epsilon)^4} \nonumber \\ 
&=
\frac{1}{768\pi^2\sigma^7}
\Biggl[
-2a^2\sigma
+8\sigma^3
\nonumber\\
&\quad
+\sqrt{\pi}\,a
\left(a^2-6\sigma^2\right)
e^{-\frac{a^2}{4\sigma^2}}
\left[
\mp\ii+\operatorname{erfi}\!\left(\frac{a}{2\sigma}\right)
\right]
\Biggr].
\end{align}
Using Eq.~\eqref{eq:mathfrack_and_friends}, it is straightforward to show that
\begin{equation}
   \mathfrak{W}(u) = 2\mathcal{J}_{-}(u), 
\end{equation}
which is the same for both detectors, as expected. Moreover, observe that
\begin{equation}
    \mathfrak{W}(0) = \frac{1}{48 \pi^2 \sigma^4},
\end{equation}
which is consistent with Eq.~\eqref{eq:tricomi_stuff_n=1}. Therefore,
\begin{equation}
    \mathcal{L}_{jj} = 2\int_{-\infty}^{\infty}\dd u \, \Phi_{j}(u) \mathcal{J}_{-}(u).
    \label{eq:niceFjjsmeared_vacuum}
\end{equation}
Next, setting $n=1$ in Eqs.~\eqref{eq:smeared_distribution_ab_n_dim} and \eqref{eq:vartheta_ab_gaussian_compact} yields (with $d \equiv x_{\tc{b}} - x_{\tc{a}}$)
\begin{align}
    \mathfrak{W}_{\tc{ab}}(u) & = \Theta(u)[\mathcal{J}_{+}(-u-d) + \mathcal{J}_{-}(u - d)] \nonumber \\ 
    & + \Theta(-u)[\mathcal{J}_{-}(-u-d) + \mathcal{J}_{+}(u - d)].
\end{align}
Therefore, the correlation terms can be compactly written as
\begin{align}
    \mathcal{M}  & = \int_{0}^{\infty}\dd u \, \Phi_{\tc{ab}}(u)[\mathcal{J}_{+}(-u-d) + \mathcal{J}_{-}(u - d)] \nonumber \\ & +   
    \int_{-\infty}^{0}\dd u \, 
    \Phi_{\tc{ab}}(u)[\mathcal{J}_{-}(-u-d) + \mathcal{J}_{+}(u - d)].
    \label{eq:termMvacuum_smeared_nice}
\end{align}

\begin{figure}
    \centering
    \includegraphics[width=\linewidth]{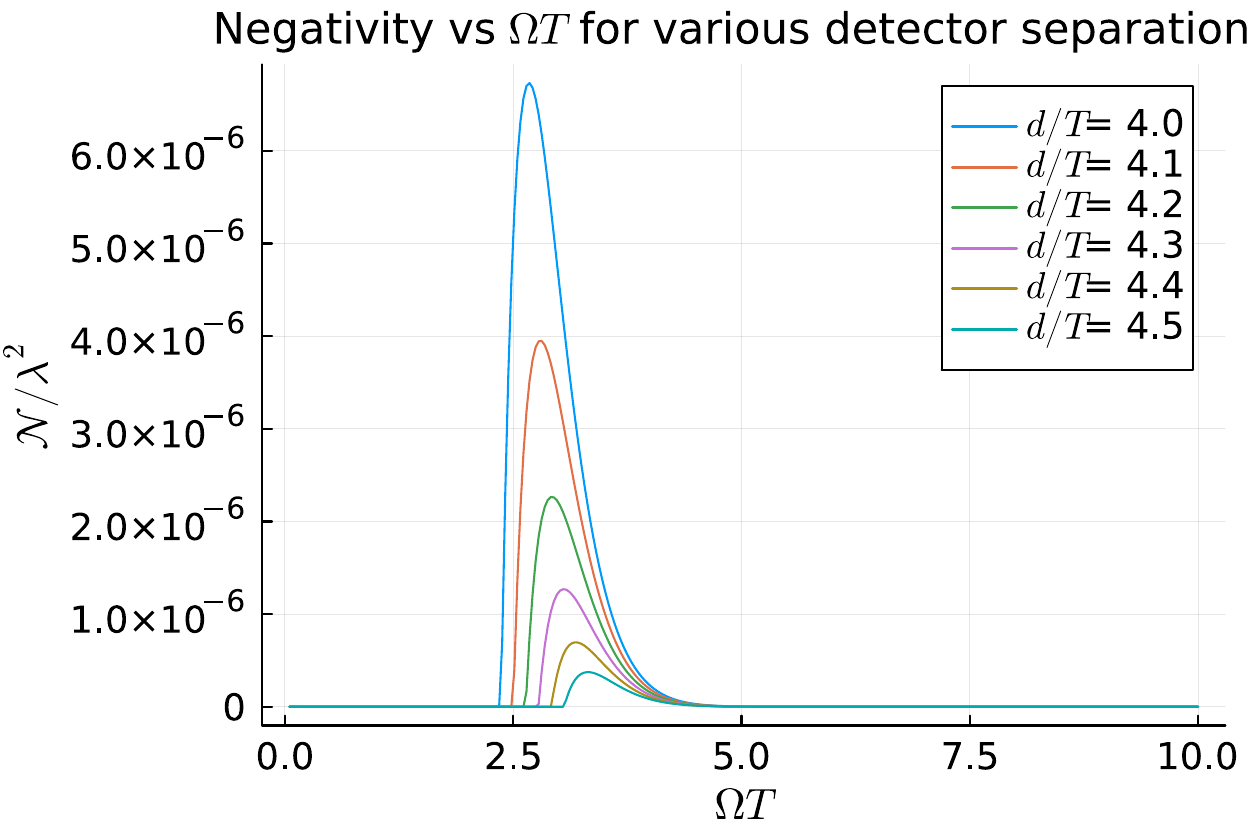}
    \caption{Negativity as a function of the detectors' gap $\Omega T$ for multiple values of the separation between the detectors, $d$. In each plot, we choose  the separation $\delta t/T = 0$, and the \eduadd{width of the Gaussian spatial profile} $\sigma/T = 0.1$}
    \label{fig:SmearedVacuumNegativity}
\end{figure}

To quantify the entanglement acquired by the detectors, we consider a Gaussian switching of the form
\begin{equation}
    \chi_{j}(t) = e^{-\frac{(t - t_{j})^2}{T^2}}.
    \label{eq:gaussian_switching}
\end{equation}
In this case, the convolution product $\Phi_{j}(u)$ reduces to
\begin{equation}
    \Phi(u) = 
    \sqrt{\frac{\pi}{2}}\,T \, e^{-\ii \Omega u} e^{-\frac{u^{2}}{2T^{2}}},
\end{equation}
which is the same for both detectors. This means that the local noise is the same for both detectors, so that the
simplified expression for the negativity, Eq.~\eqref{eq:negativity_vanilla},
holds. Denoting the local noise simply by $\mathcal{L}_{0}$ (which explicitly emphasizes that this is the vacuum contribution), we have
\begin{align}
      \mathcal{L}_{\textsc{aa}}&= \mathcal{L}_{\textsc{bb}}\equiv \mathcal{L}_{0} =  
    \frac{
        T^{2}
        \left(\Omega^{2}T^{4}+2T^{2}+4\sigma^{2}\right)
    }{
        24\pi\left(T^{2}+2\sigma^{2}\right)^{3}
    }
    e^{-T^{2}\Omega^{2}/2}
    \nonumber\\
    &\quad
    -
    \frac{
        T^{4}\Omega
        \left[\Omega^{2}T^{4}+3\left(T^{2}+2\sigma^{2}\right)\right]
    }{
        24\sqrt{2\pi}\left(T^{2}+2\sigma^{2}\right)^{7/2}
    }
    e^{-\frac{\sigma^{2}T^{2}\Omega^{2}}{T^{2}+2\sigma^{2}}}
    \nonumber\\
    &\quad\times
    \operatorname{erfc}\!\left(
        \frac{\Omega T^{2}}
        {\sqrt{2\left(T^{2}+2\sigma^{2}\right)}}
    \right).
    \label{eq:integral_result}
\end{align}
On the other hand,
\begin{equation}
    \Phi_{\tc{ab}}(u)
    =
    \sqrt{\frac{\pi}{2}}\,T\,
    e^{-\frac{(\delta t+u)^2}{2T^2}}\,
    e^{\ii\Omega(t_{\tc{a}}+t_{\tc{b}})}\,
    e^{-\frac{T^2\Omega^2}{2}} .
    \label{eq:Phi_ab_gaussian}
\end{equation}
Using this expression, the non-local correlations in
Eq.~\eqref{eq:termMvacuum_smeared_nice} can be evaluated numerically. In Fig.~\ref{fig:SmearedVacuumNegativity}, we show the resulting negativity as a function of the detector gap $\Omega$, for different spatial separations $d$. We fix $\delta t=0$, \eduadd{so that the switchings are centred at the same time. The detector centres at those times are spacelike separated.}

Because the Gaussian profiles have non-compact support, the detector profiles are not strictly spacelike separated. However, the overlap between detectors separated by the values of $d$ shown in Fig.~\ref{fig:SmearedVacuumNegativity} is
exponentially suppressed (in our case, we set $\sigma = 0.1$).\footnote{We apply the methods of Refs.~\cite{ericksonNew,MatheusAdamEdu2024} to separate the contribution of pre-existing vacuum correlations from that of field-mediated communication. For $d/T=4$, $\sigma/T=0.1$, $\delta t=0$, and $\Omega T=2.7$, the leading-order negativity remains positive when the commutator contribution associated with communication is omitted. This certifies that the detectors harvest vacuum entanglement at these parameters. Communication further increases the resulting negativity.}

The example illustrates how the spatial smearing completely removes the UV obstruction in 1+1 dimensions while preserving the characteristic harvesting behaviour, namely an optimal detector gap and a monotonic suppression of entanglement with increasing spatial separation \eduadd{over the parameter range shown, consistently with} Ref.~\cite{hectorMass}.

\section{Arbitrary  Hadamard states in \texorpdfstring{$1 + 1$}{1 + 1}-dimensional Minkowski spacetime}
\label{sec:general_hadamard}

The results of the previous sections established that, for Gaussian smeared detectors in $1+1$ dimensions, the vacuum contribution to the harvesting protocol is free of UV divergences. We now show that this extends to arbitrary Hadamard, zero-mean Gaussian states. The key observation is that every Hadamard two-point function differs from the Minkowski vacuum only by a smooth biscalar. Consequently, the singular structure of the harvesting protocol is entirely determined by the vacuum contribution, while all state-dependent corrections are either manifestly regular or involve only mild cross terms that can also be treated distributionally.

We now consider scenarios in which the field is prepared in a Hadamard, zero-mean Gaussian state $\hat{\rho}_{\phi}$. Recall that, in this case, because the field is massless, the corresponding Wightman function admits the decomposition
\begin{equation}
    W(\mf x,\mf x')
    =
    W_{0}(\mf x,\mf x')
    +
    w(\mf x,\mf x') ,
\end{equation}
where $W_{0}(\mf x,\mf x')$ denotes the Minkowski vacuum contribution, while the smooth bi-scalar $w(\mf x,\mf x')$ encodes the state-dependent part of the two-point function. The renormalized energy density reads
\begin{equation}
    \mathcal{E}_{\phi}(\mf x)
    =
    \frac{1}{2}
    \lim_{\mf x'\to \mf x}
    \sum_{\mu=0}^{n}
    \partial_{\mu}\partial_{\mu'}w(\mf x,\mf x').
    \label{eq:energy_density_w_again}
\end{equation}
Once a state is chosen, the terms 
\begin{equation}
    \mathcal{X}_{j}
    =
    \int \dd V\,
    e^{\ii \Omega t}
    \Lambda_{j}(\mf x)\mathcal{E}_{\phi}(\mf x),
    \label{eq:detector_j_coherence}
\end{equation}
can be evaluated for each detector. As for the local noise and the non-local correlations, observe that the energy density two-point function, Eq.~\eqref{eq:two_point_energy_density_concrete}, can be split into four terms, namely,
\begin{align}
    \Xi_{\phi}(\mf x, \mf x') & = \Xi_{0}(\mf x, \mf x') + \Xi_{w}(\mf x, \mf x') + \Xi_{\tc{c}}(\mf x, \mf x') \nonumber \\ & \quad + \Xi_{\tc{e}}(\mf x, \mf x'),
    \label{eq:split_energy_two_point}
\end{align}
where
\begin{equation}
    \Xi_{w}(\mf x, \mf x') \equiv \frac{1}{2}\sum_{\mu, \nu = 0}^{n}(\partial_{\mu}\partial_{\nu'}w(\mf x, \mf x'))^2,
\end{equation}
\begin{equation}
    \Xi_{\tc{c}}(\mf x, \mf x') \equiv \sum_{\mu, \nu = 0}^{n}\partial_{\mu}\partial_{\nu'}W_{0}(\mf x, \mf x')\partial_{\mu}\partial_{\nu'}w(\mf x, \mf x'),
\end{equation}
and
\begin{equation}
    \Xi_{\tc{e}}(\mf x, \mf x') \equiv \mathcal{E}_{\phi}(\mf x)\mathcal{E}_{\phi}(\mf x').
\end{equation}
In the same token, the local noise for each detector can be schematically written in the form
\begin{equation}
    \mathcal{L}_{jj} = \mathcal{L}_{jj, 0} + \mathcal{L}_{jj, w} + \mathcal{L}_{jj, \tc{c}} + \mathcal{L}_{jj, \tc{e}} \quad ,
\end{equation}
where, for $\Upsilon \in \{0, w, \tc{c} \}$, we define
\begin{equation}
    \mathcal{L}_{jj, \Upsilon}  \equiv\int \dd V \dd V' \, e^{-\ii \Omega (t -t')}\Lambda_{j}(\mf x)\Lambda_{j}(\mf x')\Xi_{\Upsilon}(\mf x, \mf x').
\end{equation}
Similarly,
\begin{equation}
    \mathcal{M} = \mathcal{M}_{0} + \mathcal{M}_{w} + \mathcal{M}_{\tc{c}} + \mathcal{M}_{\tc{e}},
    \label{eq:Mterm_split_hadamard}
\end{equation}
with the definitions
\begin{align}
    \mathcal{M}_{\Upsilon} & \equiv \int \dd V \dd V' \, e^{\ii \Omega (t +t')}\Lambda_{\tc{a}}(\mf x)\Lambda_{\tc{b}}(\mf x')\mathcal{G}_{\Upsilon}(\mf x, \mf x'),
\end{align}
\begin{equation}
    \mathcal{G}_{\Upsilon}(\mf x, \mf x') \equiv  \Theta(t-t')\Xi_{\Upsilon}(\mf x, \mf x') + \Theta(t'-t)\Xi_{\Upsilon}(\mf x', \mf x).
\end{equation}
With all the necessary definitions in place, we now specialize to $1+1$ dimensions and consider smeared detectors with Gaussian spatial profiles, as in Eq.~\eqref{eq:gaussian_and_friends}. In this setting, the entanglement harvesting protocol is UV safe independently of the choice of switching function: the purely vacuum contributions, $\mathcal{L}_{0,jj}$ and $\mathcal{M}_{0}$, are finite and are given by Eqs.~\eqref{eq:niceFjjsmeared_vacuum} and~\eqref{eq:termMvacuum_smeared_nice}, respectively.

The terms $\mathcal{L}_{jj,w}$, $\mathcal{M}_{w}$,
$\mathcal{L}_{jj,\tc{e}}$, and $\mathcal{M}_{\tc{e}}$ depend solely on the smooth bi-scalar $w(\mf x, \mf x')$, and, as such, they can be evaluated without any technical challenges once a state is chosen. In particular, observe that
\begin{equation}
    \mathcal{L}_{jj, \tc{e}} = |\mathcal{X}_{j}|^2,
    \label{eq:mathcalF_e_general}
\end{equation}
and
\begin{equation}
    \mathcal{M}_{\tc{e}} = \mathcal{X}_{\tc{a}} \mathcal{X}_{\tc{b}}.
    \label{eq:mathcalM_e_general}
\end{equation}
As for the cross terms $\mathcal{L}_{jj, \tc{c}}$ and $\mathcal{M}_{\tc{c}}$, they still depend on the bi-distribution arising from the vacuum Wightman function, so they require special treatment.
For convenience, we write
\begin{equation}
    w_{\mu\nu'}(\mf x,\mf x')
    \equiv
    \partial_{\mu}\partial_{\nu'}w(\mf x,\mf x').
\end{equation}
Using the expression for $W_{0}$ given in Eq.~\eqref{eq:Wightman_IR_section}, we find
\begin{align}
    \Xi_{\tc{c}}(\mf x,\mf x')
    &=
    -\frac{((\Delta t-\ii\epsilon)^{2}+\Delta x^{2})    \left[
        w_{tt'}(\mf x,\mf x')
        +
        w_{xx'}(\mf x,\mf x')
    \right]}
    {2\pi\left[(\Delta t-\ii\epsilon)^{2}-\Delta x^{2}\right]^{2}}
    \nonumber \\
    &\quad
    +
    \frac{(\Delta t-\ii\epsilon)\Delta x\left[
        w_{tx'}(\mf x,\mf x')
        +
        w_{xt'}(\mf x,\mf x')
    \right]}
    {\pi\left[(\Delta t-\ii\epsilon)^{2}-\Delta x^{2}\right]^{2}}
    .
\end{align}
This expression simplifies considerably in light-cone coordinates, because the cross-term factorizes into independent left and right-moving contributions,
\begin{equation}
    \zeta = t-x,
    \qquad
    \eta = t+x.
\end{equation}
Indeed, using
\begin{equation}
    \Delta \zeta = \Delta t-\Delta x,
    \qquad
    \Delta \eta = \Delta t+\Delta x,
\end{equation}
one obtains
\begin{equation}
    \Xi_{\tc{c}}(\mf x,\mf x')
    =
    -\frac{1}{\pi}
    \left[
        \frac{w_{\zeta\zeta'}(\mf x,\mf x')}
        {(\Delta \zeta-\ii\epsilon)^2}
        +
        \frac{w_{\eta\eta'}(\mf x,\mf x')}
        {(\Delta \eta-\ii\epsilon)^2}
    \right].
\end{equation}

So the only non-trivial contribution is the cross-term. We reduce $\mathcal{L}_{jj, \tc{c}}$ to a one-variable integral, with the remaining singular kernel interpreted distributionally. To avoid confusion with the change of variables, we define
\begin{equation}
    f_{\zeta}(\mf x,\mf x')
    \equiv
    w_{\zeta\zeta'}(\mf x,\mf x'),
    \qquad
    f_{\eta}(\mf x,\mf x')
    \equiv
    w_{\eta\eta'}(\mf x,\mf x').
    \label{eq:auxiliary_but_important}
\end{equation}
Here, the derivatives with respect to $\zeta,\zeta'$ and $\eta,\eta'$ are evaluated first; the resulting quantities are then regarded as functions of the original spacetime points $\mf x$ and $\mf x'$. In terms of these auxiliary function, the energy density can written as
\begin{equation}
    \mathcal{E}_{\phi}(\mf x) = \lim_{\mf x' \to \mf x}\bigl[f_{\zeta}(\mf x, \mf x') + f_{\eta}(\mf x, \mf x')\bigr].
    \label{eq:energy_density_null_coordinates}
\end{equation}
Just like in the previous sections, we introduce the variables $u = t-t'$, $s = t$, $y = x-x'$, and $z = x$. Then, using the following definitions,
\begin{align}
X_{j,\zeta}(u,y)
&\equiv
\int \dd s\,\dd z\,
\Lambda_{j}(s,z)\Lambda_{j}(s-u,z-y)
\nonumber \\
&\hspace{1.4cm}\times
f_{\zeta}(s,z,s-u,z-y),
\label{eq:Umbrellacorporationzeta}
\end{align}
\begin{align}
X_{j,\eta}(u,y)
&\equiv
\int \dd s\,\dd z\,
\Lambda_{j}(s,z)\Lambda_{j}(s-u,z-y)
\nonumber \\
&\hspace{1.4cm}\times
f_{\eta}(s,z,s-u,z-y),
\label{eq:Umbrellacorporationeta}
\end{align}
we obtain
\begin{align}
    \mathcal{L}_{jj, \tc{c}}
    &=
    -\frac{1}{\pi}
    \int \dd u\,\dd y \,
    e^{-\ii \Omega u}
    \Biggl[
        \frac{X_{j,\zeta}(u,y)}
        {(u-y-\ii\epsilon)^2}
        \nonumber \\
    &\hspace{3.2cm}
        +
        \frac{X_{j,\eta}(u,y)}
        {(u+y-\ii\epsilon)^2}
    \Biggr].
\end{align}
Finally, we change variables separately in the two terms: in the first term we set $\rho=u-y$, while in the second term, we set $\rho=u+y$. In both changes, $u$ is kept fixed. Therefore, we obtain
\begin{equation}
    \mathcal{L}_{jj, \tc{c}}
    =
    -\frac{1}{\pi}
    \int_{-\infty}^{\infty}\dd \rho\,
    \frac{\mathcal{K}_{j}(\rho)}
    {(\rho-\ii\epsilon)^2},
\end{equation}
where
\begin{align}
    \mathcal{K}_{j}(\rho)
    &\equiv
    \int_{-\infty}^{\infty}\dd u\,
    e^{-\ii\Omega u}
    \bigl[
        X_{j,\zeta}(u,u-\rho)
        \nonumber \\
    &\hspace{3.0cm}
        +
        X_{j,\eta}(u,\rho-u)
    \bigr].
    \label{eq:defintion_33}
\end{align}
Then, using Eq.~\eqref{eq:ladyD} for $n = 2$, the distributional limit $\epsilon \to 0^{+}$ can be evaluated as follows:
\begin{align}
    \mathcal{L}_{jj, \tc{c}} & = -\ii \mathcal{K}'_{j}(0) - \frac{1}{\pi}\operatorname{PV}\int_{-\infty}^{\infty} \dd \rho \,  \frac{\mathcal{K}'_{j}(\rho)}{\rho} \nonumber \\ & = -\ii \mathcal{K}'_{j}(0) - \frac{1}{\pi}\int_{0}^{\infty} \dd \rho \, \frac{\mathcal{K}'_{j}(\rho) - \mathcal{K}'_{j}(-\rho)}{\rho}.
    \label{eq:cross_local_noise_general}
\end{align}
As for the term $\mathcal{M}_{\tc{c}}$, a similar procedure allows us to write
\begin{align}
    \mathcal{M}_{\tc{c}} & = 
    -\frac{1}{\pi}\int \dd u\,\dd y \,
    \Theta(u)e^{-\ii \Omega u}
    \Biggl[
        \frac{X_{\tc{ab},\zeta}(u,y)}
        {(u-y-\ii\epsilon)^2}
        \nonumber \\
    &\hspace{3.2cm}
        +
        \frac{X_{\tc{ab},\eta}(u,y)}
        {(u+y-\ii\epsilon)^2}
    \Biggr] \nonumber \\ &
    -\frac{1}{\pi}\int \dd u\,\dd y \,
    \Theta(-u)e^{-\ii \Omega u}
    \Biggl[
        \frac{X_{\tc{ab},\zeta}(u,y)}
        {(u-y+\ii\epsilon)^2}
        \nonumber \\
    &\hspace{3.2cm}
        +
        \frac{X_{\tc{ab},\eta}(u,y)}
        {(u+y+\ii\epsilon)^2}
    \Biggr].
\end{align}
with the definitions
\begin{align}
    X_{\tc{ab}, \zeta}(u, y) &  \equiv \int \dd s\,\dd z\,
\Lambda_{\tc{a}}(s,z)\Lambda_{\tc{b}}(s-u,z-y)
\nonumber \\
&\hspace{1.4cm}\times
e^{2\ii \Omega s}f_{\zeta}(s,z,s-u,z-y),
\label{eq:auxiliaryfunction67}
\end{align}
\begin{align}
    X_{\tc{ab}, \eta}(u, y) &  \equiv \int \dd s\,\dd z\,
\Lambda_{\tc{a}}(s,z)\Lambda_{\tc{b}}(s-u,z-y)
\nonumber \\
&\hspace{1.4cm}\times
e^{2\ii \Omega s}f_{\eta}(s,z,s-u,z-y).
\label{eq:auxiliaryfunction68}
\end{align}
To reduce this expression to a one-variable integral to be evaluated distributionally, we introduce two kernels, namely
\begin{equation}
    \mathcal{K}^{+}_{\tc{ab}}(\rho)
    \equiv
    \int_{0}^{\infty}\dd u\,
    e^{-\ii\Omega u}
    \left[
        X_{\tc{ab},\zeta}(u,u-\rho)
        +
        X_{\tc{ab},\eta}(u,\rho-u)
    \right],
    \label{eq:kernalabplus}
    \end{equation}
    \begin{equation}
            \mathcal{K}^{-}_{\tc{ab}}(\rho)
    \equiv
    \int_{-\infty}^{0}\dd u\,
    e^{-\ii\Omega u}
    \left[
        X_{\tc{ab},\zeta}(u,u-\rho)
        +
        X_{\tc{ab},\eta}(u,\rho-u)
    \right].
    \label{eq:kernalabminus}
    \end{equation}
Therefore,
\begin{equation}
    \mathcal{M}_{\tc{c}}
    =
    -\frac{1}{\pi}
    \int_{-\infty}^{\infty}\dd \rho\,
    \left[
        \frac{\mathcal{K}^{+}_{\tc{ab}}(\rho)}
        {(\rho-\ii\epsilon)^2}
        +
        \frac{\mathcal{K}^{-}_{\tc{ab}}(\rho)}
        {(\rho+\ii\epsilon)^2}
    \right].
\end{equation}
Fortunately, this integral can also be evaluated using identity
\eqref{eq:ladyD} for $n = 2$. In terms of the kernels
$\mathcal{K}_{\tc{ab}}^{\pm}(\rho)$, the result reads
\begin{align}
    \mathcal{M}_{\tc{c}}
    &=
    -\ii \mathcal{K}_{\tc{ab}}^{+ \, \prime}(0)
    +
    \ii \mathcal{K}_{\tc{ab}}^{- \, \prime}(0)
    \nonumber \\
    &\quad
    -
    \frac{1}{\pi}
    \operatorname{PV}
    \int_{-\infty}^{\infty}\dd \rho\,
    \frac{
        \mathcal{K}_{\tc{ab}}^{+ \, \prime}(\rho)
        +
        \mathcal{K}_{\tc{ab}}^{- \, \prime}(\rho)
    }{\rho}.
    \label{eq:cross_correlations_general}
\end{align}
Let us briefly recapitulate the results obtained so far. When the field is
prepared in a Hadamard, zero-mean Gaussian state, the local noise and
correlation term decomposes into four distinct contributions. Two of them depend
only on the detector profiles and on the smooth state-dependent biscalar
$w(\mf x,\mf x')$. Their evaluation is therefore insensitive to the
$\ii\epsilon$ prescription and involves no distributional subtleties.

The vacuum contribution is precisely the one studied in
Sec.~\ref{sec:vacuum_smeared_examples}. In particular, for smeared detectors with Gaussian
spatial profiles, this contribution is free from UV divergences. The cross terms $\mathcal{L}_{jj,\tc{c}}$ and
$\mathcal{M}_{\tc{c}}$ involve both the smooth bi-scalar $w(\mf x,\mf x')$
and the vacuum Wightman function $W_0(\mf x,\mf x')$. As shown in
Eqs.~\eqref{eq:cross_local_noise_general} and
\eqref{eq:cross_correlations_general}, these terms can also be written in a
finite form after the distributional limit $\epsilon\to0^+$ is taken.

We therefore conclude that, for Gaussian smeared detectors in $1+1$ dimensions, every contribution to the harvesting protocol associated with an arbitrary Hadamard Gaussian zero-mean state is finite. Since the singular structure is entirely inherited from the Minkowski vacuum and the latter has already been shown to be UV finite after spatial smearing, no additional UV divergences arise from changing the quantum state of the field.

\subsection*{Example: Thermal states}
\label{sec:thermal_states}

To illustrate the formalism developed above, we consider a setup where the field is prepared in a thermal state \cite{Petar, PhysRevA.88.062336}. For a quantum system whose Hamiltonian $\hat H$ acts on a finite-dimensional
Hilbert space, a thermal state at temperature $(k_{\tc{b}}\beta)^{-1}>0$ is
defined by the Gibbs density matrix
\begin{equation}
    \hat{\rho}_{\beta}
    =
    \frac{e^{-\beta \hat{H}}}{\mathcal{Z}_{\beta}},
    \label{eq:Gibbs}
\end{equation}
where $\mathcal{Z}_{\beta}$ is the corresponding partition function,
\begin{equation}
    \mathcal{Z}_{\beta}
    =
    \operatorname{Tr}\!\left(e^{-\beta \hat{H}}\right).
\end{equation}
For quantum fields, however, this
characterization is, in general, not the appropriate way of defining
thermal equilibrium. Instead, thermal states are characterized by the
Kubo-Martin-Schwinger (KMS) condition \cite{Kubo, MartinSchwinger}. 

We will assume that the field is prepared in a KMS state $\hat{\rho}_{\beta}$ at
inverse temperature $\beta$. In $1+1$ dimensions, the corresponding thermal
Wightman function can then be written as \cite{derivativeJorma}
\begin{align}
    W_{\beta}(\mf x,\mf x')
    &=
    -\frac{1}{4\pi}
    \log\!\left[
        \frac{\beta\Lambda}{\pi}
        \sinh\!\left(\frac{\pi}{\beta}(\Delta \zeta-\ii\epsilon)\right)
    \right]
    \nonumber \\
    &\quad
    -\frac{1}{4\pi}
    \log\!\left[
        -\frac{\beta\Lambda}{\pi}
        \sinh\!\left(\frac{\pi}{\beta}(\Delta \eta-\ii\epsilon)\right)
    \right],
    \label{eq:thermal_wightman_1p1}
\end{align}
where natural units are assumed ($k_{\tc{b}} = 1$), and we recall that $\Lambda$ is the same IR regulator as in the vacuum Wightman, Eq.~\eqref{eq:Wightman_IR_section}. In the limit of zero temperature, or, equivalently, $\beta \to \infty$, one can verify that $ W_{\beta}(\mf x,\mf x')$ reduces to the vacuum Wightman function, Eq.~\eqref{eq:Wightman_IR_section}. A straightforward algebraic manipulation  allows us to cast $W_{\beta}(\mf x, \mf x')$ into
\begin{equation}
    W_{\beta}(\mf x, \mf x') = W_{0}(\mf x, \mf x') + w_{\beta}(\mf x, \mf x'),
\end{equation}
through which we can identify the state-dependent bi-scalar as
\begin{align}
    w_{\beta}(\mf x,\mf x') = -\frac{1}{4\pi}\log \left[\frac{\beta^2}{\pi^2}\frac{\sinh\left(\frac{\pi\Delta \zeta}{\beta}\right)\sinh\left(\frac{\pi \Delta \eta}{\beta}\right)}{\Delta \zeta \Delta \eta} \right],
\end{align}
where the limit $\epsilon \to 0^{+}$ was taken because this expression is regular in the coincidence limit $\mf x' \to \mf x$ (or, equivalently, when $\Delta \zeta, \Delta \eta \to 0$). Then, the auxiliary functions in Eq.~\eqref{eq:auxiliary_but_important} can be evaluated as
\begin{equation}
    f_{\zeta}(\mf x, \mf x'; \beta) = g_{\beta}(\Delta \zeta),
    \label{eq:f_zeta_beta}
\end{equation}
and
\begin{equation}
    f_{\eta}(\mf x, \mf x'; \beta) = g_{\beta}(\Delta \eta),
    \label{eq:f_eta_beta}
\end{equation}
with the definition
\begin{equation}
    g_{\beta}(a) \equiv \frac{1}{4\pi a^{2}}
    -
    \frac{\pi}{4\beta^{2}}
    \csch^{2}\!\left(
        \frac{\pi a}{\beta}
    \right).
    \label{eq:aqui_estou_mais_um_dia}
\end{equation}
Then, using Eq.~\eqref{eq:energy_density_null_coordinates}, we find the following expression for the energy density:
\begin{equation}
    \mathcal{E}_{\beta}  =\frac{\pi}{6 \beta^2}.
\end{equation}
Therefore, the thermal state $\hat{\rho}_{\beta}$ has constant, positive energy density. Let us now evaluate each one of the terms required to compute the negativity. We begin with 
\begin{align}
    \mathcal{X}_{j} & = \frac{\pi}{6 \beta^2}\int \dd t \dd x \, e^{\ii \Omega t} \chi_{j}(t) F_{j}(x) \nonumber \\ &
    = \frac{\pi^{3/2} T}{6 \beta^2}e^{\ii t_{j}\Omega}e^{-\frac{T^{2}\Omega^{2}}{4}},
\end{align}
where we recall that both the switching and the spatial profile are Gaussian, as in Eqs.~\eqref{eq:gaussian_switching} and \eqref{eq:gaussian_and_friends}. Next, we evaluate the different contributions to the energy density two-point function, as in Eq.~\eqref{eq:split_energy_two_point}. We have
\begin{equation}
    \Xi_{\tc{e}} = \frac{\pi^2}{36 \beta^4},
    \label{eq:W_e_thermal}
\end{equation}
\begin{equation}
    \Xi_{\tc{c}}(\mf x,\mf x')
    =
    -\frac{1}{\pi}
    \left[
        \frac{g_{\beta}(\Delta \zeta)}
        {(\Delta \zeta-\ii\epsilon)^2}
        +
        \frac{g_{\beta}(\Delta \eta)}
        {(\Delta \eta-\ii\epsilon)^2}
    \right].
\end{equation}
and
\begin{equation}
    \Xi_{w}(\mf x, \mf x')  = 2 [g_{\beta}(\Delta \zeta)^2 + g_{\beta}(\Delta \eta)^2].
\end{equation}
Using those, we can set up the integrals for the local noise terms. Starting with the simplest one, we have
\begin{equation}
    \mathcal{L}_{\tc{e}} = \frac{\pi^3 T^2}{36 \beta^4}e^{-\frac{(\Omega T)^2}{2}},
\end{equation}
where we dropped the subscript $jj$ as this quantity is the same for both detectors. It turns out that the state-dependent term $\mathcal{L}_{jj, w}$, is also detector-independent in this case, so we simply denote it by $\mathcal{L}_{w}$. In Appendix \ref{sec:garrafa_termica}, we show that it can be simplified as
\begin{equation}
    \mathcal{L}_{w}
    =
    \frac{
        4\sqrt{2\pi}\,T^2
        e^{-\frac{T^2\sigma^2\Omega^2}{D}}
    }{
        \sqrt{D}
    }
    \int_{0}^{\infty}\dd s\,
    e^{-\frac{s^2}{2D}}
    \cos\!\left(
        \frac{T^2\Omega}{D}s
    \right)
    g_{\beta}(s)^2,
\end{equation}
with $D \equiv T^2 + 2 \sigma^2$. Lastly, we need to evaluate all the auxiliary functions required to evaluate the cross term $\mathcal{L}_{jj, \tc{c}}$ using Eq.~\eqref{eq:cross_local_noise_general}. Starting with the integrals from Eqs.~\eqref{eq:Umbrellacorporationzeta}-\eqref{eq:Umbrellacorporationeta}, a straightforward calculation yields (where, once more, the detector subscript is dropped as the result is the same for $j  = \tc{a}$ and $j = \tc{b}$)
  \begin{equation}
    X_{\zeta}(u,y)
    =
    \frac{T}{2\sqrt{2}\,\sigma}
    e^{-\frac{u^{2}}{2T^{2}}}e^{-\frac{y^{2}}{4\sigma^{2}}}
    g_{\beta}(\Delta\zeta),
\end{equation}
and
\begin{equation}
    X_{\eta}(u,y)
    =
    \frac{T}{2\sqrt{2}\,\sigma}
    e^{-\frac{u^{2}}{2T^{2}}}e^{-\frac{y^{2}}{4\sigma^{2}}}
    g_{\beta}(\Delta\eta),
\end{equation}
where we recall that $\Delta \zeta = u - y$, and $\Delta \eta = u + y$. Using these functions, the integral in Eq.~\eqref{eq:defintion_33} can be evaluated as
\begin{equation}
    \mathcal{K}(\rho) = T^2\sqrt{\frac{2\pi}{D}}\,
    \exp\!\left[
        -\frac{
        \rho^{2}
        +2\ii T^{2}\Omega\rho
        +2T^{2}\sigma^{2}\Omega^{2}}
        {2D}
    \right]
    g_{\beta}(\rho).
\end{equation}
Then, using this expression, the cross term contribution to the local noise, $\mathcal{L}_{\tc{c}}$, can be evaluated using Eq.~\eqref{eq:cross_local_noise_general}. The integral requires numerical evaluation, whereas the constant term arises from
\begin{equation}
    \mathcal{K}'(0) = -\frac{\ii \pi^{3/2}T^{4}\Omega}{6\sqrt{2}\,\beta^{2}D^{3/2}}
\exp\!\left(
-\frac{T^{2}\sigma^{2}\Omega^{2}}{D}
\right).
\end{equation}
Let us now turn our attention to the evaluation of the non-local correlations, using the split proposed in Eq.~\eqref{eq:Mterm_split_hadamard}. The vacuum term, $\mathcal{M}_{0}$ is the same as in Eq.~\eqref{eq:termMvacuum_smeared_nice}, requiring numerical evaluation. As for $\mathcal{M}_{\tc{e}}$, one can use Eq.~\eqref{eq:W_e_thermal} to write
\begin{align}
    \mathcal{M}_{\tc{e}} & = \frac{\pi^2}{36\beta^4} \int \dd V \dd V' \, e^{\ii \Omega(t + t')}\Lambda_{\tc{a}}(\mf x)\Lambda_{\tc{b}}(\mf x') \nonumber \\ 
    & = \frac{\pi^2}{36\beta^4}\int \dd t \dd t' e^{\ii \Omega(t + t')}\chi_{\tc{a}}(t)\chi_{\tc{b}}(t') \nonumber \\ &
    =
    \frac{\pi^3T^2}{36\beta^4}
    e^{\ii\Omega(t_{\tc a}+t_{\tc b})}
    e^{-\frac{T^2\Omega^2}{2}}.
\end{align}
Next, because $\Xi_{w}(\mf x, \mf x')$ is symmetric, we have
\begin{equation}
    \mathcal{M}_{w} = \int \dd V \dd V' \, e^{\ii \Omega (t + t')}\Lambda_{\tc{a}}(\mf x)\Lambda_{\tc{b}}(\mf x')\Xi_{w}(\mf x, \mf x').
\end{equation}
Observe that this integral has the same structure as $\mathcal{L}_{w}$.
The only differences are the phase factor $e^{\ii\Omega(t+t')}$, instead of
$e^{-\ii\Omega(t-t')}$, and the fact that the two spacetime smearings are
centered at different detectors. Therefore, by following the same change of
variables and Gaussian convolutions used in Appendix~\ref{sec:garrafa_termica},
we can simplify $\mathcal{M}_{w}$ to
\begin{align}
    \mathcal M_w
    &=
    \frac{\sqrt{2\pi}\,T^2}{\sqrt{D}}\,
    e^{\ii\Omega(t_{\tc a}+t_{\tc b})}
    e^{-\frac{T^2\Omega^2}{2}}
    \int_{-\infty}^{\infty}\dd s\,
    g_\beta(s)^2
    \\
    &\quad\times
    \left(e^{-\frac{(s + \delta t - d)^2}{2D}} + e^{-\frac{(s + \delta t + d)^2}{2D}} 
    \right),
\end{align}
where we recall the notation $d = x_{\tc{b}} - x_{\tc{a}}$, $\delta t = t_{\tc{b}} - t_{\tc{a}}$, and $D = T^2 + 2 \sigma^2$.
For evaluating the last term, $\mathcal{M}_{\tc{c}}$, we first need to compute the auxiliary functions from Eqs.~\eqref{eq:auxiliaryfunction67}-\eqref{eq:auxiliaryfunction68}. In the present setup, they read
\begin{equation}
    X_{\tc{ab}, \zeta}(u, y) = 
    \frac{T e^{-\frac{T^2\Omega^2}{2}}}{2\sqrt{2}\,\sigma}\,
    e^{\ii\Omega(t_{\tc a}+t_{\tc b} + u)}
    e^{-\frac{(u+\delta t)^2}{2T^2}}
      e^{-\frac{(y+d)^2}{4\sigma^2}}
    g_{\beta}(\Delta \zeta),
\end{equation}
and
\begin{equation}
    X_{\tc{ab}, \eta}(u, y) = \frac{T e^{-\frac{T^2\Omega^2}{2}}}{2\sqrt{2}\,\sigma}\,
    e^{\ii\Omega(t_{\tc a}+t_{\tc b} + u)}
    e^{-\frac{(u+\delta t)^2}{2T^2}
      -\frac{(y+d)^2}{4\sigma^2}}
    g_{\beta}(\Delta \eta) .
\end{equation}
Then, the kernels $\mathcal{K}_{\tc{ab}}^{\pm}$ can be numerically evaluated via Eqs.~\eqref{eq:kernalabplus}-\eqref{eq:kernalabminus}, from which $\mathcal{M}_{\tc{c}}$ is numerically evaluated using Eq.~\eqref{eq:cross_correlations_general}.

Now that we have all the terms that compose the local noise and the non-local correlations, the analytical expressions can be combined with the numerical evaluations to determine $\mathcal{L} = \mathcal{L}_{\tc{aa}} = \mathcal{L}_{\tc{bb}}$ and $\mathcal{M}$. Moreover, we use $|\mathcal{X}| \equiv  |\mathcal{X}_{\tc{a}}| = |\mathcal{X}_{\tc{b}}|$, with
\begin{equation}
|\mathcal{X}| = \frac{\pi^{3/2} T}{6 \beta^2}e^{-\frac{T^{2}\Omega^{2}}{4}} = \sqrt{\mathcal{L}_{\tc{e}}}.
\end{equation}
Then, the negativity of the reduced state of the detectors to leading order in $\lambda$ is evaluated using the expression
\begin{equation}
    \mathcal{N}(\hat{\rho}_{\tc{ab}}) = \lambda^2 \text{max}\{0, |\widetilde{\mathcal{M}}| - \widetilde{\mathcal{L}}\},
\end{equation}
where
\begin{equation}
    \widetilde{\mathcal{M}} = \mathcal{M} - \mathcal{M}_{\tc{e}} = \mathcal{M}_{0} + \mathcal{M}_{\tc{c}} + \mathcal{M}_{w}
\end{equation}
and
\begin{equation}
        \widetilde{\mathcal{L}} = \mathcal{L} - \mathcal{L}_{\tc{e}} = \mathcal{L}_{0} + \mathcal{L}_{\tc{c}} + \mathcal{L}_{w},
\end{equation}
with $\mathcal{L}_{0}$, $\mathcal{M}_{0}$ the vacuum contributions that were evaluated in Sec.~\ref{sec:vacuum_smeared_examples}. In particular, $\mathcal{L}_{0}$ has the closed form expression of Eq.~\eqref{eq:integral_result}.

\begin{figure}[h!]
    \centering
    \includegraphics[width=\linewidth]{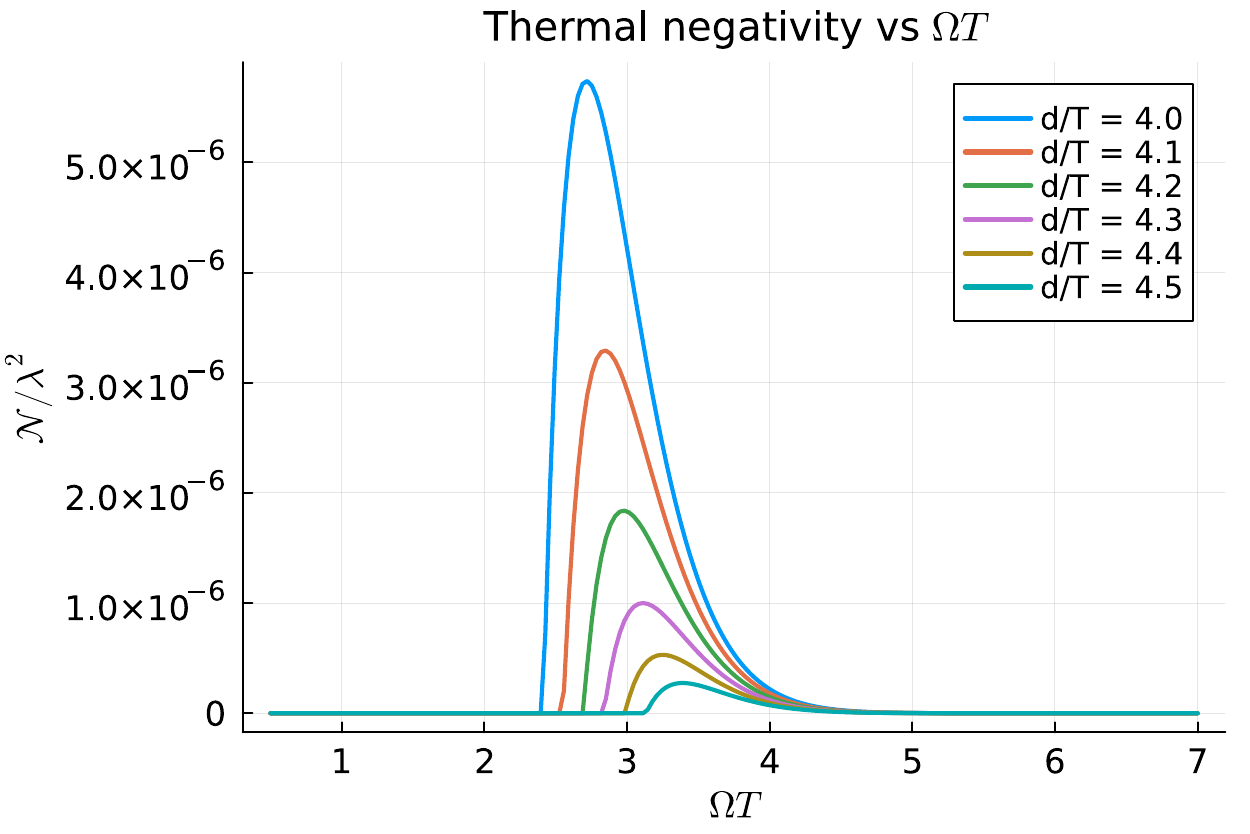}
    \caption{Negativity as a function of the detectors' gap $\Omega$ for multiple values of the separation between the detectors, $d$, when the field is prepared in a thermal state with inverse temperature $\beta = 50$. In each plot, we have fixed the scale by setting $T=1$, $\delta t = 0$, and $\sigma = 0.1$.}
    \label{fig:ThermalNegVOmega}
\end{figure}
In Fig.~\ref{fig:ThermalNegVOmega}, we plot the negativity between the two detectors as a function of their energy gap $\Omega$ for several values of the spatial separation $d$, while keeping the inverse temperature fixed at $\beta = 50$. The behaviour is qualitatively similar to that observed in the previous examples (see Figs.~\ref{fig:PointlikeVacuumNegativity} and \ref{fig:SmearedVacuumNegativity}). For each separation, the detectors only become entangled after a certain threshold value of $\Omega$, and as the distance between the detectors increases, the correlations are no longer strong enough to compensate for the local noise, and the negativity is strongly suppressed.
\begin{figure}[h!]
    \centering
    \includegraphics[width=\linewidth]{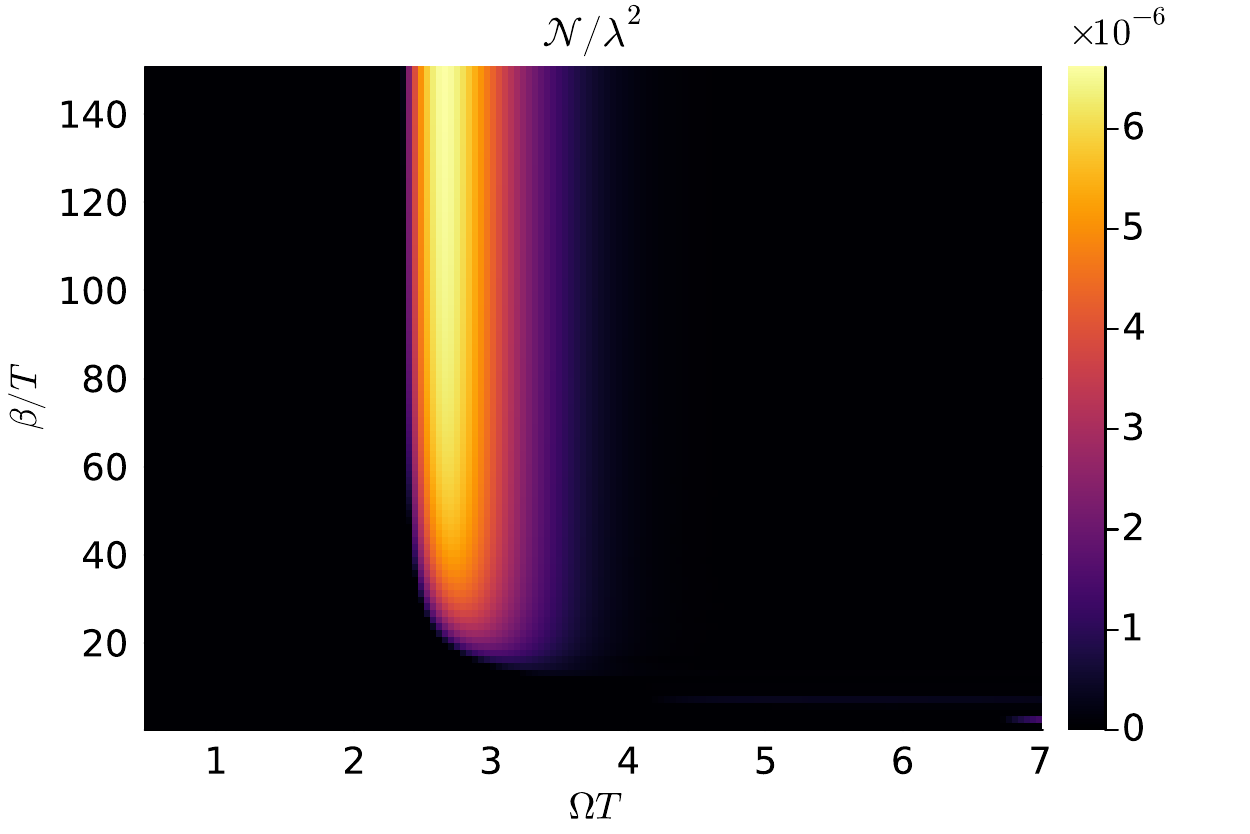}
    \caption{Negativity as a function of the detectors' gap $\Omega$ and the inverse temperature $\beta$. In this plot, we have fixed the scale by setting $T=1$, $d = 4.0$, $\delta t = 0$, and $\sigma = 0.1$}
    \label{fig:ThermalNegHeatmap}
\end{figure}

Fig.~\ref{fig:ThermalNegHeatmap} shows the negativity acquired by the detectors as a function of both the inverse temperature $\beta$ and the detector energy gap $\Omega$. The detector separation is fixed to $d/T = 4$, and the switching functions are taken to be centered at the same time, so that $\delta t = 0$. This plot makes clear that, for a fixed spatial separation, entanglement harvesting is only possible within a relatively narrow window of detector gaps. Outside this window, the correlations extracted from the field are not strong enough to overcome the local noise, and hence no entanglement is acquired between the detectors

Moreover, this example highlights the role of temperature in entanglement harvesting: Once the temperature becomes sufficiently large,  local noise overwhelms the local correlations that the detectors would otherwise extract from the field, preventing entanglement from being generated. Conversely, as the temperature is lowered, the local thermal noise becomes less dominant, and the negativity in the harvesting region is gradually restored. This suggests that temperature is detrimental to harvesting, in agreement with the behaviour previously observed in Ref.~\cite{Petar} for the standard entanglement harvesting setup, where detectors couple linearly to the field amplitude.

\begin{figure}[h!]
    \centering
    \includegraphics[width=\linewidth]{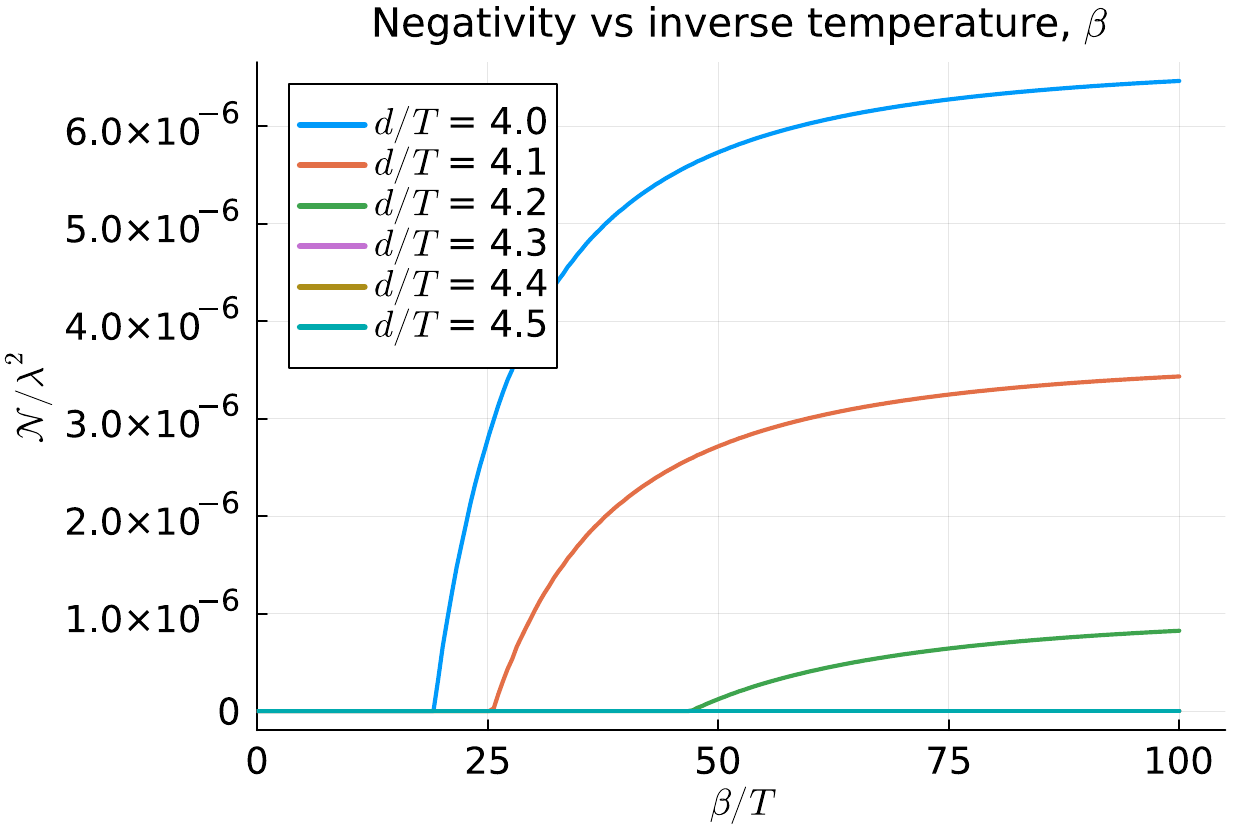}
    \caption{Negativity as a function of the inverse temperature $\beta$, with $\Omega T = 2.7$ fixed, for different detector separations. In this plot, we set $T=1$, $\delta t=0$, and $\sigma=0.1$.}
    \label{fig:ThermalNegVBeta}
\end{figure}

The same thermal suppression can be seen more directly in Fig.~\ref{fig:ThermalNegVBeta}, where we plot the negativity as a function of the inverse temperature $\beta$, with the detector gap fixed at $\Omega T = 2.7$, for several detector separations. As anticipated from Fig.~\ref{fig:ThermalNegHeatmap}, the detectors are unable to harvest entanglement at sufficiently high temperatures. For each fixed separation, harvesting only becomes possible once a threshold value of $\beta$ is crossed; beyond this point, the negativity increases monotonically with the inverse temperature.

Overall, these results reinforce the conclusion that temperature is detrimental to entanglement harvesting, in agreement with~\cite{Petar} and the fact that mixedness in general is detrimental to entanglement harvesting~\cite{GrimmerBrunoEdu_2021}, showing that this is also the case for this non-linear coupling.

\section{Conclusions}
\label{sec:conclusions}

In this work, we investigated entanglement harvesting with particle detectors coupled to the renormalized energy density of a massless scalar field. While entanglement harvesting is most commonly studied within the standard Unruh-DeWitt (UDW) model, in which the detector couples linearly to the field, couplings to composite observables are known to present additional ultraviolet (UV) subtleties. In particular, persistent UV divergences have previously been identified in the quadratic-coupling model \cite{SachsMannEdu,sachs2018entanglement}. Our goal was to determine whether analogous difficulties arise for the energy density coupling and, more importantly, to identify their mathematical origin.

To this end, we developed a distributional treatment of the model entirely in position space, allowing us to characterize the UV structure of both the local noise and the non-local correlation terms. For \eduadd{spatially coincident} pointlike detectors, we showed that the potentially divergent contributions originate from the pullback of the time-ordered energy-density two-point function to the detectors' worldlines. More precisely, \eduadd{in this case,} the UV behaviour is completely controlled by the behaviour of the switching correlation near coincident interaction times. As a consequence, all \eduadd{persistent} UV divergences disappear whenever the switching functions \eduadd{are sufficiently regular, compactly supported, and} have non-overlapping supports. \eduadd{At fixed nonzero separation, the correlation term is UV finite even for overlapping smooth, rapidly decreasing switchings.}

We then extended the analysis to spatially smeared detectors. We showed that spatial smearing modifies the short-distance structure of the sampled energy-density correlations in a dimension-dependent manner. For Gaussian spatial profiles, the non-local correlation term is always UV finite in $1+1$ and $2+1$ dimensions. In contrast, for $3+1$ dimensions and higher, spatial smearing alone \eduadd{does not generally suffice} to remove all divergences.\eduadd{The remaining UV-sensitive contributions depend on local derivatives of the spatial overlap as well as on the switching correlation at coincident interaction times.} Consequently, they are also removed by choosing \eduadd{compactly supported} switching functions with non-overlapping supports. \eduadd{For Gaussian profiles, two exceptional separations cancel the leading singularity and give finite correlations in $3+1$ and $4+1$ dimensions even for overlapping smooth switchings.}


These results clarify the physical origin of the persistent divergences encountered in energy-density couplings.  \eduadd{The obstruction is the sampling of nonintegrable singularities at equal interaction times in the time-ordered correlation term.} This perspective also clarifies the relation with the quadratic-coupling model of Refs.~\cite{SachsMannEdu,sachs2018entanglement}, where the persistent divergences likewise originate from sampling coincident-time singularities with overlapping switching functions. More generally, our analysis provides a unified distributional framework for understanding UV divergences in detector models involving composite field observables.

Finally, we derived general expressions for the local noise and correlations when the field is prepared in an arbitrary Hadamard zero-mean Gaussian state and illustrated the formalism with the example of thermal states \eduadd{in $1+1$ dimensions}. The resulting behaviour is consistent with previous studies of linearly coupled detectors, showing that increasing the field temperature suppresses the harvested entanglement \eduadd{in the configurations studied}. More broadly, the general framework developed here opens the door to studying entanglement harvesting from a wide variety of physically relevant quantum states, including squeezed states, Casimir vacua, and states exhibiting nontrivial or even locally negative energy-density distributions, both in flat and curved spacetimes. \eduadd{These results establish finite entanglement-harvesting protocols for energy-density coupling and identify how temporal and spatial sampling control their ultraviolet behaviour.}

\acknowledgements

MHZ thanks Prof. Achim Kempf for funding through his Dieter Schwarz grant. BR acknowledges the support of the Natural Sciences and Engineering Research Council of Canada (NSERC), 611676. Cette recherche a \'et\'e financ\'ee par le Conseil de recherches en sciences naturelles et en g\'enie du Canada (CRSNG), 611676. EMM acknowledges support through the Discovery Grant Program of the Natural Sciences and Engineering Research Council of Canada (NSERC). EMM thanks the support from his Ontario Early Researcher award. Research at Perimeter Institute is supported in part by the Government of Canada through the Department of Innovation, Science and Industry Canada and by the Province of Ontario through the Ministry of Colleges and Universities.

\newpage

\onecolumngrid

\appendix

\section{Derivation of the detectors' reduced density matrix}
\label{sec:derivation_matrix}

In this appendix, we explicitly show how to derive the detector's reduced matrix to leading order in $\lambda$ for the energy density coupling. On the following, we use inertial coordinates $\mf x = (t, \boldsymbol{x})$, with volume element $\dd V = \dd t\, \dd^n \boldsymbol{x}$.

We start from the perturbative expansion of the time evolution operator,
\begin{equation}
    \hat{U}_{\tc{i}}
    =
    \mathds{1}
    +
    \hat{U}^{(1)}_{\tc{i}}
    +
    \hat{U}^{(2)}_{\tc{i}}
    +
    \mathcal{O}(\lambda^3),
\end{equation}
where
\begin{equation}
    \hat{U}^{(1)}_{\tc{i}}
    =
    -\ii \int \dd V \, \hat{h}_{\tc{i}}(\mf x),
\end{equation}
and
\begin{equation}
    \hat{U}^{(2)}_{\tc{i}}
    =
    - \int_{-\infty}^{\infty} \dd t
    \int_{-\infty}^{t} \dd t'
    \int_{\mathbb{R}^{n}} \dd^{n}\boldsymbol{x}
    \int_{\mathbb{R}^{n}} \dd^{n}\boldsymbol{x}'
    \hat{h}_{\tc{i}}(\mf x)\hat{h}_{\tc{i}}(\mf x').
\end{equation}
Recall that the initial state of the joint system detector-field is assumed to be of the form
\begin{equation}
    \hat{\rho}_{0}
    =
\hat{\rho}_{0, \tc{ab}} \otimes \hat{\rho}_{\phi}.
\end{equation}
Then, the reduced state of the detectors after the interaction can be perturbatively written as
\begin{equation}
    \hat{\rho}_{\tc{ab}}
    =
    \text{Tr}_{\phi}
    \bigl[
        \hat{U}_{\tc{i}}\hat{\rho}_{0}\hat{U}^{\dagger}_{\tc{i}}
    \bigr]=
    \hat{\rho}_{0, \tc{ab}} 
    +
    \hat{\rho}^{(1)}_{\tc{ab}}
    +
    \hat{\rho}^{(2)}_{\tc{ab}}
    +
    \mathcal{O}(\lambda^3).
\end{equation}
The first-order term reads
\begin{equation}
    \hat{\rho}_{\tc{ab}}^{(1)}
    =
    \text{Tr}_{\phi}
    \Bigl[
        \hat{\rho}_{0}\hat{U}^{(1)\dagger}_{\tc{i}}
        +
        \hat{U}^{(1)}_{\tc{i}}\hat{\rho}_{0}
    \Bigr]=
    \ii \lambda
    \int \dd V \,
    [\hat{\rho}_{0,\tc{ab}},\hat{M}_{\tc{ab}}(\mf x)]
    \,\mathcal{E}_{\phi}(\mf x),
\end{equation}
with
\begin{equation}
    \mathcal{E}_{\phi}(\mf x)
    \equiv
    \text{Tr}
    \bigl[
        \hat{\rho}_{\phi}\mathopen{:}\hat{T}_{00}(\mf x)\mathopen{:}
    \bigr].
\end{equation}
As for the second-order term, we have
\begin{align}
    \hat{\rho}_{\tc{ab}}^{(2)}
    & =
    \text{Tr}_{\phi}
    \Bigl[
        \hat{U}^{(1)}_{\tc{i}}\hat{\rho}_{0}\hat{U}^{(1)\dagger}_{\tc{i}}
        +
        \hat{U}^{(2)}_{\tc{i}}\hat{\rho}_{0}
        +
        \hat{\rho}_{0}\hat{U}^{(2)\dagger}_{\tc{i}}
    \Bigr]
   \nonumber \\ & =
    \lambda^2
    \int \dd V \, \dd V'\,
    \hat{M}_{\tc{ab}}(\mf x')
    \hat{\rho}_{0,\tc{ab}}
    \hat{M}_{\tc{ab}}(\mf x)
    \Xi_{\phi}(\mf x,\mf x')
    -\lambda^2
    \int \dd V \, \dd V'\,
    \Theta(t-t')\,
    \hat{M}_{\tc{ab}}(\mf x)
    \hat{M}_{\tc{ab}}(\mf x')
    \hat{\rho}_{0,\tc{ab}}
    \Xi_{\phi}(\mf x,\mf x')
       \nonumber\\
    &-\lambda^2
    \int \dd V \, \dd V'\,
    \Theta(t-t')\,
    \hat{\rho}_{0,\tc{ab}}
    \hat{M}_{\tc{ab}}(\mf x')
    \hat{M}_{\tc{ab}}(\mf x)
    \Xi_{\phi}(\mf x',\mf x).
\end{align}
where
\begin{equation}
    \Xi_{\phi}(\mf x,\mf x')
    \equiv
    \text{Tr}
    \bigl[
        \hat{\rho}_{\phi}
        \mathopen{:}\hat{T}_{00}(\mf x)\mathopen{:}
        \mathopen{:}\hat{T}_{00}(\mf x')\mathopen{:}
    \bigr].
    \label{eq:energy_density_2_point}
\end{equation}
Let us now consider the particular case where both detectors start on their respective ground states,
\begin{equation}
    \hat{\rho}_{0,\tc{ab}}
    =
    \proj{g_{\tc{a}}}{g_{\tc{a}}}
    \otimes
    \proj{g_{\tc{b}}}{g_{\tc{b}}}.
\end{equation}
In this case, we can evaluate
\begin{equation}
\begin{aligned}
[\hat{\rho}_{0,\tc{ab}},\hat{M}_{\tc{ab}}(\mf x)]
&=
e^{-\ii\Omega t}
\Bigl[
\Lambda_{\tc a}(\mf x)\proj{g_{\tc a}g_{\tc b}}{e_{\tc a}g_{\tc b}}
+
\Lambda_{\tc b}(\mf x)\proj{g_{\tc a}g_{\tc b}}{g_{\tc a}e_{\tc b}}
\Bigr]
\\
&\quad
-
e^{\ii\Omega t}
\Bigl[
\Lambda_{\tc a}(\mf x)\proj{e_{\tc a}g_{\tc b}}{g_{\tc a}g_{\tc b}}
+
\Lambda_{\tc b}(\mf x)\proj{g_{\tc a}e_{\tc b}}{g_{\tc a}g_{\tc b}}
\Bigr],
\end{aligned}
\end{equation}
\begin{align}
    \hat{M}_{\tc{ab}}(\mf x')
    \hat{\rho}_{0,\tc{ab}}
    \hat{M}_{\tc{ab}}(\mf x)
    &=
    e^{-\ii\Omega(t-t')}
    \Bigl[
        \Lambda_{\tc{a}}(\mf x')\Lambda_{\tc{a}}(\mf x)
        \proj{e_{\tc{a}}g_{\tc{b}}}{e_{\tc{a}}g_{\tc{b}}}
        +
        \Lambda_{\tc{a}}(\mf x')\Lambda_{\tc{b}}(\mf x)
        \proj{e_{\tc{a}}g_{\tc{b}}}{g_{\tc{a}}e_{\tc{b}}}
    \nonumber\\
    &\qquad\quad
        +
        \Lambda_{\tc{b}}(\mf x')\Lambda_{\tc{a}}(\mf x)
        \proj{g_{\tc{a}}e_{\tc{b}}}{e_{\tc{a}}g_{\tc{b}}}
        +
        \Lambda_{\tc{b}}(\mf x')\Lambda_{\tc{b}}(\mf x)
        \proj{g_{\tc{a}}e_{\tc{b}}}{g_{\tc{a}}e_{\tc{b}}}
    \Bigr],
\end{align}
\begin{align}
    \hat{M}_{\tc{ab}}(\mf x)
    \hat{M}_{\tc{ab}}(\mf x')
    \hat{\rho}_{0,\tc{ab}}
    &=
    e^{-\ii\Omega(t-t')}
    \Bigl[
        \Lambda_{\tc{a}}(\mf x)\Lambda_{\tc{a}}(\mf x')
        +
        \Lambda_{\tc{b}}(\mf x)\Lambda_{\tc{b}}(\mf x')
    \Bigr]
    \proj{g_{\tc{a}}g_{\tc{b}}}{g_{\tc{a}}g_{\tc{b}}}
    \nonumber\\
    &\quad
    +
    e^{\ii\Omega(t+t')}
    \Bigl[
        \Lambda_{\tc{a}}(\mf x)\Lambda_{\tc{b}}(\mf x')
        +
        \Lambda_{\tc{b}}(\mf x)\Lambda_{\tc{a}}(\mf x')
    \Bigr]
    \proj{e_{\tc{a}}e_{\tc{b}}}{g_{\tc{a}}g_{\tc{b}}},
\end{align}
and
\begin{align}
    \hat{\rho}_{0,\tc{ab}}
    \hat{M}_{\tc{ab}}(\mf x')
    \hat{M}_{\tc{ab}}(\mf x)
    &=
    e^{\ii\Omega(t-t')}
    \Bigl[
        \Lambda_{\tc{a}}(\mf x)\Lambda_{\tc{a}}(\mf x')
        +
        \Lambda_{\tc{b}}(\mf x)\Lambda_{\tc{b}}(\mf x')
    \Bigr]
    \proj{g_{\tc{a}}g_{\tc{b}}}{g_{\tc{a}}g_{\tc{b}}}
    \nonumber\\
    &\quad
    +
    e^{-\ii\Omega(t+t')}
    \Bigl[
        \Lambda_{\tc{a}}(\mf x)\Lambda_{\tc{b}}(\mf x')
        +
        \Lambda_{\tc{b}}(\mf x)\Lambda_{\tc{a}}(\mf x')
    \Bigr]
    \proj{g_{\tc{a}}g_{\tc{b}}}{e_{\tc{a}}e_{\tc{b}}}.
\end{align}
To explicitly write a matricial representation for $\hat{\rho}_{\tc{ab}}$, we evaluate the matrix elements on the canonical basis spanned by the ground and excited states of each detector. The results are the following:
\begin{equation}
    \langle g_{\tc a} g_{\tc b}| \hat{\rho}_{\tc{ab}} | g_{\tc a} g_{\tc b} \rangle = 1 - \lambda^2\int \dd V \dd V' \, e^{-\ii \Omega (t - t')}(\Lambda_{\tc{a}}(\mf x)\Lambda_{\tc{a}}(\mf x') + \Lambda_{\tc{b}}(\mf x)\Lambda_{\tc{b}}(\mf x')) \Xi_{\phi}(\mf x, \mf x') + \mathcal{O}(\lambda^3),
\end{equation}
\begin{equation}
    \langle g_{\tc a} g_{\tc b}| \hat{\rho}_{\tc{ab}} | g_{\tc a} e_{\tc b} \rangle = \ii \lambda \int \dd V e^{-\ii \Omega t}\Lambda_{\tc{b}}(\mf x) \mathcal{E}_{\phi}(\mf x) + \mathcal{O}(\lambda^3),
\end{equation}
\begin{equation}
    \langle g_{\tc a} g_{\tc b}| \hat{\rho}_{\tc{ab}} | e_{\tc a} g_{\tc b} \rangle =\ii \lambda
    \int \dd V \,
    e^{-\ii \Omega t}
    \Lambda_{\tc a}(\mf x)
    \mathcal{E}_{\phi}(\mf x)
    +
    \mathcal{O}(\lambda^3),
\end{equation}
\begin{align}
    \langle g_{\tc a} g_{\tc b}| \hat{\rho}_{\tc{ab}} | e_{\tc a} e_{\tc b} \rangle & =  
    -\lambda^2
    \int \dd V \dd V' \,
    \Theta(t-t')\,
    e^{-\ii \Omega (t+t')}
    \bigl(
        \Lambda_{\tc a}(\mf x)\Lambda_{\tc b}(\mf x')
        +
        \Lambda_{\tc b}(\mf x)\Lambda_{\tc a}(\mf x')
    \bigr)
    \Xi_{\phi}(\mf x',\mf x)
    +
    \mathcal{O}(\lambda^3) \nonumber \\ & = - \lambda^2 \int \dd V \dd V' e^{-\ii \Omega(t + t')}\Lambda_{\tc{a}}(\mf x)\Lambda_{\tc{b}}(\mf x') \mathcal{G}^{*}_{\phi}(\mf x, \mf x'), \quad  \mathcal{G}_{\phi}(\mf x, \mf x') \equiv \Theta(t - t')\Xi_{\phi}(\mf x, \mf x') + 
    \Theta(t' - t)\Xi_{\phi}(\mf x', \mf x'),
\end{align}
\begin{equation}
    \langle g_{\tc a} e_{\tc b}| \hat{\rho}_{\tc{ab}} | g_{\tc a} g_{\tc b} \rangle =-\ii \lambda
    \int \dd V \,
    e^{\ii \Omega t}
    \Lambda_{\tc b}(\mf x)
    \mathcal{E}_{\phi}(\mf x)
    +
    \mathcal{O}(\lambda^3),
\end{equation}
\begin{equation}
    \langle g_{\tc a} e_{\tc b}| \hat{\rho}_{\tc{ab}} | g_{\tc a} e_{\tc b} \rangle = \lambda^2
    \int \dd V \dd V' \,
    e^{-\ii \Omega (t-t')}
    \Lambda_{\tc b}(\mf x)
    \Lambda_{\tc b}(\mf x')
    \Xi_{\phi}(\mf x,\mf x')
    +
    \mathcal{O}(\lambda^3),
\end{equation}
\begin{equation}
    \langle g_{\tc a} e_{\tc b}| \hat{\rho}_{\tc{ab}} | e_{\tc a} g_{\tc b} \rangle = \lambda^2
    \int \dd V \dd V' \,
    e^{-\ii \Omega (t-t')}
    \Lambda_{\tc a}(\mf x)
    \Lambda_{\tc b}(\mf x')
    \Xi_{\phi}(\mf x,\mf x')
    +
    \mathcal{O}(\lambda^3),
\end{equation}
\begin{equation}
    \langle g_{\tc a} e_{\tc b}| \hat{\rho}_{\tc{ab}} | e_{\tc a} e_{\tc b} \rangle =\mathcal{O}(\lambda^3),
\end{equation}

\begin{equation}
    \langle e_{\tc a} g_{\tc b}| \hat{\rho}_{\tc{ab}} | g_{\tc a} g_{\tc b} \rangle = -\ii \lambda
    \int \dd V \,
    e^{\ii \Omega t}
    \Lambda_{\tc a}(\mf x)
    \mathcal{E}_{\phi}(\mf x)
    +
    \mathcal{O}(\lambda^3),
\end{equation}
\begin{equation}
    \langle e_{\tc a} g_{\tc b}| \hat{\rho}_{\tc{ab}} | g_{\tc a} e_{\tc b} \rangle =\lambda^2
    \int \dd V \dd V' \,
    e^{-\ii \Omega (t-t')}
    \Lambda_{\tc b}(\mf x)  \Lambda_{\tc a}(\mf x')
    \Xi_{\phi}(\mf x,\mf x')
    +
    \mathcal{O}(\lambda^3),
\end{equation}
\begin{equation}
   \langle e_{\tc a} g_{\tc b}| \hat{\rho}_{\tc{ab}} | e_{\tc a} g_{\tc b} \rangle = \lambda^2
    \int \dd V \dd V' \,
    e^{-\ii \Omega (t-t')}
    \Lambda_{\tc a}(\mf x)
    \Lambda_{\tc a}(\mf x')
    \Xi_{\phi}(\mf x,\mf x')
    +
    \mathcal{O}(\lambda^3),
\end{equation}
\begin{equation}
    \langle e_{\tc a} g_{\tc b}| \hat{\rho}_{\tc{ab}} | e_{\tc a} e_{\tc b} \rangle = \mathcal{O}(\lambda^3),
\end{equation}
\begin{equation}
    \langle e_{\tc a} e_{\tc b}| \hat{\rho}_{\tc{ab}} | g_{\tc a} g_{\tc b} \rangle =-\lambda^2 \int \dd V \dd V' e^{\ii \Omega (t + t')} \Lambda_{\tc{a}}(\mf x) \Lambda_{\tc{b}}(\mf x')\mathcal{G}_{\phi}(\mf x, \mf x') + \mathcal{O}(\lambda^{3}),
\end{equation}
\begin{equation}
    \langle e_{\tc a} e_{\tc b}| \hat{\rho}_{\tc{ab}} | g_{\tc a} e_{\tc b} \rangle =\mathcal{O}(\lambda^{3}),
\end{equation}
\begin{equation}
    \langle e_{\tc a} e_{\tc b}| \hat{\rho}_{\tc{ab}} | e_{\tc a} g_{\tc b} \rangle = \mathcal{O}(\lambda^{3}),
\end{equation}
\begin{equation}
    \langle e_{\tc a} e_{\tc b}| \hat{\rho}_{\tc{ab}} | e_{\tc a} e_{\tc b} \rangle = \mathcal{O}(\lambda^{3}),
\end{equation}
Therefore, in the basis
$\{
\ket{g_{\tc{a}}g_{\tc{b}}},
\ket{g_{\tc{a}}e_{\tc{b}}},
\ket{e_{\tc{a}}g_{\tc{b}}},
\ket{e_{\tc{a}}e_{\tc{b}}}
\}$,
the representation of the reduced density matrix of the detectors reads
\begin{equation}
    \hat{\rho}_{\tc{ab}}
    =
    \begin{bmatrix}
        1-\lambda^2(\mathcal{L}_{\tc{aa}}+\mathcal{L}_{\tc{bb}})
        & \ii\lambda \mathcal{X}^{*}_{\tc{b}}
        & \ii\lambda \mathcal{X}^{*}_{\tc{a}}
        & -\lambda^2 \mathcal{M}^{*}
        \\
        -\ii \lambda \mathcal{X}_{\tc{b}}
        & \lambda^2 \mathcal{L}_{\tc{bb}}
        & \lambda^2 \mathcal{L}_{\tc{ab}}
        & 0
        \\
        -\ii \lambda \mathcal{X}_{\tc{a}}
        & \lambda^2 \mathcal{L}_{\tc{ba}}
        & \lambda^2  \mathcal{L}_{\tc{aa}}
        & 0
        \\
        -\lambda^2 \mathcal{M}
        & 0
        & 0
        & 0
    \end{bmatrix}
    +
    \mathcal{O}(\lambda^{3}),
    \label{eq:rho_AB_harvesting}
\end{equation}
where we defined
\begin{equation}
    \mathcal{L}_{ij} = \int \dd V \dd V' e^{-\ii \Omega(t - t')}\Lambda_{i}(\mf x)\Lambda_{j}(\mf x')\Xi_{\phi}(\mf x, \mf x'),
\end{equation}
\begin{equation}
    \mathcal{X}_{j} = \int \dd V e^{\ii \Omega t}\Lambda_{\tc{b}}(\mf x)\mathcal{E}_{\phi}(\mf x),
\end{equation}
and
\begin{equation}
    \mathcal{M} = \int \dd V \dd V' e^{\ii \Omega(t + t')}\Lambda_{\tc{a}}(\mf x)\Lambda_{\tc{b}}(\mf x') \mathcal{G}_{\phi}(\mf x, \mf x').
\end{equation}

\section{Evaluation of renormalized correlation functions}
\label{sec:two_point_appendix}

In this Appendix, we show how to evaluate the stress-energy density two-point function for any Hadamard, quasifree state. Applying the normal ordering prescrition, we can start by writting
\begin{align}
    \Xi_{\phi}(\mf x, \mf x')
    &=
    \text{Tr}\!\left[\hat{\rho}_{\phi}\mathopen{:} \hat{T}_{00}(\mf x) \mathclose{:}\mathopen{:} \hat{T}_{00}(\mf x') \mathclose{:}\right]
    \nonumber \\
    &=
    \text{Tr}\!\left[\hat{\rho}_{\phi}\hat{T}_{00}(\mf x)\hat{T}_{00}(\mf x')\right]
    -
    \langle 0 | \hat{T}_{00}(\mf x)| 0 \rangle
    \text{Tr}\!\left[\hat{\rho}_{\phi}\hat{T}_{00}(\mf x')\right]
    -
    \langle 0 | \hat{T}_{00}(\mf x')| 0 \rangle
    \text{Tr}\!\left[\hat{\rho}_{\phi}\hat{T}_{00}(\mf x)\right]
    +
    \langle 0 | \hat{T}_{00}(\mf x) | 0 \rangle
    \langle 0 | \hat{T}_{00}(\mf x') | 0 \rangle.
\end{align}
Using
\begin{equation}
    \hat{T}_{00}(\mf x)
    =
    \frac{1}{2}\sum_{\mu=0}^{n}\partial_{\mu}\hat{\phi}(\mf x)^2
\end{equation}
it follows that
\begin{align}
    \text{Tr}\!\left[\hat{\rho}_{\phi}\hat{T}_{00}(\mf x)\hat{T}_{00}(\mf x')\right]
    &=
    \frac{1}{4}\sum_{\mu,\nu=0}^{n}
    \text{Tr}\!\left[
        \hat{\rho}_{\phi}
        \partial_{\mu}\hat{\phi}(\mf x)^2
        \partial_{\nu'}\hat{\phi}(\mf x')^2
    \right].
\end{align}
Because $\hat{\rho}_{\phi}$ is assumed to be quasifree, Wick's theorem gives
\begin{equation}
    \text{Tr}\!\left[
        \hat{\rho}_{\phi}
        \partial_{\mu}\hat{\phi}(\mf x)^2
        \partial_{\nu'}\hat{\phi}(\mf x')^2
    \right]
    =
    2\bigl(\partial_{\mu}\partial_{\nu'}W(\mf x,\mf x')\bigr)^2
    +
    \text{Tr}\!\left[\hat{\rho}_{\phi}\partial_{\mu}\hat{\phi}(\mf x)^2\right]
    \text{Tr}\!\left[\hat{\rho}_{\phi}\partial_{\nu'}\hat{\phi}(\mf x')^2\right],
\end{equation}
where $W(\mf x,\mf x')=\text{Tr}[\hat{\rho}_{\phi}\hat{\phi}(\mf x)\hat{\phi}(\mf x')]$ is the two-point function of the state $\hat{\rho}_{\phi}$. Thus,
\begin{equation}
    \text{Tr}\!\left[\hat{\rho}_{\phi}\hat{T}_{00}(\mf x)\hat{T}_{00}(\mf x')\right]
    =
    \frac{1}{2}\sum_{\mu,\nu=0}^{n}
    \bigl(\partial_{\mu}\partial_{\nu'}W(\mf x,\mf x')\bigr)^2
    +
    \frac{1}{4}\sum_{\mu,\nu=0}^{n}
    \text{Tr}\!\left[\hat{\rho}_{\phi}\partial_{\mu}\hat{\phi}(\mf x)^2\right]
    \text{Tr}\!\left[\hat{\rho}_{\phi}\partial_{\nu'}\hat{\phi}(\mf x')^2\right].
\end{equation}
On the other hand,
\begin{equation}    \text{Tr}\!\left[\hat{\rho}_{\phi}\hat{T}_{00}(\mf x)\right]
=
\frac{1}{2}\sum_{\mu=0}^{n}
\text{Tr}\!\left[\hat{\rho}_{\phi}\partial_{\mu}\hat{\phi}(\mf x)^2\right]
=
\frac{1}{2}\sum_{\mu=0}^{n}\lim_{\mf x '\to\mf x}\partial_{\mu}\partial_{\mu'}W(\mf x,\mf x '),
\end{equation}
and similarly
\begin{equation}
    \langle 0|\hat{T}_{00}(\mf x)|0\rangle
    =
    \frac{1}{2}\sum_{\mu=0}^{n}
    \langle 0|\partial_{\mu}\hat{\phi}(\mf x)^2|0\rangle
    =
    \frac{1}{2}\sum_{\mu=0}^{n}
    \lim_{\mf x '\to\mf x}
    \partial_{\mu}\partial_{\mu'}W_{0}(\mf x,\mf x '),
\end{equation}
where $W_{0}$ denotes the vacuum two-point function. Substituting these expressions into $\Xi_{\phi}(\mf x,\mf x')$, we obtain
\begin{equation}
    \Xi_{\phi}(\mf x,\mf x')
    =
    \frac{1}{2}\sum_{\mu,\nu=0}^{n}
    \bigl(\partial_{\mu}\partial_{\nu'}W(\mf x,\mf x')\bigr)^2
    +
    \frac{1}{4}\sum_{\mu,\nu=0}^{n}
    \Bigl(
        \text{Tr}\!\left[\hat{\rho}_{\phi}\partial_{\mu}\hat{\phi}(\mf x)^2\right]
        -
        \langle 0|\partial_{\mu}\hat{\phi}(\mf x)^2|0\rangle
    \Bigr)
    \Bigl(
        \text{Tr}\!\left[\hat{\rho}_{\phi}\partial_{\nu'}\hat{\phi}(\mf x')^2\right]
        -
        \langle 0|\partial_{\nu'}\hat{\phi}(\mf x')^2|0\rangle
    \Bigr).
\end{equation}
Now, write
\begin{equation}
    W(\mf x,\mf x') = W_{0}(\mf x,\mf x') + w(\mf x,\mf x'),
\end{equation}
where $w(\mf x, \mf x')$ is the state-dependent bi-scalar, which is regular at the coincidence limit. Then
\begin{equation}
\text{Tr}\!\left[\hat{\rho}_{\phi}\partial_{\mu}\hat{\phi}(\mf x)^2\right]
    -
    \langle 0|\partial_{\mu}\hat{\phi}(\mf x)^2|0\rangle
    =
    \lim_{\mf x '\to\mf x}
    \partial_{\mu}\partial_{\mu'}w(\mf x,\mf x '),
\end{equation}
and hence
\begin{equation}
    \Xi_{\phi}(\mf x,\mf x')
    =
    \frac{1}{2}\sum_{\mu,\nu=0}^{n}
    \bigl(\partial_{\mu}\partial_{\nu'}W(\mf x,\mf x')\bigr)^2
    +
    \frac{1}{4}\sum_{\mu,\nu=0}^{n}
    \left(
        \lim_{\mf x '\to\mf x}
        \partial_{\mu}\partial_{\mu'}w(\mf x,\mf x ')
    \right)
    \left(
        \lim_{\mf x' \to\mf x}
        \partial_{\nu}\partial_{\nu'}w(\mf x,\mf x ')
    \right).
\end{equation}
Equivalently, since
\begin{equation}
  \mathcal{E}_{\phi}(\mf x) \equiv \text{Tr}[\hat{\rho}_{\phi}\mathopen{:} \hat{T}_{00}(\mf x) \mathopen{:}]
    =
    \frac{1}{2}\sum_{\mu=0}^{n}
    \lim_{\mf x '\to\mf x}
    \partial_{\mu}\partial_{\mu'}w(\mf x,\mf x '),
\end{equation}
the energy density two-point function can be cast into
\begin{equation}
    \Xi_{\phi}(\mf x,\mf x')
    =
    \frac{1}{2}\sum_{\mu,\nu=0}^{n}
    \bigl(\partial_{\mu}\partial_{\nu'}W(\mf x,\mf x')\bigr)^2
    + \mathcal{E}_{\phi}(\mf x)\mathcal{E}_{\phi}(\mf x')
\end{equation}

\section{Distributions on the half-line}
\label{sec:half_line_appendix}

In this appendix, we characterize the divergence of integrals of the form
\begin{equation}
    \mathcal{I}^{m}_{\epsilon}(f) = \int_{0}^{\infty}\dd x \, \frac{f(x)}{(x \pm \ii \epsilon)^m},
    \label{eq:defintegral_n}
\end{equation}
in the distributional limit $\epsilon \to 0^{+}$, for any for any function $f$ of class $\mathcal{C}^m([0,\infty])$ for which $ \lim_{x \to \pm\infty}f(x) = 0$.
Let us first consider the case $m = 1$. Integrating by parts,
\begin{equation}
    \mathcal{I}^{1}_{\epsilon}(f) = -f(0)\log(\pm \ii \epsilon) - \int_{0}^{\infty}\dd x f'(x)\log(x \pm \ii \epsilon).
\end{equation}
Using
\begin{equation}
    \log z = \log |z| + \ii \operatorname{arg}z,
\end{equation}
we have
\begin{equation}
    \mathcal{I}^{1}_{\epsilon}(f) = -f(0)\log(\epsilon) \mp \frac{\ii \pi}{2}f(0) - \int_{0}^{\infty}\dd x f'(x)\left(\log(\sqrt{x^2 + \epsilon^2}) + \ii \operatorname{arg}(x \pm \ii \epsilon)\right).
\label{eq:Iepsilonintermediary}
\end{equation}
On the other hand, for every $\delta > 0$, we can write
\begin{equation}
\int_{\delta}^{\infty}\dd x\frac{f(x)}{x} = -\log(\delta)f(\delta) - \int_{\delta}^{\infty} \dd x \log x f'(x).
\label{eq:integralforcomparison}
\end{equation}
For $x > 0$, $\operatorname{arg} x = 0$. Thus, in the limit $\epsilon \to 0^{+}$, the integral in the right hand side of Eq.\eqref{eq:Iepsilonintermediary} matches the one on the right hand side of Eq.\eqref{eq:integralforcomparison}. That is,
\begin{equation}
    \lim_{\epsilon \to 0^{+}}\int_{0}^{\infty}\dd x f'(x)\left(\log(\sqrt{x^2 + \epsilon^2}) + \ii \operatorname{arg}(x \pm \ii \epsilon)\right) = -\operatorname{FP}\int_{0}^{\infty}\dd x \frac{f(x)}{x},
\end{equation}
where the finite part distribution, FP, is defined in~\eqref{eq:FP_and_PV_main_text}. Therefore, in the limit $\epsilon \to 0^{+}$, the integral $\mathcal{I}^{1}_{\epsilon}(f)$ diverges logarithmically.  The logarithmic divergence in Eq.~\eqref{eq:Iepsilonintermediary} is entirely contained in the boundary term $-f(0)\log\epsilon$. Indeed, comparing Eqs.~\eqref{eq:Iepsilonintermediary} and \eqref{eq:integralforcomparison}, one sees that the remaining integral converges to the Hadamard finite-part integral, whose definition precisely subtracts the same logarithmic singularity. Thus, after adding back the divergent contribution $f(0)\log\epsilon$, the limit $\epsilon\to0^+$ is finite and consists only of the finite-part integral together with the imaginary contribution arising from the branch cut of the logarithm. When this result is written as an identity between distributions, the logarithmic divergence must be represented by a Dirac delta distribution supported at the origin. Since we adopt the convention $\Theta(0)=1/2$, we have $\Theta(x)\delta(x)=\delta(x)/2$. Therefore, in order for the distribution to reproduce the boundary contribution $f(0)\log\epsilon$ when acting on a test function, the coefficient multiplying $\delta(x)\Theta(x)$ must be $2\log\epsilon$, explaining the factor of two appearing in the distributional identity below: 
\begin{equation}
    \lim_{\epsilon \to 0^{+}}(
    \mathcal{I}^{1}_{\epsilon}(f) + f(0)\log(\epsilon)) = \mp\ii \pi \frac{f(0)}{2} + \operatorname{FP}\int_{0}^{\infty}\dd x\frac{f(x)}{x},
    \label{eq:Harveydent}
\end{equation}
which is the {\it Sokhotski–Plemelj theorem for the half-line}. In terms of the distributions, we have\footnote{Here, we are using the convention $\Theta(0) = \frac{1}{2}$.}
\begin{equation}
    \lim_{\epsilon \to 0^{+}}\Biggl[\frac{\Theta(x)}{x \pm \ii \epsilon} +2\log(\epsilon)\delta(x)\Theta(x)\Biggr] = \operatorname{FP}\left(\frac{\Theta(x)}{x} \right) \mp \ii \pi\Theta(x)\delta(x).
\end{equation}

Assuming $f(x)$ is such that $\lim_{x \to \infty} f(x) = 0$, one can apply integration by parts $m - 1$ times to the expression above. For $m \ge 2$, the result can be written as
\begin{equation}
    \mathcal{I}^{m}_{\epsilon}(f)
    =
    \sum_{k=0}^{m-2}
    \frac{(m-k-2)!}{(m-1)!}
    \frac{f^{(k)}(0)}{(\pm i\epsilon)^{m-k-1}}
    +
    \frac{1}{(m-1)!}
    \int_{0}^{\infty}\dd x\,
    \frac{f^{(m-1)}(x)}{x \pm \ii \epsilon}.
\end{equation}
The last term can be identified as $\mathcal{I}^{1}_{\epsilon}f^({m-1})$. Thus, in the limit $\epsilon \to 0^{+}$, we obtain
\begin{equation}
    \lim_{\epsilon \to 0^{+}} \Biggl[ \mathcal{I}^{m}_{\epsilon}(f) - \sum_{k=0}^{m-2}
    \frac{(m-k-2)!}{(m-1)!}
    \frac{f^{(k)}(0)}{(\pm i\epsilon)^{m-k-1}} + \frac{f^{(m - 1)}(0) \log(\epsilon)}{(m - 1)!}  \Biggr] = \frac{1}{(m - 1)!}\Biggl[\mp\frac{\ii \pi f^{(m - 1)}(0)}{2} + \operatorname{FP}\int_{0}^{\infty}\dd x \, \frac{f^{(m - 1)}(x)}{x} \Biggr].
    \label{eq:brazilianBBQ}
\end{equation}

\section{Evaluation of the smeared distribution $\mathfrak{W}_{AB}(u)$}
\label{sec:appendix_smeared_distribution}

In this appendix, we derive a closed-form expression for the smeared vacuum energy-density two-point distribution appearing in Eqs.~\eqref{eq:mathfrack_and_friends}and \eqref{eq:smeared_distribution_ab_n_dim}. This distribution encodes the effect of averaging the vacuum energy-density correlator over the finite spatial extent of Gaussian detectors and constitutes the fundamental ingredient entering both the local noise and the non-local correlation terms. Our goal is therefore to obtain an analytic expression for
\begin{equation}
    \mathfrak{W}_{\boldsymbol d}(u)
    =
    \int_{\mathbb R^n} d^n\mathbf y\,
    \vartheta_{\mathrm{ab}}(\mathbf y)\,
    \Xi_0(u,\mathbf y),
    \label{eq:app_Wd_def}
\end{equation}
for arbitrary detector separation $\mathbf d$. The corresponding distribution relevant for the local noise $\mathfrak{W}(u)$, is recovered by setting $d=0$.

For the Gaussian profiles used in the main text, the convolution of the two spatial profiles is
\begin{equation}
    \vartheta_{\tc{ab}}(\boldsymbol{y})
    =
    \frac{e^{-\frac{|\boldsymbol{y} + \boldsymbol{d}|^2}{4\sigma^2}}}{(4\pi\sigma^2)^{n/2}}
    ,
\label{eq:vartheta_ab_gaussian_compact_appendix}
\end{equation}
Recall that the vacuum two-point function of the energy density is given by
\begin{equation}
    \Xi_{0}(u,\boldsymbol{y})
    =
    A_{n}
    \frac{
        n\left(u_{\epsilon}^{4}+|\boldsymbol{y}|^{4}\right)
        +
        2(n+2)u_{\epsilon}^{2}|\boldsymbol{y}|^{2}
    }{
        \left(|\boldsymbol{y}|^{2}-u_{\epsilon}^{2}\right)^{n+3}
    },
    \label{eq:app_W0_energy_density}
\end{equation}
with the notation $u_{\epsilon} = u - \ii \epsilon$. Then, the integral in Eq.~\eqref{eq:app_Wd_def} becomes
\begin{equation}
    \mathfrak{W}_{\boldsymbol{d}}(u)
    =
    \frac{e^{-d^{2}/(4\sigma^{2})}}{(4\pi\sigma^{2})^{n/2}}
    \int_{\mathbb{R}^{n}} \dd^{n}\boldsymbol{y}\,
    e^{-|\boldsymbol{y}|^{2}/(4\sigma^{2})}
    e^{-\boldsymbol{y}\cdot\boldsymbol{d}/(2\sigma^{2})}
    \Xi_{0}(u,\boldsymbol{y}) .
    \label{eq:app_Wd_second}
\end{equation}
We now introduce spherical coordinates in $\mathbb{R}^{n}$ with the polar axis aligned with $\boldsymbol{d}$. Thus,
\begin{equation}
    r=|\boldsymbol{y}|,
    \qquad
    \boldsymbol{y}\cdot\boldsymbol{d}=rd\cos\theta,
    \qquad
    \dd^{n}\boldsymbol{y}
    =
    r^{n-1}\dd r\,\dd\Omega_{n-1}.
\end{equation}
where $d = |\boldsymbol{d}|$. Since $\Xi_{0}(u,\boldsymbol{y})$ is radial in $\boldsymbol{y}$, we can write it as $\Xi_{0}(u,r)$. Moreover, since the angular integral is invariant under $\cos\theta\mapsto-\cos\theta$, the sign in the exponential can be reversed. Therefore,
\begin{equation}
    \mathfrak{W}_{\boldsymbol{d}}(u)
    =
    \frac{e^{-d^{2}/(4\sigma^{2})}}{(4\pi\sigma^{2})^{n/2}}
    \int_{0}^{\infty}\dd r\,
    r^{n-1}
    e^{-r^{2}/(4\sigma^{2})}
    \Xi_{0}(u,r)
    \int_{\mathbb{S}^{n-1}}\dd\Omega_{n-1}\,
   e^{\frac{rd \cos \theta}{2 \sigma^2}}
    \label{eq:app_Wd_angular_step}
\end{equation}
Let $\kappa \equiv \frac{rd}{2\sigma^2}$. The angular integral can then be cast into
\begin{equation}
    \int_{\mathbb{S}^{n-1}}\dd\Omega_{n-1}\,
    e^{\kappa\cos\theta} = |\mathbb{S}^{n-2}| \int_{0}^{\pi}\dd \theta \, (\sin \theta)^{n-2}e^{\kappa \cos \theta},
\end{equation}
where the surface area of the unit $(n-2)$-sphere reads
\begin{equation}
    |\mathbb{S}^{n-2}| = \frac{2 \pi^{(n-1)/2}}{\Gamma \left(\frac{n-1}{2}\right)}.
\end{equation}
Then, by using the integral representation of
the modified Bessel function of the first kind~\cite{NIST:DLMF}
\begin{equation}
    I_{\nu}(z)
    =
    \frac{
        \left(\frac{z}{2}\right)^{\nu}
    }{
        \pi^{1/2}\Gamma\!\left(\nu+\frac{1}{2}\right)
    }
    \int_{0}^{\pi}  \dd\theta \, 
    e^{\pm z\cos\theta}
    (\sin\theta)^{2\nu},
\end{equation}
we obtain
\begin{equation}
    \int_{\mathbb{S}^{n-1}}\dd\Omega_{n-1}\,
    e^{\kappa\cos\theta}
    =
    2\pi^{n/2}
    \left(\frac{2}{\kappa}\right)^{\frac{n}{2}-1}
    I_{\frac{n}{2}-1}(\kappa),
    \label{eq:app_angular_integral}
\end{equation}
It is useful to introduce the normalized angular factor
\begin{equation}
    \mathcal{I}_{n}(\kappa)
    =
    \Gamma\!\left(\frac{n}{2}\right)
    \left(\frac{2}{\kappa}\right)^{\frac{n}{2}-1}
    I_{\frac{n}{2}-1}(\kappa),
    \label{eq:app_In_def}
\end{equation}
which satisfies $\mathcal{I}_{n}(0)=1$. Putting everything together, we obtain
\begin{equation}
    \mathfrak{W}_{\boldsymbol{d}}(u)
    =
    \frac{
        A_{n}e^{-d^{2}/(4\sigma^{2})}
    }{
        2^{n-1}\sigma^{n}\Gamma\!\left(\frac{n}{2}\right)
    }
    \int_{0}^{\infty}\dd r\,
    r^{n-1}
    e^{-r^{2}/(4\sigma^{2})}
    \mathcal{I}_{n}\!\left(
        \frac{rd}{2\sigma^{2}}
    \right)
    \frac{
        n\left(u_{\epsilon}^{4}+r^{4}\right)
        +
        2(n+2)u_{\epsilon}^{2}r^{2}
    }{
        \left(r^{2}-u_{\epsilon}^{2}\right)^{n+3}
    } .
    \label{eq:app_Wd_final}
\end{equation}

Observe that when $d=0$, we obtain the smeared distribution used in the evaluation of the local noise terms, Eq.~\eqref{eq:smeared_distribution_single_detector} in the main text. For $d \neq 0$, the remaining obstacle is the angular factor $\mathcal I_n$, which couples the detector separation d to the radial integration. Expanding $\mathcal I_n$ into its power series separates the dependence on $d$ from the radial integral, reducing the problem to a family of one-dimensional integrals that can each be evaluated analytically. Namely,
\begin{equation}
    \mathcal{I}_{n}(\kappa)
    =
    \Gamma\!\left(\frac{n}{2}\right)
    \sum_{m=0}^{\infty}
    \frac{
        \kappa^{2m}
    }{
        2^{2m}m!\,\Gamma\!\left(m+\frac{n}{2}\right)
    } .
    \label{eq:app_In_series}
\end{equation}
Since the series for $I_\nu$ is absolutely convergent for every finite $\kappa$, the series in Eq.~\eqref{eq:app_In_series} is absolutely convergent. For fixed $\epsilon>0$, the Gaussian damping factor in Eq.~\eqref{eq:app_Wd_final} then justifies interchanging the summation and the radial integration. We then obtain
\begin{equation}
     \mathfrak{W}_{\boldsymbol{d}}(u)
    =
    \frac{
        A_{n}e^{-d^{2}/(4\sigma^{2})}
    }{
        2^{n-1}\sigma^{n}
    }   \sum_{m=0}^{\infty}\int_{0}^{\infty}\dd r\,
    e^{-r^{2}/(4\sigma^{2})}
    \frac{
        n\left(u_{\epsilon}^{4}+r^{4}\right)
        +
        2(n+2)u_{\epsilon}^{2}r^{2}
    }{
        \left(r^{2}-u_{\epsilon}^{2}\right)^{n+3}
    } 
    \frac{
        r^{2m + n -1}d^{2m}
    }{
        16^{m}\sigma^{4m}m!\,
        \Gamma\!\left(m+\frac{n}{2}\right)
    } .
    \label{eq:LeonSKontheVillage}
\end{equation}
By considering integrals of the form
\begin{equation}
    T_{\alpha}(q) = \int_{0}^{\infty}\dd r\,
    \frac{
        r^{2\alpha-1}e^{-r^{2}/(4\sigma^{2})}
    }{
        (r^{2}+q)^{n+3}
    },
    \label{eq:auxuliary_integral}
\end{equation}
one can rewrite Eq.~\eqref{eq:LeonSKontheVillage} as (with $\alpha_{m} \equiv m + n/2$)
\begin{equation}
    \mathfrak{W}_{\boldsymbol{d}}(u) 
    =
    \frac{
        A_{n}e^{-d^{2}/(4\sigma^{2})}
    }{
        2^{n-1}\sigma^{n}
    }
    \sum_{m=0}^{\infty}
    \frac{
        d^{2m}
    }{
        16^{m}\sigma^{4m}m!\Gamma(\alpha_{m})
    }
    \Bigl[
        n u_{\epsilon}^{4}T_{\alpha_{m}}(-u_{\epsilon}^2)
        +
        2(n+2)u_{\epsilon}^{2}T_{\alpha_{m}+1}(-u_{\epsilon}^2)
        +
        nT_{\alpha_{m}+2}(-u_{\epsilon}^2)
    \Bigr].
\end{equation}
The integral in Eq.~\eqref{eq:auxuliary_integral} can be evaluated in closed form by introducing the change of variables
\begin{equation}
t=\frac{r^{2}}{4\sigma^{2}},
\end{equation}
which brings it into the standard integral representation of the Tricomi confluent hypergeometric function $U(a,b,z)$ (see Eq.~13.4.4 of the DLMF~\cite{NIST:DLMF}). One then obtains 

\begin{equation}
    T_{\alpha}(q) =     \frac{1}{2}
    q^{\alpha-n-3}
    \Gamma(\alpha)
    U\!\left(
        \alpha,
        \alpha-n-2,
        \frac{q}{4\sigma^{2}}
    \right).
\end{equation}
Therefore, the closed form expression for the smeared distribution is
\begin{equation}
    \mathfrak{W}_{\boldsymbol{d}}(u)
    =
    \frac{
        A_{n}e^{-d^{2}/(4\sigma^{2})}
    }{
        2^{n-1}\sigma^{n}
    }
    \sum_{m=0}^{\infty}
    \frac{1}{m!}
    \left(
        \frac{d^{2}}{16\sigma^{4}}
    \right)^{m}
    (-u_{\epsilon}^2)^{m-\frac{n}{2}-1}
    \mathcal{U}_{m}\left(-\frac{u_{\epsilon}^2}{4\sigma^2}\right),
    \label{eq:Wd_closed}
\end{equation}
where
\begin{equation}
    \mathcal{U}_{m}(z)
    =
    \frac{n}{2}
    U\!\left(
        \alpha_{m},
        m-\frac{n}{2}-2,
        z
    \right)
    -
    (n+2)\alpha_{m}
    U\!\left(
        \alpha_{m}+1,
        m-\frac{n}{2}-1,
        z
    \right) +
    \frac{n}{2}\alpha_{m}(\alpha_{m}+1)
    U\!\left(
        \alpha_{m}+2,
        m-\frac{n}{2},
        z
    \right).
    \label{eq:app_Um_def}
\end{equation}
For $d = 0$, since every term with $m>0$ carries an explicit factor $d^{2m}$, setting $d=0$ leaves only the $m=0$ contribution:
\begin{equation}
\mathfrak{W}(u)
=
\frac{A_{n} n}{2^{n}\sigma^{n}}
\left(-u_{\epsilon}^{2}\right)^{-\frac{n}{2}-1}
\Biggl[
U\!\left(
\frac{n}{2},
-\frac{n}{2}-2,
-\frac{u_{\epsilon}^{2}}{4\sigma^{2}}
\right)
-(n+2)\,
U\!\left(
\frac{n}{2}+1,
-\frac{n}{2}-1,
-\frac{u_{\epsilon}^{2}}{4\sigma^{2}}
\right)
+\frac{n(n+2)}{4}\,
U\!\left(
\frac{n}{2}+2,
-\frac{n}{2},
-\frac{u_{\epsilon}^{2}}{4\sigma^{2}}
\right)
\Biggr],
\label{eq:smeared_tricomi_appendix}
\end{equation}
which is Eq.~\eqref{eq:smeared_tricomi} presented in the main text.

\section{Leading singularity behavior of $\mathfrak{W}_{AB}(u)$}
\label{sec:new_appendix}

In this appendix, we show explicitly how to derive the singularity structure of the smeared distribution $\mathfrak{W}_{\tc{ab}}$ for different spacetime dimensions, as in Eqs.~\eqref{eq:n=1}-\eqref{eq:n_3}.

To that end, recall that
\begin{equation}
    \mathfrak{W}_{\tc{ab}}(u) = \Theta(u)\mathfrak{W}_{\boldsymbol{d}}(u) + \Theta(-u)\mathfrak{W}_{\boldsymbol{d}}(u)^{*},
    \label{eq:mathfrak_ab_appendix}
\end{equation}
where
\begin{equation}
\mathfrak{W}_{\boldsymbol{d}}(u)
    =
    \frac{
        A_{n}e^{-d^{2}/(4\sigma^{2})}
    }{
        2^{n-1}\sigma^{n}
    }
    \sum_{m=0}^{\infty}
    \frac{1}{m!}
    \left(
        \frac{d^{2}}{16\sigma^{4}}
    \right)^{m}
    (-u_{\epsilon}^{2})^{m-\frac{n}{2}-1}
    \mathcal{U}_{m}\!\left(
        -\frac{u_{\epsilon}^{2}}{4\sigma^{2}}
    \right),
\end{equation}
and (with $\alpha_{m} \equiv m + n/2$)
\begin{equation}
    \begin{aligned}
    \mathcal{U}_{m}(z)
    &=
    \frac{n}{2}
    U\!\left(
        \alpha_{m},
        m-\frac{n}{2}-2,
        z
    \right) -
    (n+2)\alpha_{m}
    U\!\left(
        \alpha_{m}+1,
        m-\frac{n}{2}-1,
        z
    \right) \\
    &\quad+
    \frac{n}{2}\alpha_{m}(\alpha_{m}+1)
    U\!\left(
        \alpha_{m}+2,
        m-\frac{n}{2},
        z
    \right).
    \end{aligned}
    \label{eq:mathcalUm_app}
\end{equation}
To ease the notation, let
\begin{equation}
    \rho \equiv \frac{d^2}{4 \sigma^2}, \qquad z \equiv -\frac{u_{\epsilon}^2}{4 \sigma^2}.
\end{equation}
Then, $\mathfrak{W}_{\boldsymbol{d}}(u)$ can be recast as
\begin{equation}
    \mathfrak{W}_{\boldsymbol d}(u)
    =\frac{A_n e^{-\rho}z^{-n/2-1}}{2^{2n+1}\sigma^{2n+2}}
    \sum_{m=0}^{\infty}\frac{(\rho z)^m}{m!}\mathcal U_m(z),
    \label{eq:mathfrakWd_app}
\end{equation}
where we recall that the constant $A_{n}$ is defined in Eq.~\eqref{eq:def_An_main}. 

Now, it turns out that all the auxiliary functions in Eq.~\eqref{eq:mathcalUm_app} can be obtained from $\mathcal{U}_{0}(z)$. Indeed, the derivative identity for the Tricomi function~\cite{NIST:DLMF},
\begin{equation}
    \frac{\dd}{\dd z}U(A,B,z)=-A\,U(A+1,B+1,z),
\end{equation}
applied to each term in Eq.~\eqref{eq:mathcalUm_app} gives
\begin{equation}
    \mathcal U_m'(z)=-\alpha_{m}\mathcal U_{m+1}(z).
    \label{eq:recursive_Um_app}
\end{equation}
Consequently,
\begin{equation}
    \mathcal U_m(z)
    =\frac{(-1)^m \Gamma(n/2)}{\Gamma(\alpha_{m})}
    \frac{\dd^m}{\dd z^m}\mathcal U_0(z).
\end{equation}
Hence, to determine the leading singularity structure of Eq.~\eqref{eq:mathfrakWd_app}, it is sufficient to consider the asymptotic behavior of $\mathcal{U}_{0}(z)$ as $z \to 0$. Notice that we will derive the results for positive $z$, and then continue them to $z=-u_{\epsilon}^2/(4\sigma^2)$ using the principal branches. For $m=0$, the following identity is useful:
\begin{equation}
    \mathcal{U}_{0}(z) = \frac{n}{2} z^2 U(n/2 + 2, 2-n/2, z).
\end{equation}
Let us now establish the leading singularity behavior of the smeared distribution $\mathfrak{W}_{\tc{ab}}$ for each dimension.
\subsection{1 + 1 dimensions}
For $n=1$, the small argument expansion of $\mathcal{U}_{0}$
yields
\begin{equation}
    \mathcal{U}_{0}(z) = \frac23z^{3/2}-\sqrt\pi\,z^2
    +\mathcal O(z^{5/2}).
\end{equation}
Thus, according to the recursive relation \eqref{eq:recursive_Um_app}, the leading behavior of each $\mathcal{U}_{m}$ is $z^{3/2-m}$. In the series of Eq.~\eqref{eq:mathfrakWd_app}, each $\mathcal{U}_{m}$ multiplies $z^{m -3/2}$. Hence, all values of $m$ will contribute a constant to the leading behavior of $\mathfrak{W}_{\boldsymbol{d}}(u)$ for small $z$. Concretely,
\begin{equation}
    \lim_{z\to0}z^{m-3/2}\mathcal U_m(z)
    =\frac{2}{(2m-1)(2m-3)}.
\end{equation}
Therefore, at leading order, 
\begin{align}
  \mathfrak{W}_{\boldsymbol{d}}(u) & = \frac{A_1 e^{-\rho}}{2^{3}\sigma^{4}}
    \sum_{m=0}^{\infty}\frac{\rho^m}{m!} \frac{2}{(2m -1)(2m - 3)} + \mathcal{O}(z^{1/2}) \nonumber \\ & = \frac{e^{-d^2/(4\sigma^2)}}{48\pi^2\sigma^4}
    {}_1F_1\!\left(-\frac32;\frac12;
        \frac{d^2}{4\sigma^2}\right) + \mathcal{O}(|u_{\epsilon}|),
\end{align}
where $ {}_1F_1(a; b; c)$ is the Kummer confluent hypergeometric function~\cite{NIST:DLMF}. Here $A_1=1/(4\pi^2)$ is understood as the continuous $n\to1$ limit of $A_n$. Moreover, using the principal branch of the square root, $z^{1/2}=(\epsilon+\ii u)/(2\sigma)$. Therefore, using Eq.~\eqref{eq:mathfrak_ab_appendix} and taking the limit $\epsilon\to0^+$, it follows that
\begin{equation}
\mathfrak{W}_{\tc{ab}}(u) = \frac{e^{-d^2/(4\sigma^2)}}{48\pi^2\sigma^4}
    {}_1F_1\!\left(-\frac32;\frac12;
        \frac{d^2}{4\sigma^2}\right) + \mathcal O(|u|).
\end{equation}

\subsection{2 + 1 dimensions}
For $n=2$, we can write
\begin{equation}
    \mathcal{U}_{0}(z) = -\frac{z^2}{2}\left(\log z+\gamma+\frac32\right)
    +\mathcal O(z^3\log z),
\end{equation}
where $\gamma$ is the Euler--Mascheroni constant. In Eq.~\eqref{eq:mathfrakWd_app} for $n=2$, each $\mathcal{U}_{m}$ is multiplied by $z^{m-2}$. Since
$\mathcal U_1=-\mathcal U_0'$ and
$\mathcal U_2=\mathcal U_0''/2$, we obtain
\begin{align}
    z^{-2}\mathcal U_0(z)
    &=-\frac12\log z-\frac\gamma2-\frac34
    +\mathcal O(z\log z),
    \nonumber\\
    z^{-1}\mathcal U_1(z)
    &=\log z+\gamma+2+\mathcal O(z\log z),
    \nonumber\\
    \mathcal U_2(z)
    &=-\frac12\log z-\frac{\gamma+3}{2}
    +\mathcal O(z\log z).
\end{align}
For each $m \ge 3$, we have
\begin{equation}
    \frac{\dd^m}{\dd z^m}(z^2\log z)
    =2(-1)^{m-3}(m-3)!z^{2-m}.
\end{equation}
Thus,
\begin{equation}
    \lim_{z\to0}z^{m-2}\mathcal U_m(z)
    =\frac{1}{m(m-1)(m-2)}.
\end{equation}
This means that each term with $m \ge 3$ contributes a constant to the finite part of $\mathfrak{W}_{\boldsymbol{d}}$, whereas $m=0,1,2$ contribute to the logarithmic singularity. Hence,
\begin{equation}
    \mathfrak{W}_{\boldsymbol{d}}(u) = -\frac{3e^{-\rho}}{1024\pi^2\sigma^6}
    \left(\frac{1}{2} -\rho + \frac{\rho^2}{4}\right)\log z+\mathcal{O}(1),
\end{equation}
where the notation $\mathcal{O}(1)$ indicates constant terms, as well as higher order terms that vanish in the limit $z \to 0$, and as such do not contribute to the leading singularity. Then, using Eq.~\eqref{eq:mathfrak_ab_appendix}, we obtain
\begin{align}
    \mathfrak W_{\tc{ab}}(u)
    &={} -\frac{3e^{-d^2/(4\sigma^2)}}{1024\pi^2\sigma^6} \left( \frac{1}{2} -\frac{d^2}{4\sigma^2} + \frac{d^4}{64\sigma^4}\right)
    \nonumber\\
    &\quad\times\Biggl[
    \Theta(u)\log\!\left(-\frac{(u-\ii\epsilon)^2}{4\sigma^2}\right)
    +\Theta(-u)\log\!\left(-\frac{(u+\ii\epsilon)^2}{4\sigma^2}\right)
    \Biggr]+\mathcal O(1).
\end{align}

\subsection{Higher dimensions}
Finally, for $n\geq3$, the small argument expansion of the Tricomi function gives
\begin{equation}
    U(n/2+2,2-n/2,z) \sim \frac{\Gamma(n/2-1)}{n!},
\end{equation}
with the omitted terms vanishing in the limit $z \to 0$. Thus, $\mathcal U_0(z)$ is proportional to $z^2$ at leading order. Using the recursive relation \eqref{eq:recursive_Um_app}, the first three terms in the series for $\mathfrak{W}_{\boldsymbol d}$ have the following leading behavior:
\begin{align}
    z^{-n/2-1}\mathcal U_0(z)
    &\sim\frac{\Gamma(n/2-1)}{2(n-1)!}z^{1-n/2},
    \nonumber\\
    z^{-n/2}\mathcal U_1(z)
    &\sim-\frac{2\Gamma(n/2-1)}{n!}z^{1-n/2},
    \nonumber\\
    z^{1-n/2}\mathcal U_2(z)
    &\sim\frac{4\Gamma(n/2-1)}{(n+2)n!}z^{1-n/2}.
\end{align}
Since the leading $z^2$ term in $\mathcal U_0(z)$ vanishes after three derivatives, the terms with $m\geq3$ do not contribute at this order. Including the weights $1$, $\rho$, and $\rho^2/2$ of the first three terms in Eq.~\eqref{eq:mathfrakWd_app}, we obtain
\begin{equation}
    \mathfrak W_{\boldsymbol d}(u)
    =\frac{A_n e^{-\rho}\Gamma(n/2-1)}{2^{2n+2}(n-1)!\sigma^{2n+2}}
    \left(1-\frac{4\rho}{n}+\frac{4\rho^2}{n(n+2)}\right)
    z^{1-n/2}+\mathcal R_n(z),
\end{equation}
where the subleading terms satisfy
\begin{equation}
    \mathcal R_n(z)=
    \begin{cases}
        \mathcal O(1),&n=3,\\
        \mathcal O(\log z),&n=4,\\
        \mathcal O(z^{2-n/2}),&n\geq5.
    \end{cases}
\end{equation}
Restoring the original variables and using $(-u_\epsilon^2)^{1/2}=\epsilon+\ii u=\ii u_\epsilon$, we can write
\begin{equation}
    \mathfrak W_{\boldsymbol d}(u)
    =e^{-d^2/(4\sigma^2)}
    \left(1-\frac{d^2}{n\sigma^2}+\frac{d^4}{4n(n+2)\sigma^4}\right)
    \frac{B_n}{(u-\ii\epsilon)^{n-2}}+\mathcal R_n(z),
\end{equation}
with
\begin{equation}
    B_n=\frac{\ii^{\,2-n}A_n\Gamma(n/2-1)}{2^{n+4}(n-1)!\sigma^{n+4}}.
\end{equation}
Finally, using Eq.~\eqref{eq:mathfrak_ab_appendix}, the leading behavior of the time-ordered distribution is
\begin{equation}
    \mathfrak W_{\tc{ab}}(u)
    \sim e^{-d^2/(4\sigma^2)}
    \left(1-\frac{d^2}{n\sigma^2}+\frac{d^4}{4n(n+2)\sigma^4}\right)\left[
        \frac{B_n\Theta(u)}{(u-\ii\epsilon)^{n-2}}
        +\frac{B_n^*\Theta(-u)}{(u+\ii\epsilon)^{n-2}}
    \right]
\end{equation}
where the omitted terms follow by applying the same time-ordering prescription to $\mathcal R_n(z)$.


\subsection{Dependence on general spatial profiles}
\label{sec:edu_general_profiles}

To determine which properties of the spatial profiles control the singularity at coincident times, we assume that $F_{\tc a}$ and $F_{\tc b}$ are real Schwartz functions. They and all their derivatives are smooth and decrease faster than any inverse power at infinity. This class includes smooth compactly supported profiles. Their overlap, defined in Eq.~\eqref{eq:edu_spatial_overlap_definition}, is
\begin{equation}
 \vartheta_{\tc{ab}}(\boldsymbol y)
 =\int_{\mathbb R^n}\dd^n\boldsymbol z\,
 F_{\tc a}(\boldsymbol z)F_{\tc b}(\boldsymbol z-\boldsymbol y).
 \label{eq:edu_overlap_recalled}
\end{equation}
It measures the overlap after a relative displacement $\boldsymbol y$ and is also a Schwartz function. We consider $n\geq3$, where nonintegrable singularities can remain after spatial smearing.

The first step is to express the energy-density correlator as a spatial derivative of a simpler kernel. Write $s=|\boldsymbol y|^2$, $q=-u_\epsilon^2$, and $u_\epsilon=u-\ii\epsilon$. For a radial function $g(s)$, the chain rule gives $\Delta_{\boldsymbol y}g=4s g''+2n g'$, where $\Delta_{\boldsymbol y}=\sum_i\partial_{y_i}^2$. Applying this operator again yields
\begin{equation}
 \Delta_{\boldsymbol y}^{\,2}g(s)
 =16s^2g^{(4)}(s)+16(n+2)s g^{(3)}(s)+4n(n+2)g''(s).
 \label{eq:edu_radial_bilaplacian}
\end{equation}
Substituting $g(s)=(s+q)^{1-n}$ reproduces the rational function in Eq.~\eqref{eq:app_W0_energy_density}, giving
\begin{equation}
 \Xi_0(u,\boldsymbol y)=\kappa_n\Delta_{\boldsymbol y}^{\,2}
 (|\boldsymbol y|^2+q)^{1-n},
 \qquad \kappa_n=\frac{A_n}{4n(n-1)(n+2)}.
 \label{eq:edu_spatial_derivative_identity}
\end{equation}
At finite $\epsilon>0$, the kernel is smooth at real $\boldsymbol y$, while the overlap decreases rapidly at infinity. We may therefore transfer both Laplacians onto the overlap by integration by parts, with no boundary terms or overall minus sign. Thus
\begin{equation}
 \mathfrak W_{\boldsymbol d}(u)
 =\kappa_n\int_{\mathbb R^n}\dd^n\boldsymbol y\,
 \frac{h(\boldsymbol y)}{(|\boldsymbol y|^2+q)^{n-1}},
 \qquad h\equiv\Delta_{\boldsymbol y}^{\,2}\vartheta_{\tc{ab}}.
 \label{eq:edu_general_spatial_sampling}
\end{equation}
This identifies the bi-Laplacian of the overlap as the spatial quantity whose local behaviour we need to examine.

We first take $q>0$. Since the denominator is radial, only the spherical average $\bar h(r)$ contributes. Rotational symmetry gives $\langle\omega_i\rangle=0$ and $\langle\omega_i\omega_j\rangle=\delta_{ij}/n$, so the Taylor expansion of $h(r\boldsymbol\omega)$ implies
\begin{equation}
 \bar h(r)\equiv\frac1{S_{n-1}}\int_{\mathbb S^{n-1}}\dd\Omega_{n-1}\,
 h(r\boldsymbol\omega)
 =h(\boldsymbol0)+\frac{r^2}{2n}\Delta_{\boldsymbol y}h(\boldsymbol0)+\mathcal O(r^4),
 \label{eq:edu_spatial_average}
\end{equation}
where $S_{n-1}=2\pi^{n/2}/\Gamma(n/2)$ is the area of the unit sphere. The spatial integral becomes
\begin{equation}
 \mathfrak W_{\boldsymbol d}(u)
 =\kappa_n S_{n-1}\int_0^\infty\dd r\,
 \frac{r^{n-1}\bar h(r)}{(r^2+q)^{n-1}}.
 \label{eq:edu_radial_sampling}
\end{equation}
Split this integral at a fixed radius $R>0$. For $r\geq R$ and $|q|<R^2/2$, expansion in $q/r^2$ converges uniformly and gives an outer contribution analytic at $q=0$. The singular terms therefore come entirely from $r<R$, where the local expansion applies.

For the higher terms, write a finite expansion
\begin{equation}
 \bar h(r)=\sum_{j=0}^{J}a_j r^{2j}+r^{2J+2}b_J(r^2),
 \qquad
 a_j=\frac{\Delta_{\boldsymbol y}^{\,j+2}\vartheta_{\tc{ab}}(\boldsymbol0)}
 {4^j j!(n/2)_j},
 \label{eq:edu_angular_coefficients}
\end{equation}
with $b_J$ smooth and $(a)_j=\Gamma(a+j)/\Gamma(a)$. The coefficients follow by applying the radial Laplacian repeatedly: it maps $r^{2j}$ to $4j(j+n/2-1)r^{2j-2}$ and commutes with angular averaging. Evaluating after $j$ applications at $r=0$ selects $a_j$.

Each Taylor term contributes $\kappa_n S_{n-1}a_j I_j(q,R)$, with
\begin{equation}
 I_j(q,R)=\int_0^R\dd r\,\frac{r^{n-1+2j}}{(r^2+q)^{n-1}}
 =q^{\alpha_j}\int_0^{R/\sqrt q}\dd t\,
 \frac{t^{n-1+2j}}{(1+t^2)^{n-1}},
 \qquad \alpha_j=j+1-\frac n2.
 \label{eq:edu_radial_moments}
\end{equation}
The substitution $r=\sqrt q\,t$ isolates the possible power of $q$. If $n>2j+2$, the large-$t$ behaviour $t^{2j+1-n}$ is integrable. Extending the upper limit to infinity changes only an analytic term, and $v=t^2$ reduces the integral to the beta integral (Eq.~5.12.3 of the DLMF~\cite{NIST:DLMF}). Hence
\begin{equation}
 I_j(q,R)=\frac{\Gamma(n/2+j)\Gamma(n/2-j-1)}{2\Gamma(n-1)}
 q^{\alpha_j}+\text{a function analytic at }q=0.
 \label{eq:edu_radial_power}
\end{equation}
Taking $j=0$ and inserting $\kappa_n S_{n-1}a_0$ gives the leading term
\begin{equation}
 \mathfrak W_{\boldsymbol d}(u)
 =\frac{A_n\pi^{n/2}\Gamma(n/2-1)}{4n(n+2)\Gamma(n)}
 \Delta_{\boldsymbol y}^{\,2}\vartheta_{\tc{ab}}(\boldsymbol0)
 (-u_\epsilon^2)^{1-n/2}+\text{subleading terms}.
 \label{eq:edu_general_leading_coefficient}
\end{equation}
At $n=2j+2$, the large-$t$ integrand instead behaves as $t^{-1}$, giving
\begin{equation}
 I_j(q,R)=-\frac12\log\!\left(\frac q{R^2}\right)+\mathcal O(1).
 \label{eq:edu_radial_logarithm}
\end{equation}
The factor $1/2$ comes from the upper limit $R/\sqrt q$. For $n<2j+2$, the original radial integrand at $q=0$ has exponent $2j+1-n>-1$ and is integrable, so $I_j$ is bounded as $q\to0^+$.

To return to real time, we use the principal branch, for which $q=(\epsilon+\ii u)^2$ and $\sqrt q=\epsilon+\ii u$. Thus $q$ approaches the negative real axis from above for $u>0$ and below for $u<0$, and
\begin{equation}
 \lim_{\epsilon\to0^+}\log\!\left(\frac q{R^2}\right)
 =2\log\!\left(\frac{|u|}{R}\right)+\ii\pi\operatorname{sgn}(u),
 \qquad u\neq0.
 \label{eq:edu_log_boundary}
\end{equation}
The exact power terms and analytic outer contributions continue with this prescription. For the Taylor remainder, choosing $J\geq n$, setting $s=r^2$, and integrating by parts $n-2$ times reduces the potentially singular part to a constant times
\begin{equation}
 \int_0^{R^2}\dd s\,\frac{\psi_J(s)}{s+q},
 \qquad
 \psi_J(s)=\frac{\dd^{n-2}}{\dd s^{n-2}}\bigl[s^{n/2+J}b_J(s)\bigr].
 \label{eq:edu_taylor_remainder}
\end{equation}
The boundary terms at $R^2$ are analytic in $q$, and those at zero vanish. Here $\psi_J$ is continuously differentiable with $\psi_J(0)=0$. Near the cut, write $q=-a+\ii b$ with $a>0$ and subtract $\psi_J(a)$ from the numerator. The mean value theorem gives $|\psi_J(s)-\psi_J(a)|\leq C|s-a|\leq C|s+q|$, so the resulting integral is uniformly bounded. The subtracted constant gives $\psi_J(a)\log[(R^2+q)/q]$, which remains bounded because $\psi_J(a)=\mathcal O(a)$. For $\operatorname{Re}q\geq0$, boundedness follows directly from $|s+q|\geq s$. The same argument covers the Taylor terms with $\alpha_j>0$: after integration by parts their numerators are proportional to $s^{\alpha_j}$, whose subtracted difference quotients are integrable. Thus the remainder and these terms stay bounded as the regulator is removed. No additional nonintegrable contribution is concealed by the continuation from positive $q$.

We can now determine which spatial coefficients must vanish to make the time integral finite. A power term has magnitude $|u|^{2\alpha_j}=|u|^{2j+2-n}$ in the boundary limit. It is locally integrable precisely when $2j+2-n>-1$. The potentially nonintegrable indices are therefore $0\leq j\leq\lfloor(n-3)/2\rfloor$. Using Eq.~\eqref{eq:edu_angular_coefficients} and setting $k=j+2$ gives the sufficient condition
\begin{equation}
 \Delta_{\boldsymbol y}^{\,k}\vartheta_{\tc{ab}}(\boldsymbol0)=0,
 \qquad k=2,\ldots,\left\lfloor\frac{n+1}{2}\right\rfloor,
 \qquad n\geq3.
 \label{eq:edu_spatial_sufficient_condition}
\end{equation}
After these coefficients vanish, the kernel is bounded for odd $n$ and at most logarithmic for even $n$. The logarithm is locally integrable, and $|q|=u^2+\epsilon^2$ supplies an integrable bound uniform in small $\epsilon$. Together with the remainder estimate, this permits the regulator limit under the temporal integral. Multiplication by the step functions implementing time ordering preserves this local integrability. The condition is sufficient for any smooth temporal sampling, with suitable decay at large times. It is not necessary for a particular switching, whose correlation may itself vanish near $u=0$ or cancel the divergent coefficients.

For Gaussian profiles, substituting $g(s)\propto e^{-s/(4\sigma^2)}$, with $s=|\boldsymbol y+\boldsymbol d|^2$, into Eq.~\eqref{eq:edu_radial_bilaplacian} gives
\begin{equation}
 \Delta_{\boldsymbol y}^{\,2}\vartheta_{\tc{ab}}(\boldsymbol0)
 =\frac{n(n+2)}{4\sigma^4}\vartheta_{\tc{ab}}(\boldsymbol0)
 \left(1-\frac{d^2}{n\sigma^2}
 +\frac{d^4}{4n(n+2)\sigma^4}\right).
\end{equation}
This recovers the separation polynomial in Eq.~\eqref{eq:n_3}, whose zeros are the exceptional separations in Eq.~\eqref{eq:exceptional_separations}. For $n=3,4$, only the bi-Laplacian must vanish. The next term has $j=1$, so it is bounded for $n=3$ and logarithmic for $n=4$, establishing finiteness in both cases. For $n\geq5$, that term can also be nonintegrable, and further derivatives enter Eq.~\eqref{eq:edu_spatial_sufficient_condition}.

Finally, if compact spatial supports are separated by a positive distance $\ell$, the two factors in Eq.~\eqref{eq:edu_overlap_recalled} cannot both be nonzero for $|\boldsymbol y|<\ell$. The overlap and all its derivatives vanish there, leaving only a regular outer contribution near $u=0$. Together with the Gaussian cancellations, this shows why the local derivatives of the overlap, rather than non-compact tails alone, determine which singularities survive.

\section{Auxiliary calculations for the thermal state case}
\label{sec:garrafa_termica}

In this appendix, we derive the expression for the contribution $\mathcal L_{jj,w}$ to the local noise appearing in the thermal-state analysis. The derivation consists of a sequence of changes of variables which reduce the original four-dimensional integral to a single integral
over the thermal correlation function $g_\beta(s)$.Let us start by showing how to simplify the integral
\begin{align}
    \mathcal{L}_{jj, w}
    &=
    \int_{-\infty}^{\infty}\dd t\,\dd x\,\dd t'\,\dd x'\,
    e^{-\ii\Omega(t-t')}
    \chi_j(t)\chi_j(t')
    F_j(x)F_j(x')
    \Xi_{w}(\mf x,\mf x') .
\end{align}
 For convenience, we introduce the function
\begin{equation}
    g_{\beta}(s)
    =
    \frac{1}{4\pi s^2}
    -
    \frac{\pi}{4\beta^2}
    \csch^2\!\left(
        \frac{\pi s}{\beta}
    \right),
\end{equation}
we can write
\begin{equation}
    \Xi_{w}(\mf x,\mf x')
    =
    2\left[
        g_{\beta}(\Delta\zeta)^2
        +
        g_{\beta}(\Delta\eta)^2
    \right].
\end{equation}
Let us introduce the variables
\begin{equation}
    u=t-t',
    \qquad
    v=\frac{t+t'}{2},
    \qquad
    y=x-x',
    \qquad
    X=\frac{x+x'}{2}.
\end{equation}
Equivalently,
\begin{equation}
    t=v+\frac{u}{2},
    \qquad
    t'=v-\frac{u}{2},
    \qquad
    x=X+\frac{y}{2},
    \qquad
    x'=X-\frac{y}{2},
\end{equation}
and the volume element transforms as $\dd t\,\dd t'\,\dd x\,\dd x'
    =
\dd u\,\dd v\,\dd y\,\dd X $.
In terms of these variables, we have
\begin{equation}
    \Xi_{w}(\mf x,\mf x')
    =
    2\left[
        g_{\beta}(u-y)^2
        +
        g_{\beta}(u+y)^2
    \right].
\end{equation}
The integral over $v$ gives
\begin{equation}
    \int_{-\infty}^{\infty}\dd v\,
    \chi_j\!\left(v+\frac{u}{2}\right)
    \chi_j\!\left(v-\frac{u}{2}\right)
    =
    \sqrt{\frac{\pi}{2}}\,T\,e^{-\frac{u^2}{2T^2}} .
\end{equation}
Similarly,
\begin{equation}
    \int_{-\infty}^{\infty}\dd X\,
    F_j\!\left(X+\frac{y}{2}\right)
    F_j\!\left(X-\frac{y}{2}\right)
    =
    \frac{e^{-\frac{y^2}{4\sigma^2}}}{2\sqrt{\pi}\sigma}.
\end{equation}
Substituting these two Gaussian convolutions into $\mathcal{L}_{jj, w}$ gives (where we write $\mathcal L_w \equiv \mathcal L_{jj,w}$ as the result is independent of the detector label)
\begin{equation}
    \mathcal{L}_{w}
    =
    \frac{T}{\sqrt{2}\,\sigma}
    \int_{-\infty}^{\infty}\dd u
    \int_{-\infty}^{\infty}\dd y\,
    e^{-\ii\Omega u}
    e^{-\frac{u^2}{2T^2}}
    e^{-\frac{y^2}{4\sigma^2}}
    \left[
        g_{\beta}(u-y)^2
        +
        g_{\beta}(u+y)^2
    \right].
\end{equation}
This can be further simplified by introducing
\begin{equation}
    s=u-y,
    \qquad
    r=u+y.
\end{equation}
Thus,
\begin{equation}
    u=\frac{r+s}{2},
    \qquad
    y=\frac{r-s}{2},
\end{equation}
and the volume element transforms as
\begin{equation}
    \dd u\,\dd y
    =
    \frac{1}{2}\dd r\,\dd s.
\end{equation}
Therefore,
\begin{equation}
    \mathcal{L}_{w}
    =
    \frac{T}{2\sqrt{2}\,\sigma}
    \int_{-\infty}^{\infty}\dd r
    \int_{-\infty}^{\infty}\dd s\,
    \exp \left[
        -\frac{(r+s)^2}{8T^2}
        -\frac{(r-s)^2}{16\sigma^2}
        -\frac{\ii\Omega(r+s)}{2}
    \right]
    \left[
        g_{\beta}(s)^2
        +
        g_{\beta}(r)^2
    \right].
\end{equation}
 The exponential factor is invariant under the interchange
$r\leftrightarrow s$, and so is the integration domain
$\mathbb R^2$. Since the two terms inside the brackets are exchanged
under this transformation, they contribute equally to the integral.
Therefore,
\begin{equation}
    \mathcal{L}_{w}
    =
    \frac{T}{\sqrt{2}\,\sigma}
    \int_{-\infty}^{\infty}\dd s\,
    g_{\beta}(s)^2
    \int_{-\infty}^{\infty}\dd r\,
    \exp \left[
        -\frac{(r+s)^2}{8T^2}
        -\frac{(r-s)^2}{16\sigma^2}
        -\frac{\ii\Omega(r+s)}{2}
    \right].
\end{equation}
The remaining integral over $r$ is Gaussian. To simplify the notation, we define
\begin{equation}
    D \equiv T^2+2\sigma^2 .
\end{equation}
Expanding the exponent as a quadratic polynomial in $r$ and evaluating the
resulting Gaussian integral gives
\begin{align}
    &\int_{-\infty}^{\infty}\dd r\,
    \exp \left[
        -\frac{(r+s)^2}{8T^2}
        -\frac{(r-s)^2}{16\sigma^2}
        -\frac{\ii\Omega(r+s)}{2}
    \right]
    \nonumber \\
    &\qquad =
    \frac{4\sqrt{\pi}T\sigma}{\sqrt{D}}\,
    \exp \left[
        -\frac{s^2}{2D}
        -\frac{\ii T^2\Omega}{D}s
        -\frac{T^2\sigma^2\Omega^2}{D}
    \right].
\end{align}
Hence,
\begin{equation}
    \mathcal{L}_{w}
    =
    \frac{
        2\sqrt{2\pi}\,T^2
        e^{-\frac{T^2\sigma^2\Omega^2}{D}}
    }{
        \sqrt{D}
    }
    \int_{-\infty}^{\infty}\dd s\,
    e^{-\frac{s^2}{2D}}
    e^{-\frac{\ii T^2\Omega}{D}s}
    g_{\beta}(s)^2 .
\end{equation}
Since $g_\beta(s)^2$ and the Gaussian factor are both even functions of
$s$, the integrand multiplying the sine is odd. Its contribution
therefore vanishes after integration over the symmetric domain. The local noise contribution therefore reduces to
\begin{equation}
    \mathcal{L}_{w}
    =
    \frac{
        4\sqrt{2\pi}\,T^2
        e^{-\frac{T^2\sigma^2\Omega^2}{D}}
    }{
        \sqrt{D}
    }
    \int_{0}^{\infty}\dd s\,
    e^{-\frac{s^2}{2D}}
    \cos\!\left(
        \frac{T^2\Omega}{D}s
    \right)
    g_{\beta}(s)^2 .
\end{equation}

\twocolumngrid

\bibliography{references}

@article{Plenio_2005,
   title={Logarithmic Negativity: A Full Entanglement Monotone That is not Convex},
   volume={95},
   ISSN={1079-7114},
   url={http://dx.doi.org/10.1103/PhysRevLett.95.090503},
   number={9},
   journal={Phys. Rev. Letters},
   publisher={American Physical Society (APS)},
   author={Plenio, M. B.},
   year={2005},
   month=Aug }

@Article{Louko2014,
author={Louko, Jorma},
title={Unruh-DeWitt detector response across a Rindler firewall is finite},
journal={J. High Energy Phys.},
year={2014},
month={Sep},
day={24},
volume={2014},
number={9},
pages={142},
issn={1029-8479},
doi={10.1007/JHEP09(2014)142},
url={https://doi.org/10.1007/JHEP09(2014)142}
}

@article{Petar_Edu_coherent,
   title={All coherent field states entangle equally},
   volume={96},
   ISSN={2470-0029},
   url={http://dx.doi.org/10.1103/PhysRevD.96.025020},
   number={2},
   journal={Phys. Rev. D},
   publisher={American Physical Society (APS)},
   author={Simidzija, Petar and Martín-Martínez, Eduardo},
   year={2017},
   month=July }

@article{Unruh1976,
  title = {Notes on black-hole evaporation},
  author = {Unruh, W. G.},
  journal = {Phys. Rev. D},
  volume = {14},
  issue = {4},
  pages = {870--892},
  numpages = {0},
  year = {1976},
  month = {Aug},
  publisher = {American Physical Society},
  doi = {10.1103/PhysRevD.14.870},
  url = {https://link.aps.org/doi/10.1103/PhysRevD.14.870}
}

@book{DeWitt,
	Address = {Cambridge, UK},
	Author = {B. DeWitt},
	Publisher = {Cambridge University Press},
	Title = {General Relativity; an Einstein Centenary Survey},
	Year = {1980}}

@article{TalesBrunoEdu2020,
  title = {General relativistic quantum optics: Finite-size particle detector models in curved spacetimes},
  author = {Mart\'{i}n-Mart\'{i}nez, Eduardo and Perche, T. Rick and de S. L. Torres, Bruno},
  journal = {Phys. Rev. D},
  volume = {101},
  issue = {4},
  pages = {045017},
  numpages = {10},
  year = {2020},
  month = {Feb},
  publisher = {American Physical Society},
  doi = {10.1103/PhysRevD.101.045017},
  url = {https://link.aps.org/doi/10.1103/PhysRevD.101.045017}
}

@article{MorrisThorneWormholes,
    author = {Morris, Michael S. and Thorne, Kip S.},
    title = "{Wormholes in spacetime and their use for interstellar travel: A tool for teaching general relativity}",
    journal = {Am. J. Phys.},
    volume = {56},
    number = {5},
    pages = {395-412},
    year = {1988},
    month = {05},
}

@article{Langlois_2006,
   title={Causal particle detectors and topology},
   volume={321},
   ISSN={0003-4916},
   url={http://dx.doi.org/10.1016/j.aop.2006.01.013},
   DOI={10.1016/j.aop.2006.01.013},
   number={9},
   journal={Ann. Phys.},
   publisher={Elsevier BV},
   author={Langlois, Paul},
   year={2006},
   month=Sept, pages={2027–2070} }

@article{Hollands_2002,
   title={The State Space of Perturbative Quantum Field Theory in Curved Spacetimes},
   volume={3},
   ISSN={1424-0637},
   url={http://dx.doi.org/10.1007/s00023-002-8629-2},
   DOI={10.1007/s00023-002-8629-2},
   number={4},
   journal={Ann. Henri Poincare},
   publisher={Springer Science and Business Media LLC},
   author={Hollands, S. and Ruan, W.},
   year={2002},
   month=Aug, pages={635–657} }

@book{GelfandShilov2016,
  author    = {Gelfand, I. M. and Shilov, G. E.},
  title     = {Generalized Functions, Volume 1: Properties and Operations},
  publisher = {American Mathematical Society},
  series    = {AMS Chelsea Publishing},
  address   = {Providence, RI},
  year      = {2016},
  isbn      = {9781470426583}
}

@article{MatheusAdamEdu2026,
author  = {Zambianco, Matheus H. and Teixidó-Bonfill, Adam and Martín-Martínez, Eduardo},
title   = {Cosmological Expansion Induces Interference Between Communication and Entanglement Harvesting},
journal = {Phys. Rev. D},
volume  = {113},
number  = {8},
pages   = {085023},
year    = {2026}
}

@article{Adametalexperimental,
  title = {Towards an experimental implementation of entanglement harvesting in superconducting circuits: Effect of detector gap variation on entanglement harvesting},
  author = {Teixid\'o-Bonfill, Adam and Dai, Xi and Lupascu, Adrian and Mart\'{\i}n-Mart\'{\i}nez, Eduardo},
  journal = {Phys. Rev. A},
  volume = {113},
  issue = {4},
  pages = {043732},
  numpages = {36},
  year = {2026},
  month = {Apr},
  publisher = {American Physical Society},
  doi = {10.1103/wv9n-k3jj},
  url = {https://link.aps.org/doi/10.1103/wv9n-k3jj}
}

@article{WarpDriveAlcubierre,
doi = {10.1088/0264-9381/11/5/001},
url = {https://dx.doi.org/10.1088/0264-9381/11/5/001},
year = {1994},
month = {may},
publisher = {},
volume = {11},
number = {5},
pages = {L73},
author = {Miguel Alcubierre},
title = {The warp drive: hyper-fast travel within general relativity},
journal = {Class. Quantum Grav.}
}

@article{AdamEduderivativeharvesting,
  title = {Derivative coupling enables genuine entanglement harvesting in causal communication},
  author = {Teixid\'o-Bonfill, Adam and Mart\'{\i}n-Mart\'{\i}nez, Eduardo},
  journal = {Phys. Rev. D},
  volume = {110},
  issue = {10},
  pages = {105016},
  numpages = {8},
  year = {2024},
  month = {Nov},
  publisher = {American Physical Society},
  doi = {10.1103/PhysRevD.110.105016},
  url = {https://link.aps.org/doi/10.1103/PhysRevD.110.105016}
}

@article{SachsMannEdu,
  title = {Entanglement harvesting and divergences in quadratic Unruh-DeWitt detector pairs},
  author = {Sachs, Allison and Mann, Robert B. and Mart\'{\i}n-Mart\'{\i}nez, Eduardo},
  journal = {Phys. Rev. D},
  volume = {96},
  issue = {8},
  pages = {085012},
  numpages = {17},
  year = {2017},
  month = {Oct},
  publisher = {American Physical Society},
  doi = {10.1103/PhysRevD.96.085012},
  url = {https://link.aps.org/doi/10.1103/PhysRevD.96.085012}
}

@article{Jorma,
	doi = {10.1088/0264-9381/23/22/015},
	url = {https://doi.org/10.10882F0264-93812F232F222F015},
	year = 2006,
	month = {oct},
	publisher = {{IOP} Publishing},
	volume = {23},
	number = {22},
	pages = {6321--6343},
	author = {Jorma Louko and Alejandro Satz},
	title = {How often does the {U}nruh-{{D}e{W}itt} detector click? Regularization by a spatial profile},
	journal = {	Class. Quantum Gravity}
}

@article{EduTalesBruno2021,
  title = {Broken covariance of particle detector models in relativistic quantum information},
  author = {Mart\'{\i}n-Mart\'{\i}nez, Eduardo and Perche, T. Rick and Torres, Bruno de S. L.},
  journal = {Phys. Rev. D},
  volume = {103},
  issue = {2},
  pages = {025007},
  numpages = {14},
  year = {2021},
  month = {Jan},
  publisher = {American Physical Society},
  doi = {10.1103/PhysRevD.103.025007},
  url = {https://link.aps.org/doi/10.1103/PhysRevD.103.025007}
}

@article{FordRoman1997,
  author  = {Ford, L. H. and Roman, Thomas A.},
  title   = {Restrictions on Negative Energy Density in Flat Spacetime},
  journal = {Phys. Rev. D},
  volume  = {55},
  pages   = {2082--2089},
  year    = {1997},
  doi     = {10.1103/PhysRevD.55.2082}
}

@article{EGlaserJaffe1965,
  author  = {Epstein, H. and Glaser, V. and Jaffe, A.},
  title   = {Nonpositivity of the Energy Density in Quantized Field Theories},
  journal = {Nuovo Cim.},
  volume  = {36},
  pages   = {1016--1022},
  year    = {1965},
  doi     = {10.1007/BF02749799}
}

@article{FORD_2010,
   title={NEGATIVE ENERGY DENSITIES IN QUANTUM FIELD THEORY},
   volume={25},
   ISSN={1793-656X},
   url={http://dx.doi.org/10.1142/S0217751X10049633},
   DOI={10.1142/s0217751x10049633},
   number={11},
   journal={Int. J. Mod. Phys. A},
   publisher={World Scientific Pub Co Pte Lt},
   author={Ford, L. H.},
   year={2010},
   month=Apr, pages={2355–2363} }

@misc{MatheusEduFirewallArxiv,
      title={Nonlinear particle detectors across the Rindler firewall}, 
      author={Matheus H. Zambianco and Eduardo Martín-Martínez},
      year={2026},
      eprint={2607.09660},
      archivePrefix={arXiv},
      primaryClass={quant-ph},
      url={https://arxiv.org/abs/2607.09660}, 
}

@article{fewster1,
title = {Quantum Fields and Local Measurements},
author = {Fewster, Christopher J. and Verch, Rainer},
journal = {Commun. Math. Phys.},
volume = {378},
number = {2},
pages = {851--889},
year = {2020},
month = sep,
doi = {10.1007/s00220-020-03800-6},
url = {https://doi.org/10.1007/s00220-020-03800-6}
}

@article{VidalNegativity,
  title = {Computable measure of entanglement},
  author = {Vidal, G. and Werner, R. F.},
  journal = {Phys. Rev. A},
  volume = {65},
  issue = {3},
  pages = {032314},
  numpages = {11},
  year = {2002},
  month = {Feb},
  publisher = {American Physical Society},
  doi = {10.1103/PhysRevA.65.032314},
  url = {https://link.aps.org/doi/10.1103/PhysRevA.65.032314}
}

@article{ericksonNew,
  title = {When entanglement harvesting is not really harvesting},
  author = {Tjoa, Erickson and Mart\'{\i}n-Mart\'{\i}nez, Eduardo},
  journal = {Phys. Rev. D},
  volume = {104},
  issue = {12},
  pages = {125005},
  numpages = {21},
  year = {2021},
  month = {Dec},
  publisher = {American Physical Society},
  doi = {10.1103/PhysRevD.104.125005},
  url = {https://link.aps.org/doi/10.1103/PhysRevD.104.125005}
}

@article{boris,
  title = {Harvesting entanglement from the gravitational vacuum},
  author = {Perche, T. Rick and Ragula, Boris and Mart\'{\i}n-Mart\'{\i}nez, Eduardo},
  journal = {Phys. Rev. D},
  volume = {108},
  issue = {8},
  pages = {085025},
  numpages = {58},
  year = {2023},
  month = {Oct},
  publisher = {American Physical Society},
  doi = {10.1103/PhysRevD.108.085025},
  url = {https://link.aps.org/doi/10.1103/PhysRevD.108.085025}
}

@article{fewsterNecessityHadamard,
	doi = {10.1088/0264-9381/30/23/235027},
	url = {https://doi.org/10.1088/0264-9381/30/23/235027},
	year = 2013,
	month = {nov},
	publisher = {{IOP} Publishing},
	volume = {30},
	number = {23},
	pages = {235027},
	author = {Christopher J Fewster and Rainer Verch},
	title = {The necessity of the Hadamard condition},
	journal = {Class. Quantum Grav. }
}

@Article{ericksonBH,
author={Tjoa, Erickson
and Mann, Robert B.},
title={{Harvesting correlations in Schwarzschild and collapsing shell spacetimes}},
journal={Jour. High Energy Phys.},
year={2020},
month={Aug},
day={28},
volume={2020},
number={8},
pages={155},
issn={1029-8479},
doi={10.1007/JHEP08(2020)155},
url={https://doi.org/10.1007/JHEP08(2020)155}
}

@misc{sachs2018entanglement,
      title={Entanglement harvesting from multiple massless scalar fields and divergences in {U}nruh-{D}e{W}itt detector models}, 
      author={Allison M. Sachs and Robert B. Mann and Eduardo Martin-Martinez},
      year={2018},
      eprint={1808.05980},
      archivePrefix={arXiv},
      primaryClass={quant-ph}
}

@article{Sabin2010Dynamics,
  title = {Dynamics of entanglement via propagating microwave photons},
  author = {Sab\'{\i}n, C. and Garc\'{\i}a-Ripoll, J. J. and Solano, E. and Le\'on, J.},
  journal = {Phys. Rev. B},
  volume = {81},
  issue = {18},
  pages = {184501},
  numpages = {6},
  year = {2010},
  month = {May},
  publisher = {American Physical Society},
  doi = {10.1103/PhysRevB.81.184501},
  url = {https://link.aps.org/doi/10.1103/PhysRevB.81.184501}
}

@article{HarvestingGravitationalWaves,
   title={Harvesting entanglement from cylindrical gravitational wave spacetime},
   volume={49},
   ISSN={2058-6132},
   url={http://dx.doi.org/10.1088/1674-1137/adcc89},
   DOI={10.1088/1674-1137/adcc89},
   number={8},
   journal={Chin. Phys. C},
   publisher={IOP Publishing},
   author={He, Feifan and Pan, Yongjie and Zhang, Baocheng},
   year={2025},
   month=Aug, pages={085103} }

@article{FermionicHarvestingSCZ,
   title={Harvesting fermionic field entanglement in Schwarzschild spacetime},
   volume={112},
   ISSN={2470-0029},
   url={http://dx.doi.org/10.1103/1mwx-7jmf},
   number={2},
   journal={Phys. Rev. D},
   publisher={American Physical Society (APS)},
   author={Dubey, Nitesh K. and Kolekar, Sanved},
   year={2025},
   month=July }

@article{HarvestingMultiqubitCQE,
  title = {Harvesting Multiqubit Entanglement from Ultrastrong Interactions in Circuit Quantum Electrodynamics},
  author = {Armata, F. and Calajo, G. and Jaako, T. and Kim, M. S. and Rabl, P.},
  journal = {Phys. Rev. Lett.},
  volume = {119},
  issue = {18},
  pages = {183602},
  numpages = {6},
  year = {2017},
  month = {Nov},
  publisher = {American Physical Society},
  doi = {10.1103/PhysRevLett.119.183602},
  url = {https://link.aps.org/doi/10.1103/PhysRevLett.119.183602}
}

@article{Sabin2011Fermi,
  title = {Fermi problem with artificial atoms in circuit {QED}},
  author = {Sab{\'i}n, C. and del Rey, M. and Garc{\'i}a-Ripoll, J. J. and Le{\'o}n, J.},
  journal = {Phys. Rev. Letters},
  volume = {107},
  pages = {150402},
  year = {2011},
  doi = {10.1103/PhysRevLett.107.150402}
}

@article{Sabin2012PastFuture,
  title = {Extracting Past-Future Vacuum Correlations Using Circuit {QED}},
  author = {Sab{\'i}n, C. and Peropadre, B. and del Rey, M. and Mart{\'i}n-Mart{\'i}nez, E.},
  journal = {Phys. Rev. Letters},
  volume = {109},
  pages = {033602},
  year = {2012},
  doi = {10.1103/PhysRevLett.109.033602}
}

@article{Petar,
  title = {Harvesting correlations from thermal and squeezed coherent states},
  author = {Simidzija, Petar and Mart\'{\i}n-Mart\'{\i}nez, Eduardo},
  journal = {Phys. Rev. D},
  volume = {98},
  issue = {8},
  pages = {085007},
  numpages = {17},
  year = {2018},
  month = {Oct},
  publisher = {American Physical Society},
  doi = {10.1103/PhysRevD.98.085007},
  url = {https://link.aps.org/doi/10.1103/PhysRevD.98.085007}
}

@article{hectorMass,
  title = {Entanglement harvesting: Detector gap and field mass optimization},
  author = {Maeso-Garc\'{\i}a, H\'ector and Perche, T. Rick and Mart\'{\i}n-Mart\'{\i}nez, Eduardo},
  journal = {Phys. Rev. D},
  volume = {106},
  issue = {4},
  pages = {045014},
  numpages = {15},
  year = {2022},
  month = {Aug},
  publisher = {American Physical Society},
  doi = {10.1103/PhysRevD.106.045014},
  url = {https://link.aps.org/doi/10.1103/PhysRevD.106.045014}
}

@article{Pozas2016,
  title = {Entanglement harvesting from the electromagnetic vacuum with hydrogenlike atoms},
  author = {Pozas-Kerstjens, Alejandro and Mart\'{i}n-Mart\'{i}nez, Eduardo},
  journal = {Phys. Rev. D},
  volume = {94},
  issue = {6},
  pages = {064074},
  numpages = {27},
  year = {2016},
  month = {Sep},
  publisher = {American Physical Society},
  doi = {10.1103/PhysRevD.94.064074},
  url = {https://link.aps.org/doi/10.1103/PhysRevD.94.064074}
}

@article{Nick,
	Author = {Greg VerSteeg and Nicolas C. Menicucci},
	Journal = {Phys. Rev. D},
	Pages = {044027},
	Title = {Entangling power of an expanding universe},
	Volume = {79},
	Year = {2009}}

@article{Pozas-Kerstjens:2015,
	Author = {Pozas-Kerstjens, Alejandro and Mart\'{i}n-Mart\'{i}nez, Eduardo},
	Doi = {10.1103/PhysRevD.92.064042},
	Issue = {6},
	Journal = {Phys. Rev. D},
	Month = {Sep},
	Numpages = {18},
	Pages = {064042},
	Publisher = {American Physical Society},
	Title = {Harvesting correlations from the quantum vacuum},
	Url = {http://link.aps.org/doi/10.1103/PhysRevD.92.064042},
	Volume = {92},
	Year = {2015}}

@article{HarvestingBHLaura,
	doi = {10.1088/1361-6382/aae27e},
	url = {https://doi.org/10.1088%2F1361-6382%2Faae27e},
	year = 2018,
	month = {oct},
	publisher = {{IOP} Publishing},
	volume = {35},
	number = {21},
	pages = {21LT02},
	author = {Laura J Henderson and Robie A Hennigar and Robert B Mann and Alexander R H Smith and Jialin Zhang},
	title = {Harvesting entanglement from the black hole vacuum},
	journal = {Class. Quantum Gravity}
}

@article{Reznik2005BellInequalities,
	Author = {Benni Reznik and Alex Retzker and Jonathan Silman},
	Eid = {042104},
	Journal = {Phys. Rev. A},
	Number = {4},
	Numpages = {4},
	Pages = {042104},
	Publisher = {APS},
	Title = {Violating Bell's inequalities in vacuum},
	Url = {http://link.aps.org/abstract/PRA/v71/e042104},
	Volume = {71},
	Year = {2005}}

@article{Salton:2014jaa,
	Author = {Salton, Grant and Mann, Robert B. and Menicucci, Nicolas C.},
	Doi = {10.1088/1367-2630/17/3/035001},
	Journal = {New J. Phys.},
	Number = {3},
	Pages = {035001},
	Title = {{Acceleration-assisted entanglement harvesting and rangefinding}},
	Volume = {17},
	Year = {2015}}

@article{Valentini1991,
	Author = {Antony Valentini},
	Doi = {http://dx.doi.org/10.1016/0375-9601(91)90952-5},
	Issn = {0375-9601},
	Journal = {Phys. Lett. A},
	Number = {6-7},
	Pages = {321 - 325},
	Title = {Non-local correlations in quantum electrodynamics},
	Url = {http://www.sciencedirect.com/science/article/pii/0375960191909525},
	Volume = {153},
	Year = {1991}}

@article{Reznik2003,
	Author = {Reznik, Benni},
	Doi = {10.1023/A:1022875910744},
	Issn = {0015-9018},
	Journal = {Found. Phys.},
	Language = {English},
	Number = {1},
	Pages = {167-176},
	Publisher = {Kluwer Academic Publishers-Plenum Publishers},
	Title = {Entanglement from the Vacuum},
	Url = {http://dx.doi.org/10.1023/A%3A1022875910744},
	Volume = {33},
	Year = {2003}}

@Article{fullingHadamard,
author={Fulling, Stephen A.
and Sweeny, Mark
and Wald, Robert M.},
title={Singularity structure of the two-point function in quantum field theory in curved spacetime},
journal={Commun. Math. Phys},
year={1978},
month={Oct},
day={01},
volume={63},
number={3},
pages={257-264},
issn={1432-0916},
doi={10.1007/BF01196934},
url={https://doi.org/10.1007/BF01196934}
}

@article{KeithRobEdu2018,
  title = {New techniques for entanglement harvesting in flat and curved spacetimes},
  author = {Ng, Keith K. and Mann, Robert B. and Mart\'{\i}n-Mart\'{\i}nez, Eduardo},
  journal = {Phys. Rev. D},
  volume = {97},
  issue = {12},
  pages = {125011},
  numpages = {8},
  year = {2018},
  month = {Jun},
  publisher = {American Physical Society},
  doi = {10.1103/PhysRevD.97.125011},
  url = {https://link.aps.org/doi/10.1103/PhysRevD.97.125011}
}

@misc{wurtz2026,
      title         = {Entanglement harvesting in conformal field theory},
      author        = {Kelly Wurtz and Caroline Lima and Robert C. Myers and Eduardo Martín-Martínez},
      year          = {2026},
      eprint        = {2602.07112},
      archivePrefix = {arXiv},
      primaryClass  = {quant-ph},
      url           = {https://arxiv.org/abs/2602.07112},
      note          = {In press in J. High Energy Phys.},
}

@article{Takagi,
    author = {Takagi, Shin},
    title = {On the Response of a Rindler Particle Detector. III},
    journal = {Progress of Theoretical Physics},
    volume = {74},
    number = {3},
    pages = {501-510},
    year = {1985},
    month = {09},
    issn = {0033-068X},
    doi = {10.1143/PTP.74.501},
    url = {https://doi.org/10.1143/PTP.74.501},
    eprint = {https://academic.oup.com/ptp/article-pdf/74/3/501/5353230/74-3-501.pdf},
}

@article{ampEntBH2020,
doi = {10.1088/1361-6382/ac08a8},
url = {https://dx.doi.org/10.1088/1361-6382/ac08a8},
year = {2021},
month = {dec},
publisher = {IOP Publishing},
volume = {39},
number = {2},
pages = {02LT01},
author = {Matthew P G Robbins and Laura J Henderson and Robert B Mann},
title = {Entanglement amplification from rotating black holes},
journal = {Class. Quantum Grav.}
}

@article{Membrere_2023,
author  = {Membrere, Ireneo James and Gallock-Yoshimura, Kensuke and Henderson, Laura J. and Mann, Robert B.},
title   = {Tripartite Entanglement Extraction from the Black Hole Vacuum},
journal = {Adv. Quantum Technol.},
volume  = {6},
number  = {9},
pages   = {2300125},
year    = {2023},
doi     = {10.1002/qute.202300125}
}

@misc{morotebalboa2026,
      title={Optimization of entanglement harvesting with arbitrary temporal profiles: the limit of second order perturbation theory}, 
      author={Marcos Morote-Balboa and T. Rick Perche},
      year={2026},
      eprint={2604.06303},
      archivePrefix={arXiv},
      primaryClass={quant-ph},
      url={https://arxiv.org/abs/2604.06303}, 
}

@article{PhysRevA.88.062336,
  title = {Thermal amplification of field-correlation harvesting},
  author = {Brown, Eric G.},
  journal = {Phys. Rev. A},
  volume = {88},
  issue = {6},
  pages = {062336},
  numpages = {15},
  year = {2013},
  month = {Dec},
  publisher = {American Physical Society},
  doi = {10.1103/PhysRevA.88.062336},
  url = {https://link.aps.org/doi/10.1103/PhysRevA.88.062336}
}

@misc{TalesFrancescoMarkus,
      title={Bose polarons as relativistic Unruh-DeWitt detectors: Entanglement harvesting from Bose-Einstein condensates}, 
      author={T. Rick Perche and Francesco Gozzini and Markus K. Oberthaler},
      year={2026},
      eprint={2512.21381},
      archivePrefix={arXiv},
      primaryClass={quant-ph},
      url={https://arxiv.org/abs/2512.21381}, 
}

@article{Gooding_2024,
author  = {Gooding, Cisco and Sachs, Allison and Mann, Robert B. and Weinfurtner, Silke},
title   = {Vacuum Entanglement Probes for Ultracold Atom Systems},
journal = {New J. Phys.},
volume  = {26},
number  = {10},
pages   = {105001},
year    = {2024}
}

@article{GrimmerBrunoEdu_2021,
author  = {Grimmer, Daniel and Torres, Bruno de S. L. and Martín-Martínez, Eduardo},
title   = {Measurements in {QFT}: Weakly Coupled Local Particle Detectors and Entanglement Harvesting},
journal = {Phys. Rev. D},
volume  = {104},
number  = {8},
pages   = {085014},
year    = {2021},
doi     = {10.1103/PhysRevD.104.085014}
}

@misc{NIST:DLMF,
  key = "{\relax DLMF}",
  title = "{\it NIST Digital Library of Mathematical Functions}",
  howpublished = "\url{https://dlmf.nist.gov/}, Release 1.2.6 of 2026-03-15",
  url = "https://dlmf.nist.gov/",
  note = "F.~W.~J. Olver, A.~B. {Olde Daalhuis}, D.~W. Lozier, B.~I. Schneider,
  R.~F. Boisvert, C.~W. Clark, B.~R. Miller, B.~V. Saunders,
  H.~S. Cohl, and M.~A. McClain, eds."
}

@article{Kubo,
    author = "Kubo, Ryogo",
    title = "{Statistical mechanical theory of irreversible processes. 1. General theory and simple applications in magnetic and conduction problems}",
    doi = "10.1143/JPSJ.12.570",
    journal = "J. Phys. Soc. Jap.",
    volume = "12",
    pages = "570--586",
    year = "1957"
}

@article{MartinSchwinger,
  title = {Theory of Many-Particle Systems. I},
  author = {Martin, Paul C. and Schwinger, Julian},
  journal = {Phys. Rev.},
  volume = {115},
  issue = {6},
  pages = {1342--1373},
  numpages = {0},
  year = {1959},
  month = {Sep},
  publisher = {American Physical Society},
  doi = {10.1103/PhysRev.115.1342},
  url = {https://link.aps.org/doi/10.1103/PhysRev.115.1342}
}

@article{Unruh-Wald,
	Author = {Unruh, William G. and Wald, Robert M.},
	Doi = {10.1103/PhysRevD.29.1047},
	Issue = {6},
	Journal = {Phys. Rev. D},
	Month = {Mar},
	Pages = {1047--1056},
	Publisher = {American Physical Society},
	Title = {What happens when an accelerating observer detects a Rindler particle},
	Volume = {29},
	Year = {1984}
}

@article{Schlicht,
	doi = {10.1088/0264-9381/21/19/011},
	url = {https://doi.org/10.1088%2F0264-9381%2F21%2F19%2F011},
	year = 2004,
	month = {sep},
	publisher = {{IOP} Publishing},
	volume = {21},
	number = {19},
	pages = {4647--4660},
	author = {Sebastian Schlicht},
	title = {Considerations on the {U}nruh effect: causality and regularization},
	journal = {Class. Quantum Gravity}}

@article{HadamardBible,
   title={Hadamard renormalization of the stress-energy tensor for a quantized scalar field in a general spacetime of arbitrary dimension},
   volume={78},
   ISSN={1550-2368},
   url={http://dx.doi.org/10.1103/PhysRevD.78.044025},
   number={4},
   journal={Phys. Rev. D},
   publisher={American Physical Society (APS)},
   author={Décanini, Yves and Folacci, Antoine},
   year={2008},
   month=aug }

@article{carol,
  title = {Harvesting entanglement from complex scalar and fermionic fields with linearly coupled particle detectors},
  author = {Perche, T. Rick and Lima, Caroline and Mart\'{\i}n-Mart\'{\i}nez, Eduardo},
  journal = {Phys. Rev. D},
  volume = {105},
  issue = {6},
  pages = {065016},
  numpages = {24},
  year = {2022},
  month = {Mar},
  publisher = {American Physical Society},
  doi = {10.1103/PhysRevD.105.065016},
  url = {https://link.aps.org/doi/10.1103/PhysRevD.105.065016}
}

@article{B,
  title = {Spin Entanglement Witness for Quantum Gravity},
  author = {Bose, Sougato and Mazumdar, Anupam and Morley, Gavin W. and Ulbricht, Hendrik and Toro\ifmmode \check{s}\else \v{s}\fi{}, Marko and Paternostro, Mauro and Geraci, Andrew A. and Barker, Peter F. and Kim, M. S. and Milburn, Gerard},
  journal = {Phys. Rev. Lett.},
  volume = {119},
  issue = {24},
  pages = {240401},
  numpages = {6},
  year = {2017},
  month = {Dec},
  publisher = {American Physical Society},
  doi = {10.1103/PhysRevLett.119.240401},
  url = {https://link.aps.org/doi/10.1103/PhysRevLett.119.240401}
}

@article{MatheusAdamEdu2024,
  title = {Interference of communication and field correlations in entanglement harvesting},
  author = {Zambianco, Matheus H. and Teixid\'o-Bonfill, Adam and Mart\'{\i}n-Mart\'{\i}nez, Eduardo},
  journal = {Phys. Rev. D},
  volume = {110},
  issue = {2},
  pages = {025016},
  numpages = {25},
  year = {2024},
  month = {Jul},
  publisher = {American Physical Society},
  doi = {10.1103/PhysRevD.110.025016},
  url = {https://link.aps.org/doi/10.1103/PhysRevD.110.025016}
}

@article{derivativeJorma,
doi = {10.1088/0264-9381/31/24/245007},
url = {https://dx.doi.org/10.1088/0264-9381/31/24/245007},
year = {2014},
month = {nov},
publisher = {IOP Publishing},
volume = {31},
number = {24},
pages = {245007},
author = {Benito A Juárez-Aubry and Jorma Louko},
title = {{Onset and decay of the 1 + 1 Hawking–Unruh effect: what the derivative-coupling detector saw}},
journal = {Class. Quantum Grav.}
}

\end{document}